\documentclass{article}
\usepackage{psfig}
\usepackage{latexsym}
\begin{document}
\setlength{\oddsidemargin}{-3mm}
\setlength{\evensidemargin}{-3mm}
\newcommand {\ee}{\end{equation}}
\newcommand {\bea}{\begin{eqnarray}}
\newcommand {\eea}{\end{eqnarray}}
\newcommand {\nn}{\nonumber \\}
\newcommand {\Tr}{{\rm Tr\,}}
\newcommand {\tr}{{\rm tr\,}}
\newcommand {\e}{{\rm e}}
\newcommand {\etal}{{\it et al.}}
\newcommand {\m}{\mu}
\newcommand {\n}{\nu}
\newcommand {\pl}{\partial}
\newcommand {\p} {\phi}
\newcommand {\vp}{\varphi}
\newcommand {\vpc}{\varphi_c}
\newcommand {\al}{\alpha}
\newcommand {\be}{\beta}
\newcommand {\ga}{\gamma}
\newcommand {\Ga}{\Gamma}
\newcommand {\x}{\xi}
\newcommand {\ka}{\kappa}
\newcommand {\la}{\lambda}
\newcommand {\La}{\Lambda}
\newcommand {\si}{\sigma}
\newcommand {\Si}{\Sigma}
\newcommand {\Th}{\Theta}
\newcommand {\om}{\omega}
\newcommand {\Om}{\Omega}
\newcommand {\ep}{\epsilon}
\newcommand {\vep}{\varepsilon}
\newcommand {\na}{\nabla}
\newcommand {\del}  {\delta}
\newcommand {\Del}  {\Delta}
\newcommand {\mn}{{\mu\nu}}
\newcommand {\ls}   {{\lambda\sigma}}
\newcommand {\ab}   {{\alpha\beta}}
\newcommand {\gd}   {{\gamma\delta}}
\newcommand {\half}{ {\frac{1}{2}} }
\newcommand {\third}{ {\frac{1}{3}} }
\newcommand {\fourth} {\frac{1}{4} }
\newcommand {\sixth} {\frac{1}{6} }
\newcommand {\sqtwo} {\sqrt{2}}
\newcommand {\sqg} {\sqrt{g}}
\newcommand {\fg}  {\sqrt[4]{g}}
\newcommand {\invfg}  {\frac{1}{\sqrt[4]{g}}}
\newcommand {\sqZ} {\sqrt{Z}}
\newcommand {\sqk} {\sqrt{\kappa}}
\newcommand {\sqt} {\sqrt{t}}
\newcommand {\sql} {\sqrt{l}}
\newcommand {\reg} {\frac{1}{\epsilon}}
\newcommand {\fpisq} {(4\pi)^2}
\newcommand {\Acal}{{\cal A}}
\newcommand {\Lcal}{{\cal L}}
\newcommand {\Ocal}{{\cal O}}
\newcommand {\Dcal}{{\cal D}}
\newcommand {\Ncal}{{\cal N}}
\newcommand {\Mcal}{{\cal M}}
\newcommand {\scal}{{\cal s}}
\newcommand {\Dvec}{{\hat D}}   
\newcommand {\dvec}{{\vec d}}
\newcommand {\Evec}{{\vec E}}
\newcommand {\Hvec}{{\vec H}}
\newcommand {\Vvec}{{\vec V}}
\newcommand {\lsim}{
\vbox{\baselineskip=4pt \lineskiplimit=0pt \kern0pt 
\hbox{$<$}\hbox{$\sim$}}
                    }
\newcommand {\rpl}{{\vec \partial}}
\def\overleftarrow#1{\vbox{\ialign{##\crcr
 $\leftarrow$\crcr\noalign{\kern-1pt\nointerlineskip}
 $\hfil\displaystyle{#1}\hfil$\crcr}}}
\def\lpl{{\overleftarrow\partial}}
\newcommand {\Btil}{{\tilde B}}
\newcommand {\atil}{{\tilde a}}
\newcommand {\btil}{{\tilde b}}
\newcommand {\ctil}{{\tilde c}}
\newcommand {\dtil}{{\tilde d}}
\newcommand {\Ftil}{{\tilde F}}
\newcommand {\Ktil}  {{\tilde K}}
\newcommand {\Ltil}  {{\tilde L}}
\newcommand {\mtil}{{\tilde m}}
\newcommand {\ttil} {{\tilde t}}
\newcommand {\Qtil}  {{\tilde Q}}
\newcommand {\Rtil}  {{\tilde R}}
\newcommand {\Stil}{{\tilde S}}
\newcommand {\Ztil}{{\tilde Z}}
\newcommand {\altil}{{\tilde \alpha}}
\newcommand {\betil}{{\tilde \beta}}
\newcommand {\etatil} {{\tilde \eta}}
\newcommand {\latil}{{\tilde \lambda}}
\newcommand {\Latil}{{\tilde \Lambda}}
\newcommand {\ptil}{{\tilde \phi}}
\newcommand {\Ptil}{{\tilde \Phi}}
\newcommand {\natil} {{\tilde \nabla}}
\newcommand {\xitil} {{\tilde \xi}}
\newcommand {\Hbtil} {{\Huge {\~{b}}}}
\newcommand {\Hctil} {{\Huge {\~{c}}}}
\newcommand {\Hdtil} {{\Huge {\~{d}}}}
\newcommand {\Ahat}{{\hat A}}
\newcommand {\ahat}{{\hat a}}
\newcommand {\Rhat}{{\hat R}}
\newcommand {\Shat}{{\hat S}}
\newcommand {\ehat}{{\hat e}}
\newcommand {\mhat}{{\hat m}}
\newcommand {\shat}{{\hat s}}
\newcommand {\Dhat}{{\hat D}}   
\newcommand {\Vhat}{{\hat V}}   
\newcommand {\xhat}{{\hat x}}
\newcommand {\Zhat}{{\hat Z}}
\newcommand {\Gahat}{{\hat \Gamma}}
\newcommand {\Phihat} {{\hat \Phi}}
\newcommand {\phihat} {{\hat \phi}}
\newcommand {\vphat} {{\hat \varphi}}
\newcommand {\nah} {{\hat \nabla}}
\newcommand {\etahat} {{\hat \eta}}
\newcommand {\omhat} {{\hat \omega}}
\newcommand {\psihat} {{\hat \psi}}
\newcommand {\thhat} {{\hat \theta}}
\newcommand {\gh}  {{\hat g}}
\newcommand {\abar}{{\bar a}}
\newcommand {\Abar}{{\bar A}}
\newcommand {\cbar}{{\bar c}}
\newcommand {\bbar}{{\bar b}}
\newcommand {\gbar}{\bar{g}}
\newcommand {\Bbar}{{\bar B}}
\newcommand {\fbar}{{\bar f}}
\newcommand {\Fbar}{{\bar F}}
\newcommand {\kbar}  {{\bar k}}
\newcommand {\Kbar}  {{\bar K}}
\newcommand {\Lbar}  {{\bar L}}
\newcommand {\Qbar}  {{\bar Q}}
\newcommand {\albar}{{\bar \alpha}}
\newcommand {\bebar}{{\bar \beta}}
\newcommand {\labar}{{\bar \lambda}}
\newcommand {\psibar}{{\bar \psi}}
\newcommand {\vpbar}{{\bar \varphi}}
\newcommand {\Psibar}{{\bar \Psi}}
\newcommand {\Phibar}{{\bar \Phi}}
\newcommand {\chibar}{{\bar \chi}}
\newcommand {\sibar}{{\bar \sigma}}
\newcommand {\xibar}{{\bar \xi}}
\newcommand {\thbar}{{\bar \theta}}
\newcommand {\Thbar}{{\bar \Theta}}
\newcommand {\bbartil}{{\tilde {\bar b}}}
\newcommand {\aldot}{{\dot{\alpha}}}
\newcommand {\bedot}{{\dot{\beta}}}
\newcommand {\alp}{{\alpha'}}
\newcommand {\bep}{{\beta'}}
\newcommand {\gap}{{\gamma'}}
\newcommand {\bfZ} {{\bf Z}}
\newcommand {\BFd} {{\bf d}}
\newcommand  {\vz}{{v_0}}
\newcommand  {\ez}{{e_0}}
\newcommand  {\mz}{{m_0}}
\newcommand  {\xf}{{x^5}}
\newcommand  {\yf}{{y^5}}
\newcommand  {\Zt}{{Z$_2$}}
\newcommand {\intfx} {{\int d^4x}}
\newcommand {\intdX} {{\int d^5X}}
\newcommand {\inttx} {{\int d^2x}}
\newcommand {\change} {\leftrightarrow}
\newcommand {\ra} {\rightarrow}
\newcommand {\larrow} {\leftarrow}
\newcommand {\ul}   {\underline}
\newcommand {\pr}   {{\quad .}}
\newcommand {\com}  {{\quad ,}}
\newcommand {\q}    {\quad}
\newcommand {\qq}   {\quad\quad}
\newcommand {\qqq}   {\quad\quad\quad}
\newcommand {\qqqq}   {\quad\quad\quad\quad}
\newcommand {\qqqqq}   {\quad\quad\quad\quad\quad}
\newcommand {\qqqqqq}   {\quad\quad\quad\quad\quad\quad}
\newcommand {\qqqqqqq}   {\quad\quad\quad\quad\quad\quad\quad}
\newcommand {\lb}    {\linebreak}
\newcommand {\nl}    {\newline}

\newcommand {\vs}[1]  { \vspace*{#1 cm} }

\newcommand {\MPL}  {Mod.Phys.Lett.}
\newcommand {\NP}   {Nucl.Phys.}
\newcommand {\PL}   {Phys.Lett.}
\newcommand {\PR}   {Phys.Rev.}
\newcommand {\PRL}   {Phys.Rev.Lett.}
\newcommand {\IJMP}  {Int.Jour.Mod.Phys.}
\newcommand {\CMP}  {Commun.Math.Phys.}
\newcommand {\JMP}  {Jour.Math.Phys.}
\newcommand {\AP}   {Ann.of Phys.}
\newcommand {\PTP}  {Prog.Theor.Phys.}
\newcommand {\NC}   {Nuovo Cim.}
\newcommand {\CQG}  {Class.Quantum.Grav.}


\font\smallr=cmr5
\newcommand {\npl}  {{\frac{n\pi}{l}}}
\newcommand {\mpl}  {{\frac{m\pi}{l}}}
\newcommand {\kpl}  {{\frac{k\pi}{l}}}

\def\ocirc#1{#1^{^{{\hbox{\smallr\llap{o}}}}}}
\def\ogamma{\ocirc{\gamma}{}}
\def\oM{{\buildrel {\hbox{\smallr{o}}} \over M}}
\def\osigma{\ocirc{\sigma}{}}

\def\overleftrightarrow#1{\vbox{\ialign{##\crcr
 $\leftrightarrow$\crcr\noalign{\kern-1pt\nointerlineskip}
 $\hfil\displaystyle{#1}\hfil$\crcr}}}
\def\overnab{{\overleftrightarrow\nabslash}}

\def\va{{a}}
\def\vb{{b}}
\def\vc{{c}}
\def\tilpsi{{\tilde\psi}}
\def\tbpsi{{\tilde{\bar\psi}}}

\def\delL{{\delta_{LL}}}
\def\delG{{\delta_{G}}}
\def\delc{{\delta_{cov}}}

\newcommand {\sqxx}  {\sqrt {x^2+1}}   
\newcommand {\gago}  {\gamma^5}
\newcommand {\Pp}  {P_+}
\newcommand {\Pm}  {P_-}
\newcommand {\GfMp}  {G^{5M}_+}
\newcommand {\GfMpm}  {G^{5M'}_-}
\newcommand {\GfMm}  {G^{5M}_-}
\newcommand {\Omp}  {\Omega_+}    
\newcommand {\Omm}  {\Omega_-}
\def\Aslash{{}\hbox{\hskip2pt\vtop
 {\baselineskip23pt\hbox{}\vskip-24pt\hbox{/}}
 \hskip-11.5pt $A$}}
\def\Rslash{{}\hbox{\hskip2pt\vtop
 {\baselineskip23pt\hbox{}\vskip-24pt\hbox{/}}
 \hskip-11.5pt $R$}}

\def\kslash{
{}\hbox       {\hskip2pt\vtop
                   {\baselineskip23pt\hbox{}\vskip-24pt\hbox{/}}
               \hskip-8.5pt $k$}
           }    
\def\qslash{
{}\hbox       {\hskip2pt\vtop
                   {\baselineskip23pt\hbox{}\vskip-24pt\hbox{/}}
               \hskip-8.5pt $q$}
           }    
\def\dslash{
{}\hbox       {\hskip2pt\vtop
                   {\baselineskip23pt\hbox{}\vskip-24pt\hbox{/}}
               \hskip-8.5pt $\partial$}
           }    
\def\dbslash{{}\hbox{\hskip2pt\vtop
 {\baselineskip23pt\hbox{}\vskip-24pt\hbox{$\backslash$}}
 \hskip-11.5pt $\partial$}}

\def\Kbslash{{}\hbox{\hskip2pt\vtop
 {\baselineskip23pt\hbox{}\vskip-24pt\hbox{$\backslash$}}
 \hskip-11.5pt $K$}}
\def\Ktilbslash{{}\hbox{\hskip2pt\vtop
 {\baselineskip23pt\hbox{}\vskip-24pt\hbox{$\backslash$}}
 \hskip-11.5pt ${\tilde K}$}}
\def\Ltilbslash{{}\hbox{\hskip2pt\vtop
 {\baselineskip23pt\hbox{}\vskip-24pt\hbox{$\backslash$}}
 \hskip-11.5pt ${\tilde L}$}}
\def\Qtilbslash{{}\hbox{\hskip2pt\vtop
 {\baselineskip23pt\hbox{}\vskip-24pt\hbox{$\backslash$}}
 \hskip-11.5pt ${\tilde Q}$}}
\def\Rtilbslash{{}\hbox{\hskip2pt\vtop
 {\baselineskip23pt\hbox{}\vskip-24pt\hbox{$\backslash$}}
 \hskip-11.5pt ${\tilde R}$}}
\def\Kbarbslash{{}\hbox{\hskip2pt\vtop
 {\baselineskip23pt\hbox{}\vskip-24pt\hbox{$\backslash$}}
 \hskip-11.5pt ${\bar K}$}}
\def\Lbarbslash{{}\hbox{\hskip2pt\vtop
 {\baselineskip23pt\hbox{}\vskip-24pt\hbox{$\backslash$}}
 \hskip-11.5pt ${\bar L}$}}
\def\Rbarbslash{{}\hbox{\hskip2pt\vtop
 {\baselineskip23pt\hbox{}\vskip-24pt\hbox{$\backslash$}}
 \hskip-11.5pt ${\bar R}$}}
\def\Qbarbslash{{}\hbox{\hskip2pt\vtop
 {\baselineskip23pt\hbox{}\vskip-24pt\hbox{$\backslash$}}
 \hskip-11.5pt ${\bar Q}$}}
\def\Acalbslash{{}\hbox{\hskip2pt\vtop
 {\baselineskip23pt\hbox{}\vskip-24pt\hbox{$\backslash$}}
 \hskip-11.5pt ${\cal A}$}}

\begin{flushright}
October 2004\\
US-03-08\\
hep-th/0401011 \\
\end{flushright}

\vspace{0.5cm}

\begin{center}

{\Large\bf 
Quantum Dynamics \\
of \\
A Bulk-Boundary System
}

\vspace{1.5cm}
{\large Shoichi ICHINOSE
         \footnote{
E-mail address:\ ichinose@u-shizuoka-ken.ac.jp
                  }
}\ and\ 
{\large Akihiro MURAYAMA$^\ddag$
         \footnote{
E-mail address:\ edamura@ipc.shizuoka.ac.jp
                  }
}
\vspace{1cm}

{\large 
Laboratory of Physics, 
School of Food and Nutritional Sciences, 
University of Shizuoka, 
Yada 52-1, Shizuoka 422-8526, Japan
 }

$\mbox{}^\ddag${\large
Department of Physics, Faculty of Education, Shizuoka University,
Shizuoka 422-8529, Japan
}
\end{center}

\vfill

{\large Abstract}\nl
The quantum dynamics of a bulk-boundary theory
is closely examined by the use of the background
field method. As an example we take the 
Mirabelli-Peskin model, which is composed of
5D super Yang-Mills (bulk) and 4D Wess-Zumino
(boundary). {\it Singular interaction} terms play an 
important role of canceling the divergences
coming from the KK-mode sum. 
Some {\it new regularization} of the momentum integral
is proposed. 
An interesting background
configuration of scalar fields is found. 
It is a localized solution of the field equation. 
In this process of the {\it vacuum search}, 
we present a new treatment of 
the vacuum with respect to the {\it extra coordinate}. 
The "supersymmetric" effective potential is obtained
at the 1-loop full (w.r.t. the coupling) level. 
This is the bulk-boundary generalization of
the Coleman-Weinberg's case. 
Renormalization group analysis is done
and the correct 4D result is reproduced. 
The {\it Casimir energy} is calculated and is compared with the case of
the Kaluza-Klein model.

\vspace{1cm}

PACS NO:
\ 11.10.Kk,
\ 11.27.+d,
\ 12.60.Jv,
\ 12.10.-g,
\ 11.25.Mj,
\ 04.50.+h 



\section{Introduction}
Through recent several years of development, it looks that  
the higher-dimensional approach has obtained the citizenship 
as an important building tool in constructing the unified theories.
It appears with some names such as "Randall-Sundrum model", 
"brane world", "extra-dimension model", "orbifold model", etc.
Before the appearance of the new approach, supersymmetry 
(SUSY) was the main promising tool to
go beyond the standard model. Among many ideas in 
the higher-dimensional approach, 
the system of {\it bulk and boundary} theories becomes 
a fascinating model of the unification. 
A boundary is regarded as our world. It is inspired by 
the M, string and D-brane theories\cite{HW96}.
One pioneering paper, which concretely describes the model, 
is that by Mirabelli and Peskin\cite{MP97}. 
They take
the 5D supersymmetric Yang-Mills theory as a bulk theory
and make it couple with a boundary matter.
The boundary couplings (with the bulk world)
are uniquely fixed by the SUSY requirement. 
They demonstrated some consistency in the {\it bulk} quantum theory
by calculating {\it self-energy} of the scalar matter field.
Here we examine the {\it effective potential} 
and the {\it vacuum energy} of this system. 
We investigate further closely
the role of the bulk fields and the singular
interactions.

A field theoretical analysis of the bulk and boundary 
system was recently done in the work by 
Goldberger and Wise\cite{GW0104}. 
They try to tackle the problem by the "generalized" use of
the renormalization group. 
Randall and Schwarts\cite{RS0108a,RS0108b} also attacked the same problem
by introducing a special regularization of the
ultraviolet divergences guided by the idea of the holography. 
We take a different approach to the bulk and boundary system.
(We will see, however, some similar results.) 
It is, at present,  hard to show 
any consistency ( such as renormalizability,
unitarity, etc.) 
in the higher dimensional quantum field theory. 
It is, at least perturbatively, {\it unrenormalizable}. 
We would rather
regard the bulk world as an {\it external heat-reservoir}
which gives some "freedom" to the boundary world
and "define" or "regularize" the 4D dynamics.
The external world of the bulk {\it classically} and 
{\it quantumly} 
affect our world of 4D, and vice versa. 
In this circumstance
we focus on the renormalization properties of the 4D world.  
We examine a way to treat the linear (power) divergences coming from
the bulk quantum effect. 
Another important aspect 
is the present treatment of the extra axis.
We will find some ``freedom'' in the definition of the vacuum.
$Z^2$-symmetry plays an important role there. 
We can naturally introduce the singular behaviour for
some scalars.
One important merit of the bulk-boundary approach 
is that the anomaly phenomenon
(of the 4D world) is naturally accepted as a 
current flow which goes out through the wall or comes into 
the 4D world.\cite{CH85}.

Contrary to the motivation of 
the original work of ref.\cite{MP97}, 
we do {\it not} seek the SUSY breaking mechanism, rather
we keep the supersymmetry and 
make use of the SUSY invariant properties
in order to make the analysis as simple as possible.
The SUSY symmetry is so restrictive that we only
need to calculate some small portion of
all possible diagrams.

As the analysis of the effective potential
of the 5D model, we recall that of the Kaluza-Klein(KK) model\cite{AC83}. 
The dynamics quantumly produces the effective
potential which describes the Casimir effect. 
The situation, however, is contrastively 
different from the present case
in some points.\nl
1)\ The present approach realizes the 4D reduction
by the localized (along the extra axis) configuration
(kink, soliton, delta-function),
whereas KK does it by the shrinkage of the radius
of the extra S$_1$ space.\nl
2)\ KK does not use Z$_2$-symmetry whereas the present one
exploits it in order to make a singular structure
at $x^5=0,l$ (fixed points) where the 4D worlds are.
The discrete symmetry imposes a nontrivial boundary
condition on the vacuum.
\nl
3)\ KK takes the condition scalar field = constant 
in order to find the vacuum configuration,
whereas, in the present case, we do the {\it new} treatment of the vacuum
by allowing the extra-coordinate dependence on some scalars.\nl
4)\ The present model is supersymmetric,
 whereas KK is not. \nl
5)\ In KK, the scalar field comes from the (5,5)-component
of the 5D metric (and partially from the dilaton). 
The 5D quantum effect produces
the effective potential which can be interpreted as
the Casimir force induced by the vacuum polarization
between the $l$ separated objects. On the other hand,
in the present case, the scalar components come from
various places:\ 
the 5th component of the bulk vector, the bulk scalar
and the boundary scalar fields. Hence the vacuum structure
becomes much richer.
\nl
6)The present model has, as the characteristic length scales, 
the thickness parameter
(brane tension) besides  the period of the extra space. 
As the vacuum energy calculation, we should see
the dependence on both lengths.  

The present model shares common properties with those
of the RS-model in the points such as localization, $Z_2$-symmetry,
bulk-boundary relation, etc. 
(The comparative aspect of the KK model and the RS-model
is explained in ref.\cite{SI02PR}.) 
We could regard the present result about the Casimir energy as
some RS counter-part
of the result obtained by Appelquist and Chodos for
the case of the KK-model.

The concrete object we will obtain is the effective potential. 
The formalism itself is very orthodox. 
The new point is its application to the 5D bulk-boundary system.
The system is much extended from the ordinary field theory.
The effective potential is 
well-established in the field theory.
Especially in the middle of 70's much literature appeared. One of
the famous outcome is the Coleman-E.Weinberg potential\cite{CW73}.
In the SUSY theories, Miller proposed a useful method, 
called AFTM\cite{Mil83PL},  based on
the tadpole diagram method by S. Weinberg\cite{Wein73}. 
It was applied to unified models\cite{AM98IJMPA}.
We will take another formalism, the background
field method, by B.S. DeWitt\cite{DeW67} and G. t'Hooft\cite{tH73}.
\footnote{
The use of the background field method in the brane world analysis
is stressed by Randall and Schwarz\cite{RS0108a,RS0108b}.
They develop a perturbative treatment in the AdS$_5$ 5D bulk theory.
They try to solve the similar problems to the present ones.
Especially perturbative treatment, log versus power divergences,
regularization, renormalization group running of the coupling.
They do not use a SUSY theory. They focus on the bulk gauge field theory.
}
The new formulation of the bulk-boundary system is another aim of
this paper.

We summarize the new points as follows:\nl
(1)\ background-field formulation of the bulk-boundary theory,\nl
(2)\ $\del(0)$ singularity problem is solved,\nl
(3)\ a new proposal for resolving the UV-divergences in the 5D
quantum $S^1/Z_2$ orbifold theory,\nl
(4)\ new treatment of the vacuum in the presence of the extra-space,\nl
(5)\ Casimir energy calculation.

Some of the present results are briefly reported in ref.\cite{IMplb587}.

The paper is organized as follows. In Sec.2, we 
introduce the present formalism of the background field
method, in the analysis of the effective potential. The simple
model of Wess-Zumino is taken as an example. 
Here we explain the "supersymmetric" effective potential. 
Mirabelli-Peskin model is explained in Sec.3. It is 
a typical bulk-boundary model based on 5D SUSY. 
In Sec.4, we quantize the model
using the background field method. 
A new treatment of the background field, 
in relation to the extra coordinate, is presented. This leads to
an interesting background solution (vacuum) which describes
the field-localization. 
Feynman rules are obtained 
for the perturbative analysis in Sec.5. 
The {\it singular} vertices, which involve the delta function, 
appear. 
Some Feynman diagrams are explicitly calculated.
We take into account both bulk and boundary quantum effects.
We will find that the singular interaction terms play the role
of the "counter-terms" to cancel the divergences coming
from the KK-mode sum. 
In Sec.6, the mass matrix appearing in the 1-loop Lagrangian
is obtained. This is the preparation for the 1-loop full
calculation of the next section. 
Assumption of the form of the background field
about its extra-coordinate dependence is crucial for the
present analysis. In Sec.7, the effective potential is
obtained. Two typical cases, A and B, are considered. In Case A
we look at the potential from the vanishing vacuum of
the brane matter-field. The final 
form of the potential is similar to the 4D super QED. In the
intermediate stage, we find a {\it new} type Casimir energy
which is characteristic for the brane world. In Case B, 
we obtain the potential for the no brane configuration. 
In the intermediate stage,
we find the ordinary type Casimir energy. The effective 
potential has rich structure. 
We conclude in Sec.8. 
We relegate some important detailed explanation to three appendices. 
App. A treats the super QED which is a good reference point
in the analysis of the bulk-boundary theory in the text.
App. B provides the calculation of
the eigenvalues of the mass matrix of Sec.6.
The results are used in Sec.7.
App. C explains the concrete form of the present
background fields. They satisfy the field equation
with the required boundary condition.

\section{Effective Potential of Wess-Zumino Model}
In order to explain the background field approach
to obtain the effective potential, we take the simplest
4D SUSY theory, that is, the Wess-Zumino model:
\begin{eqnarray}
\Lcal[\psi,A,F;\la,m]=i\pl_m\psibar\sibar^m\psi
+\Abar\pl_m\pl^m A+\Fbar F\nn
+[m(AF-\half \psi\psi)+\frac{\la}{2}(AAF-\psi\psi A)+\mbox{h.c.}]
\com
\label{wz1}
\end{eqnarray}
where $(\eta^{mn})=\mbox{diag}(-1,1,1,1)$.
The notation is basically the same as the textbook by Wess-Bagger\cite{WB92}.
$\psi$ is a Majorana fermion, $A$ is a complex scalar field and $F$ is 
an (complex scalar) auxiliary field. The general background field method
\cite{DeW67, tH73, FadSlav91}
tells us that the (DeWitt-Wilsonian) {\it effective action} 
$S^{eff}[\xi,a,f]$ is given by
\begin{eqnarray}
\exp\{ iS^{eff}[\xi,a,f]\}=\int\Dcal\psi\Dcal A\Dcal F\nn
\times\exp i\intfx\left\{ \Lcal[\xi+\psi,a+A,f+F]
-\left.\frac{\del\Lcal}{\del\Phi^I} \right|_b\Phi^I
             \right\}
\com
\label{wz2}
\end{eqnarray}
where $(\Phi^I)\equiv (\psi,A,F)$ 
are the {\it quantum} fields 
and their {\it background} fields
$(\Phi^I)|_b\equiv (\xi,a,f)$. 
We define the {\it effective potential}
$V^{eff}$ as the {\it non-derivative} part of $S^{eff}$. A simple
and practical way to pick up the part is to consider
the case: 
\begin{eqnarray}
\xi=0\ ,\ a=\mbox{const.}\ ,\ 
f=\mbox{const.}\com
\label{wz2b}
\end{eqnarray}
where we put $\xi=0$ from the requirement of 
the Lorentz invariance of the vacuum
and "const." means a constant.
\begin{eqnarray}
\exp\{ -iV^{eff}[a,f]\}=
\exp i\{ -V^{eff}_0\}\nn
\times
\int\Dcal\psi\Dcal A\Dcal F 
\exp i\intfx\left\{ \Lcal_2+\mbox{order of }(\mbox{quant. field})^3\ 
             \right\}\com\nn
-V^{eff}_0=\fbar f+(ma+\frac{\la}{2}a^2)f+(m\abar+\frac{\la}{2}\abar^2)\fbar
\equiv\Lcal_0\com\nn
\left.\frac{\del\Lcal}{\del\Phi^I} \right|_b\Phi^I
=(\fbar+ma+\frac{\la}{2}a^2)F+f\chi A+\mbox{h.c.}\equiv\Lcal_1\ ,
                             \ \chi\equiv m+\la a\ ,\nn
\Lcal_2=i\pl_m\psibar\sibar^m\psi-\half (\chi\psi\psi+\chibar\psibar\psibar)
+\half Q^\dag MQ\com\nn
M=\left(
\begin{array}{cccc}
\Box & \la\fbar & 0 & \chibar \\
\la f & \Box & \chi & 0 \\
0 & \chibar & 1 & 0  \\
\chi & 0 & 0 & 1
\end{array}
\right)\com\q\Box=\pl_m\pl^m \com
\label{wz3}
\end{eqnarray}
where the scalar quantum fields are denoted
by the column matrix $Q$:\ 
$Q^T=(A,\Abar,F,\Fbar), Q^\dag=(\Abar,A,\Fbar,F)$. 
The matrix $M$ is the same as 
the matrix appearing in eq.(15) of Ref.\cite{Mil83PL}. 
There is the special case, called {\it on-shell}, of the background values $a,f$:\ 
\begin{eqnarray}
\fbar+ma+\frac{\la}{2}a^2=0\com\q
\chi f=(m+\la a)f=0
\com
\label{wz4}
\end{eqnarray}
which satisfies the field equation and makes 
$\Lcal_1$ vanish. When
the above background values ($a,f$) satisfy the 
on-shell condition above,
$V^{eff}_0$ reduces to
\begin{eqnarray}
V^{eff}_0|_{\mbox{on-shell}}=\fbar f \geq 0
\com
\label{wz4b}
\end{eqnarray}
which shows the {\it positive semi-definiteness}. 
This shows the characteristic aspect of the supersymmetric
configuration. 
The (classical) vacuum is given by:\ 
$f=\fbar=0,\ a(m+\frac{\la}{2}a)=0\ (a=0\ \mbox{or}\ a=
-\frac{2}{\la}m)$.
In the following, except when explicitly stated, 
we do {\it not} require the on-shell condition
(\ref{wz4}). We regard $a$ and $f$ not as specific constants
(specific vacuum) but as the general source (external) fields
appearing in the effective potential. 
It is {\it an} off-shell generalization but is the most natural
one based on the background field method.

Let us now evaluate the 1-loop quantum effect. First
we can integrate out the auxiliary quantum-fields $F$ and $\Fbar$
using a "squaring" equation:\ 
$\Fbar F+\chi AF+\chibar \Abar\Fbar
=(\Fbar+\chi A)(F+\chibar\Abar)-\chibar\chi\Abar A$. Then 
the quadratic-part Lagrangian $\Lcal_2$ reduces to
$\Lcal_2'$:\ 
\begin{eqnarray}
\Lcal_2'=i\pl_m\psibar_\aldot(\sibar^m)^{\aldot\be}\psi_\be
+\Abar\Box A\nn
-\half 
\left(\begin{array}{cc}
\psi^\al & \psibar_\aldot
\end{array}
\right)
\left(\begin{array}{cc}
\chi\del_\al^\be & 0 \\
0 & \chibar\del_\bedot^\aldot \end{array}
\right)
\left(\begin{array}{c}
\psi_\be \\
\psibar^\bedot \end{array}
\right)\ 
-\half 
\left(\begin{array}{cc}
\Abar & A \end{array}
\right)
{\bf M}
\left(\begin{array}{c}
A \\
\Abar \end{array}
\right) 
\com\nn
{\bf M}=
\left(\begin{array}{cc}
\chibar\chi & -\la\fbar \\
-\la f & \chibar\chi \end{array}
\right)
\pr\label{wz5}
\end{eqnarray}
The eigenvalues of ${\bf M}$ are given as
\begin{eqnarray}
m^2_+=\chibar\chi+\la\sqrt{\fbar f}\com\q
m^2_-=\chibar\chi-\la\sqrt{\fbar f}\pr
\label{wz6}
\end{eqnarray}
The contribution to the 1-loop effective potential
$V^{eff}_{1-loop}$, 
from the {\it bosonic} part ( scalar loop ), is evaluated as
\begin{eqnarray}
\int\Dcal\Abar\Dcal A\exp i\intfx \left\{
\Abar\Box A
-\half 
\left(\begin{array}{cc}
\Abar & A \end{array}
\right)
{\bf M}
\left(\begin{array}{c}
A \\
\Abar \end{array}
\right)                             \right\}\nn
=\left[
\det (\Box-m_+^2)(\Box-m_-^2)\right]^{-\half}
=\exp\left\{
-\half\Tr\sum_{i=+,-}\ln (1-\frac{m_i^2}{\Box})
\right\}\nn
=\exp \left\{ -i\intfx V^{eff}_{1-loop}\right\}
\pr\label{wz7}
\end{eqnarray}
The $V^{eff}_{1-loop}$ above 
lacks
the {\it fermionic} 1-loop contribution:
\begin{eqnarray}
\int\Dcal\psibar\Dcal\psi\exp i\intfx
               \left\{
i\pl_m\psibar_\aldot(\sibar^m)^{\aldot\be}\psi_\be
-\half 
\left(\begin{array}{cc}
\psi^\al & \psibar_\aldot
\end{array}
\right)
\left(\begin{array}{cc}
\chi\del_\al^\be & 0 \\
0 & \chibar\del_\bedot^\aldot \end{array}
\right)
\left(\begin{array}{c}
\psi_\be \\
\psibar^\bedot \end{array}
\right)\ 
               \right\}\nn
=[\det (\Box -\chibar\chi)]^{+1}
\pr\label{wz7b}
\end{eqnarray}

\begin{figure}
\centerline{
\psfig{figure=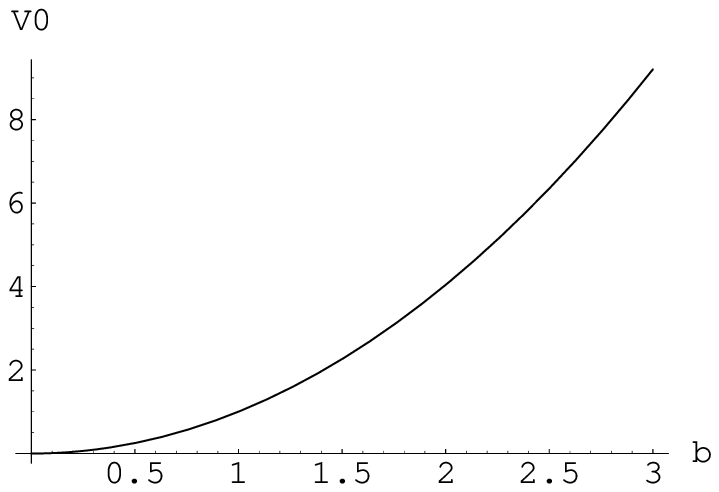,height=4cm,angle=0}
\psfig{figure=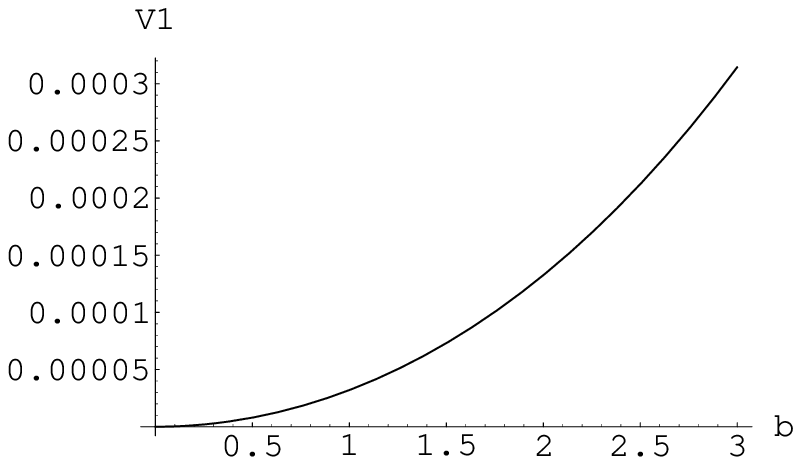,height=4cm,angle=0}
}
\centerline{
\psfig{figure=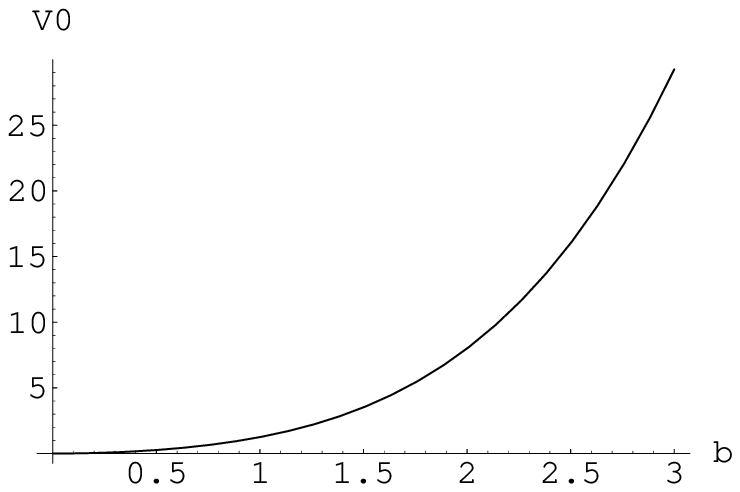,height=4cm,angle=0}
\psfig{figure=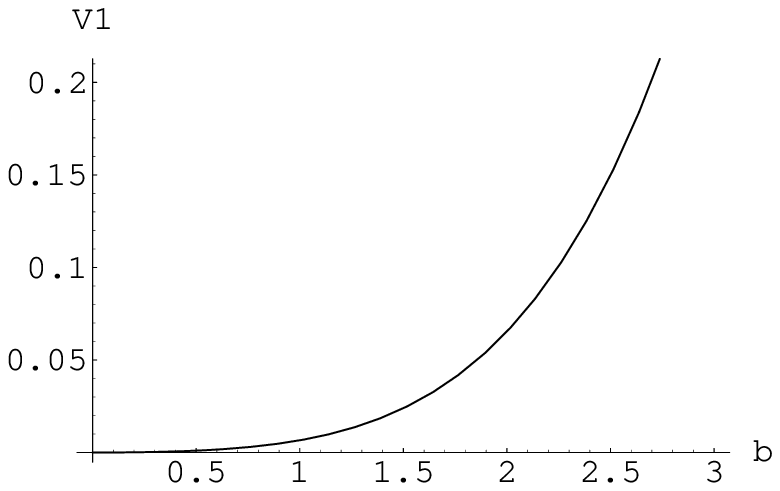,height=4cm,angle=0}
}
\caption{The effective potential of the Wess-Zumino model(\ref{wz11}). 
The case $a=i b$($b$: real) and the mass parameter $m=1$. The tree part 
($V^{eff}_0|_{\mbox{on-shell}}\equiv V_0$, left) and the 1-loop correction part 
($V^{1-loop}_R|_{\mbox{on-shell}}\equiv V_1$, right) are depicted for
the coupling parameter $\la=0.1$(up), $\la=1$(down). The horizontal axis is $b$.
The potentials are even functions of $b$.
}
\label{fig:WZ1}
\end{figure}

This part does {\it not} depend on $f$ and $\fbar$.
It says the 1-loop effective potential calculated
only by the scalar part is correct 
{\it up to the $f$-independent terms}.  
As far as the $f$-dependent part is concerned,
the scalar part result (\ref{wz7}) is sufficient. 
If we trace the source of this phenomenon, 
it is simply that
the auxiliary fields $f$ and $\fbar$ have the
{\it higher physical dimension}, $M^2$. They
{\it cannot}
 have the Yukawa coupling with fermions.
( $F\psi\psi$ has the mass dimension 5. )
This fact means that 
$dV^{eff}_{1-loop}/df$ ( or $dV^{eff}_{1-loop}/d\fbar$ )  
is definitely determined {\it only by the scalar part}.
Miller\cite{Mil83PL,Mil83NP} utilized this fact, that is, 
F-tadpole or D-tadpole \cite{Wein73} in general SUSY theories are rather
simply obtained. 
In the present case, (1-loop) F-tadpole corresponds to $dV^{eff}_{1-loop}/df$.
He noticed, if the SUSY is preserved in the quantization, 
the $f$-independent part can be fixed 
by the following {\it boundary condition}.
\footnote{
This reminds us of 
the similar situation of 2D WZNW model and 2D induced gravity. 
Polyakov and Wiegman\cite{PW83PLB}
obtained the former "effective action" not by integrating
the quantum field fluctuation but by solving 
the {\it chiral anomaly} equation in 2D QED. 
Polyakov\cite{Pol87MPL} obtained the latter "effective action"
by solving the {\it Weyl anomaly} equation. They treated
the 'gauge-field tadpole' ($\del\Ga/\del A_\mu$) 
and the 'Weyl-mode tadpole' 
($\del\Ga/\del \si=g^\mn\del\Ga/\del g^\mn$)
respectively.
          }
We follow Miller's idea.
Looking at the tree-level (on-shell) result (\ref{wz4b}), 
and taking into account the quantum stableness
of the SUSY theory, we are allowed to take the
{\it supersymmetric boundary condition}:\ 
\begin{eqnarray}
\mbox{The SUSY effective potential vanishes at}\ f=0. 
\label{wz7c}
\end{eqnarray}
Normalizing at $f=0$, 
the 1-loop effective potential is 
finally obtained as
\begin{eqnarray}
V^{eff}_{1-loop}-V^{eff}_{1-loop}|_{f=0}=
\half\int \frac{d^4k}{(2\pi)^4}\ln \left(
1-\frac{\la^2\fbar f}{(k^2+\chibar\chi)^2}
\right)\nn
\approx -\half\la^2\fbar f\int \frac{d^4k}{(2\pi)^4}
\frac{1}{(k^2+m^2)^2}+O(\la^3)
\pr\label{wz8}
\end{eqnarray}
In this simple example, we can explicitly see
the subtracting term $-V^{eff}_{1-loop}|_{f=0}$
is just given by the fermion-loop contribution (\ref{wz7b}).
The SUSY condition recovers the ignored (1-loop) contribution in $V^{eff}$.
The middle expression of (\ref{wz8}) 
is the same as eq.(26) of Ref.\cite{Mil83PL}
The {\it quadratic} divergences appearing in the intermediate stages,
as in (\ref{wz7}) and (\ref{wz7b}), cancel and the {\it logarithmic}
divergence only remain in the final expression (\ref{wz8}). 
It is absorbed by the {\it wave-function renormalization}
of the auxiliary fields as follows.
\footnote{
No divergences for the coupling operator, $\la AAF$, 
and the mass term, $mAF$, are consistent with the
{\it non-renormalization theorem}(see a textbook \cite{West90}). 
The F-term part does not receive radiative correction.
See, for example, the West's textbook\cite{West90}.
}
In order to do the renormalization, we first introduce
a {\it counterterm} $\Del\Lcal$ in the following form.
\begin{eqnarray}
V^{1-loop}_R\equiv V^{eff}_{1-loop}-V^{eff}_{1-loop}|_{f=0}
-\Del\Lcal\com\nn
V_R\equiv V^{eff}_0+V^{1-loop}_R
\com\q V^{eff}_0=-\fbar f+\cdots\nn
\Del\Lcal=\Del Z\,\fbar f
\com\label{wz8b}
\end{eqnarray}
where $Z\equiv 1+\Del Z$ is the wave function renormalization factor
of $f$ and $\fbar$. The 0th (classical) part, $V^{eff}_0$, is added.
Now we fix $\Del Z$ by {\it demanding} the following 
{\it renormalization condition}.
\footnote{
We follow ref.\cite{CW73} in the choice of the renormalization condition.
In (\ref{wz9}), by setting the coefficient in front of the term $\fbar f$ 
, appearing in the effective (renormalized) potential, we define
the present renormalization. 
No new mass parameter (such as the renormalization point) is introduced.
}
\begin{eqnarray}
-1\equiv \frac{dV_R}{d(\fbar f)}|_{f=\fbar=0, a=\abar=0}\nn
=-1-\half\la^2 
\int_{|k|\leq \La} \frac{d^4k}{(2\pi)^4}\frac{1}{(k^2+m^2)^2}+\Del Z\com\nn
\mbox{hence  }\q
Z\equiv 1+\Del Z=1-\frac{\la^2}{16\pi^2}\ln\frac{\La}{m}
\com
\label{wz9}
\end{eqnarray}
where $\La$ is the {\it momentum cut-off}- $|k^2|\leq \La^2$.
The {\it anomalous dimension} of the auxiliary field is given by
\begin{eqnarray}
F_b=\sqrt{Z}F\com\q\nn
\mbox{anomalous dimension }\ of\ F\ :\q
\ga_F=\frac{\pl}{\pl \ln \La}\ln Z=- \frac{\la^2}{16\pi^2}+O(\la^4)\pr
\label{wz10}
\end{eqnarray}
We see the quantum effect in the SUSY theory
apparently appears in the scaling behaviour of the 
{\it auxiliary field}.
It implies the structure(shape) of the {\it effective potential}
is very sensitive to the quantization. 
The final form, after the renormalization, is given by
\begin{eqnarray}
V_R|_{\mbox{on-shell}}=(V^{eff}_0+V^{1-loop}_R)|_{\mbox{on-shell}}
\com\nn
V^{eff}_0|_{\mbox{on-shell}}=\fbar f\com\nn
V^{1-loop}_R|_{\mbox{on-shell}}=
\frac{1}{64\pi^2}\left[ -\la^2\fbar f 
-(\chibar\chi)^2
\ln\left\{ \frac{(\chibar \chi)^2}
{(\chibar\chi-\la\sqrt{\fbar f})(\chibar\chi+\la\sqrt{\fbar f})}
    \right\}
                  \right.\nn
\left.+2\la\,\chibar\chi\,\sqrt{\fbar f}
\ln\left\{ \frac{\chibar\chi+\la\sqrt{\fbar f}}{\chibar\chi-\la\sqrt{\fbar f}}
   \right\} 
            +\la^2\fbar f\ln\left\{ 
\frac{(\chibar\chi-\la\sqrt{\fbar f})(\chibar\chi+\la\sqrt{\fbar f})}{m^4}
                            \right\}
\right]\com\nn
\mbox{where}\q\q 
\fbar f=|ma+\frac{\la}{2}a^2|^2\com\q
\chibar\chi=|m+\la a|^2
\pr\label{wz11}
\end{eqnarray}
For the pure imaginary case of $a=ib$ ($b$ is a real number),
the above potentials, 
$V^{eff}_0|_{\mbox{on-shell}}$ and 
$V^{1-loop}_R|_{\mbox{on-shell}}$, are depicted in Fig.\ref{fig:WZ1}.
In this case we have
$
\fbar f=m^2b^2+\frac{\la^2}{4}b^4\com\q
\chibar\chi=m^2+\la^2 b^2
$.
The precise shape of the quantum correction depends
on the renormalization condition. However
some characteristic features are considered meaningful. 
The shape of the 1-loop correction is {\it not}
the Coleman-Weinberg type.
The total shape of $V^{1-loop}_R|_{\mbox{on-shell}}$
is similar to the tree potential $V^{eff}_0|_{\mbox{on-shell}}$.
This shows that 
the SUSY invariant vacuum, $b=0$, is {\it stable} against the quantum effect.
\footnote{
Here we should be careful for the meaning of $V_R|_{\mbox{on-shell}}$. 
It is still an off-shell quantity in the sense that the true
vacuum is realized at $b=0$ only. Only at this vaccum, SUSY is preserved.
We call $V_R|_{\mbox{on-shell}}$ the SUSY-invariant effective action
because it is, at its vacuum, SUSY-invariant.
}
The {\it positive definiteness} is preserved after the
1-loop correction.
The form of the potential does not essentially change.
This result typically shows a general feature of SUSY theories.
It was confirmed before in the counter-term calculation\cite{BFMPS82}.

Super QED is similarly treated in Appendix A. 
In this case the matter sector is vector-like 
by introducing a pair of chiral multiplets. 
The {\it SUSY boundary condition} is taken at D=0. 
The anomalous dimension of the D-field and the $\be$-function
of the coupling are obtained. 
The 1-loop effective potential is explicitly obtained
and its SUSY invariant properties are confirmed. 
The result will become an important
reference in the analysis
of the bulk-boundary theory in the following sections.

\section{Mirabelli-Peskin Model}
As a toy model of a bulk-boundary model,
Mirabelli and Peskin proposed the following system.
Let us consider the 5 dimensional space-time. The space of the fifth
component is taken to be $S_1$, 
with the {\it periodicity} $2l$.
\begin{eqnarray}
\xf\ra\xf+2l\pr
\label{mp1b}
\end{eqnarray}
We also require the system to be (anti)symmetric with respect to
the $Z_2$-symmetry:
\begin{eqnarray}
Z_2\mbox{ transformation}:\ 
x^5\ra -x^5\pr
\label{mp5}
\end{eqnarray}
This makes the two points, $\xf=0$ and $\xf=l$, fixed points under
$Z_2$-transformation. The extra space is $S^1/Z_2$ orbifold. 
Let us consider 
5D bulk theory $\Lcal_{bulk}$ which is
coupled with 4D matter theory $\Lcal_{bnd}$ on a "wall" at $\xf=0$
and with $\Lcal'_{bnd}$ on the other "wall" at $\xf=l$.
The boundary Lagragians are, in the bulk action, described by
 the delta-functions along the extra axis $x^5$.
\begin{eqnarray}
S=\int d^5x\{\Lcal_{blk}+\del(x^5)\Lcal_{bnd}+\del(x^5-l){\Lcal'}_{bnd}\}
\pr\label{mp1}
\end{eqnarray}

We make use of the SUSY symmetry in order to make
the problem simple. 
Both bulk and boundary quantum effects are taken into account.

\vspace{2 mm}
(i) 5D super Yang-Mills theory\nl
\ We take, as the bulk dynamics, the 5D super YM theory
which is made of 
a vector field $A^M\ (M=0,1,2,3,5)$, 
a scalar field $\Phi$, 
a doublet of symplectic Majorana fields $\la^i\ (i=1,2)$, 
and a triplet of auxiliary scalar fields $X^a\ (a=1,2,3)$.
The metric is $(\eta_{MN})=\mbox{diag}(-1,1,1,1,1)$.
We basically follow the notation of \cite{Hebec01}.
\begin{eqnarray}
\Lcal_{SYM}=\tr \left(
-\half {F_{MN}}^2-(\na_M\Phi)^2
-i\labar_i\ga^M\na_M\la^i
+(X^a)^2+g\labar_i[\Phi,\la^i]
            \right)\com\nn
F_{MN}=\pl_MA_N-\pl_NA_M+ig[A_M,A_N]\com\q
\na_M\Phi=\pl_M\Phi+ig[A_M,\Phi]\com\nn
\na_M\la^i=\pl_M\la^i+ig[A_M,\la^i]
\com\label{mp2}
\end{eqnarray}
where all bulk fields are the {\it adjoint} representation
of the gauge group $G$. 
The SU(2)$_{\mbox{R}}$ index $i$ is raised and lowered
by the anti-symmetric tensors $\ep^{ij}$ and $\ep_{ij}$. 
\footnote{
The present notation:\ 
$[A_M,\Phi]=if^{\ab\ga}A_{M\al}\Phi_\be T^\ga
=i(A_M\times\Phi)_\ga T^\ga=iA_M\times\Phi$,\ 
$\tr\{[A_M,\Phi]\pl_5\Phi\}=
(i/2)f^{\ab\ga}A_{M\al}\Phi_\be\pl_5\Phi_\ga=
i\,\tr\{(A_M\times\Phi)\pl_5\Phi\}$.
Hence $\na_M\Phi=\pl_M\Phi-gA_M\times\Phi$. 
}
\begin{eqnarray}
A_M=A_{M\al}T^\al\com\q \Phi=\Phi_\al T^\al\com\q
\la^i=\la^i_\al T^\al\com\q
X^a=X^a_\al T^\al\com\nn
\q [T^\al,T^\be]=i f^{\ab\ga}T^\ga\com\q
\tr(T^\al T^\be)=\half \del^\ab
\com\label{mp3}
\end{eqnarray}
where $T^\al$ is the generator of the group and 
$ f^{\ab\ga}$ is the structure constant. 
(
As for the group indices $\al,\be,\cdots$, 
there is no distinction between
the upper one and the lower one.
)
The bulk Lagrangian $\Lcal_{SYM}$ of (\ref{mp2})
is invariant under the following SUSY transformation.
\begin{eqnarray}
\del_\xi A^M=i\xibar_i\ga^M\la^i\com\nn
\del_\xi\Phi=i\xibar_i\la^i\com\nn
\del_\xi\la^i=(\Si^{MN}F_{MN}+\ga^M\na_M\Phi)\xi^i
+i(X^a\si^a)^i_{~j}\xi^j\com\nn
\del_\xi X^a=\xibar_i(\si^a)^i_{~j}\ga^M\na_M\la^j
+i[\Phi,\xibar_i(\si^a)^i_{~j}\la^j]\com
\label{mp4}
\end{eqnarray}
where $\Si^{MN}=\fourth [\ga^M,\ga^N]$, and
the SUSY global parameter $\xi^i$ is the symplectic Majorana
spinor. This system has the symmetry of
8 real super charges.
\footnote 
{
Two Dirac spinors $\xi^1$ and $\xi^2$ has a "reality" condition.
The total number of the independent real SUSY-freedom
is 8, which is the same as that of one Dirac spinor
}

As the 5D gauge-fixing term, we take the Feynman gauge:
\begin{eqnarray}
\Lcal_{gauge}=-\tr (\pl_MA^M)^2=-\half (\pl_MA^M_{~\al})^2
\pr
\label{mp4b}
\end{eqnarray}
The corresponding ghost Lagrangian is given by
\begin{eqnarray}
\Lcal_{ghost}=-2\,\tr \pl_M\cbar\cdot \na^M(A)c
=-2\,\tr\pl_M\cbar\cdot (\pl^Mc+ig[A^M,c])
\com
\label{mp4c}
\end{eqnarray}
where $c$ and $\cbar$ are the complex ghost fields. 
We take the following bulk action.
\begin{eqnarray}
\Lcal_{blk}=\Lcal_{SYM}+\Lcal_{gauge}+\Lcal_{ghost}
\pr
\label{mp4d}
\end{eqnarray}

\vspace{3 mm}

(ii) $Z_2$-structure\nl
\ In order to consistently project out $\Ncal=1$ SUSY
multiplet which has 4 real super charges(4Q's), 
we make use of the $Z_2$ symmetry (\ref{mp5})
which divides the 8 $Q's$ system
into two 4Q's systems. 
We {\it assign} $Z_2$-parity, even ($P=+1$) or odd ($P=-1$) 
under the $Z_2$-transformation,
to all fields in accordance with the 5D SUSY (\ref{mp4}). 
{\it A consistent choice} is given as in Table 1.
Note that $A^m, (m=0,1,2,3)$, is the 4D components of the bulk vector$A^M$.
A symplectic Majorana field is 
expressed by two Weyl spinors. 
We can write $\la^i$ and $\xi^i$ as follows:
\begin{eqnarray}
\la^1=
\left(
\begin{array}{c}
(\la_L)_\al    \\
(\labar_R)^\aldot 
\end{array}
\right) 
\com\q
\la^2=
\left(
\begin{array}{c}
(\la_R)_\al    \\
-(\labar_L)^\aldot 
\end{array}
\right) 
\com\q
\xi^1=
\left(
\begin{array}{c}
(\xi_L)_\al    \\
(\xibar_R)^\aldot 
\end{array}
\right) 
\com\q
\xi^2=
\left(
\begin{array}{c}
(\xi_R)_\al    \\
-(\xibar_L)^\aldot 
\end{array}
\right) 
\pr\label{mp5b}
\end{eqnarray}
\nl
\nl
$$
\begin{array}{|c|c|c|}
\hline
             &   P=+1\ ,\ \xi_L         & P=-1\ ,\ \xi_R   \\
\hline
A^M  & A^m & A^5 \\
\hline
\Phi  & - & \Phi \\
\hline
\la^i   & \la_L & \la_R \\
\hline
X^a  & X^3 & X^{1,2} \\
\hline
\multicolumn{3}{c}{\q}                                                 \\
\multicolumn{3}{c}{
Table\ 1\ \ Z_2-parity~ assignment.
                  }\\
\end{array}
$$
$\Lcal_{SYM}$ of (\ref{mp2}) is invariant
under the $Z_2$-transformation (\ref{mp5}). On the wall ($x^5=0$),
all {\it odd-parity} states vanish. 
The parity odd fields, $A^5$ and $\Phi$ in Table 1, 
will play an important role in the effective potential. 
The SUSY transformation (\ref{mp4})
reduces to the following one of {\it even-parity} 
states generated by $\xi_L$:
\begin{eqnarray}
\del_\xi A^m=i\xibar_L\sibar^m\la_L+i\xi_L\si^m\labar_L\com\nn
\del_\xi\la_L=\si^{mn}F_{mn}\xi_L+i(X^3-\na_5\Phi)\xi_L\com\nn
\del_\xi (X^3-\na_5\Phi)=\xibar_L\sibar^m\na_m\la_L
-\xi_L\si^m\na_m\labar_L
\pr\label{mp6}
\end{eqnarray}
(Note that the odd parity field $\Phi$ appears in the 
$\xf$-derivative form.)
This is a $\Ncal =1$ (4D) vector multiplet 
($A^m,\la_L,X^3-\na_5\Phi$)
transformation in the {\it WZ-gauge} although all fields depend on
the extra coordinate $\xf$. 
Especially $\Dcal\equiv X^3-\na_5\Phi$ plays the role
of  D-field. The multiplet defined in the 5D world
can be expressed in the form of
the 4D vector superfied.
\begin{eqnarray}
V=-\theta\si^m\thbar A_m+i\theta^2\thbar\labar_L
-i\thbar^2\theta\la_L+\half\theta^2\thbar^2\Dcal
\com\q
\Dcal=X^3-\na_5\Phi
\pr
\label{mp6b}
\end{eqnarray}

The odd-parity fields $(\Phi+iA_5, -i\sqrt{2}\la_R, X^1+iX^2)$ 
transform as a chiral (adjoint) multiplet. 
\begin{eqnarray}
\del_\xi (\Phi+iA_5)=\sqtwo\xi_L(-i\sqtwo \la_R)\com\nn
\del_\xi \la_R=(i\si^m F_{5m}-\si^m \na_m\Phi)\xibar_L+i(X^1+iX^2)\xi_L\com\nn
\del_\xi (X^1+iX^2)=2\xibar_L(\sibar^m\na_m\la_R-i\na_5\labar_L)
-2i[\Phi,\xibar_L\labar_L]
\pr\label{mp6c}
\end{eqnarray}
This multiplet can be expressed in the 4D superspace as
\begin{eqnarray}
\Si=(\Phi+iA_5)+\sqtwo\theta (-i\sqtwo \la_R)
+\theta^2 (X^1+iX^2)
\pr\label{mp6d}
\end{eqnarray}
We will not use this multiplet in this paper.
\nl

\vspace{3 mm}

iii)\ Matter Lagrangian on the wall 

Let us introduce matter fields on the walls. 
We consider two cases:\ a) chiral matter,
\ b) vector-like matter.\nl

\vspace{2 mm}

iiia)\ Chiral matter\nl
\ We introduce, on the $\xf=0$ brane, 
a 4 dim chiral multiplet ($\phi,\psi,F$) where
$\phi$ is a complex scalar field, $\psi$ is a Weyl spinor and
$F$ is an auxiliary field of complex scalar. 
This is the simplest case as a matter candidate and was taken
in the original theory\cite{MP97}. 
The chiral superfield $\Th$
is introduced :\ 
\begin{eqnarray}
\Th=\phi+\sqtwo\theta\psi+\theta^2 F
\com\q
\Thbar=\phi^\dag+\sqtwo\thbar\psibar+\thbar^2 F^\dag
\pr
\label{mp7c}
\end{eqnarray}
Using the $\Ncal=1$ SUSY property of the bulk fields 
($A^m,\la_L,\Dcal=X^3-\na_5\Phi$),
we can find the bulk-boundary coupling.
\begin{eqnarray}
\Lcal^{(a)}_{bnd}
=\left.\Thbar\e^{2gV}\Th\right|_{\theta^2\thbar^2}
=-\na_m\phi^\dag \na^m\phi-i\psibar \sibar^m \na_m\psi+F^\dag F
\nn
+\sqrt{2}ig(\psibar \labar_L\phi-\phi^\dag\la_L\psi)
+g\phi^\dag \Dcal\phi\com
\label{mp7}
\end{eqnarray}
where $
\na_m\equiv \pl_m+igA_m,\ \Dcal=X^3-\na_5\Phi$.
All matter fields are taken to be
the {\it fundamental} representation of the internal group G. 
We may add the following superpotential term to the above
Lagrangian.
\begin{eqnarray}
\Lcal_{SupPot}=
(\half m_{\alp\bep}\Th_\alp\Th_\bep
+\frac{1}{3!}\la_{\alp\bep\gap}\Th_\alp\Th_\bep\Th_\gap)|_{\theta^2}
+\mbox{h.c.}
\com
\label{mp7b}
\end{eqnarray}
where the primed Greek suffixes ($\alp,\bep,\cdots$) show those
of the fundamental representation.

\vspace{2 mm}

iiib)\ Vector-like matter\nl
\ We introduce, as the 4D matter fields on the $\xf=0$ brane, 
one pair of 4 dim chiral multiplets, 
$\Th_S$=($\phi_S,\psi_S,F_S$) and 
$\Th_R$=($\phi_R,\psi_R,F_R$).   
\begin{eqnarray}
\Lcal^{(b)}_{bnd}=
\left.\left(\Thbar_S\e^{2gV}\Th_S+\Thbar_R\e^{-2gV}\Th_R\right)
\right|_
{\theta^2\thbar^2}
+m(\Th_S\Th_R|_{\theta^2}+\Thbar_S\Thbar_R|_{\thbar^2})
\nn
=-(\na^+_m\phi_S)^\dag (\na^m_+\phi_S)
-(\na^-_m\phi_R)^\dag (\na^m_-\phi_R)
+F^\dag_S F_S+F^\dag_R F_R
\nn
-\half(\psibar_S i\sibar^m \na^+_m\psi_S
+\psibar_R i\sibar^m \na^-_m\psi_R+\mbox{h.c.})
\nn
+\sqrt{2}g(i\psibar_S\labar_L\phi_S
-i\psibar_R\labar_L \phi_R+\mbox{h.c.})
\nn
+g(\phi^\dag_S D\phi_S-\phi^\dag_R D\phi_R)
+m\{\phi_S F_R+\phi_R F_S-\psi_S\psi_R+\mbox{h.c.}\}
\com
\label{mp8}
\end{eqnarray}
where $\na^\pm_m=\pl_m\pm igA_m$. 
This is the bulk-boundary generalization of super QED or QCD. 
We can identify the matter fermions
$(\psi_S,\psi_R)$ as one Dirac fermion ("electron, quark").\nl

On the other brane $\xf=l$, we introduce another WZ-multiplet(s),
$(\phi',\psi',F')$ for case a), and 
$(\phi_S',\psi_S',F_S')$ and $(\phi_R',\psi_R',F_R')$ for case b). 
The bulk-boundary couplings are fixed in the same way. 
The quadratic (kinetic) terms of the vector $A^m$, the gaugino spinor $\la_L$
and $\Dcal=X^3-\na_5\Phi$ are in the {\it bulk} world. 
We note here the interaction between the bulk fields and the boundary
ones is {\it definitely fixed} from SUSY.
In the ordinary standpoint of the field theory,
the boundary theory (\ref{mp7}) or (\ref{mp8})
is {\it perturbatively unrenormalizable} because
the coupling $g$ has the physical dimension
of $M^{-1/2}$ (unrenormalizable coupling).
\section{Quantization Using Background Field Method}
From the results of Sect.2 (and App.A), we may put, 
for the purpose of obtaining the 1-loop effective potential, 
the following conditions on $\Lcal_{blk}$:
\begin{eqnarray}
A^m=0\ (m=0,1,2,3)\com\q \la^i=\labar^i=0   
\pr
\label{ep1}
\end{eqnarray}
Here the extra (fifth) component of the bulk vector $A^5$ 
does {\it not} taken to be zero because it is regarded
as a {\it 4D scalar} on the wall. 
Then $\Lcal_{blk}=\Lcal^{SYM}+\Lcal_{gauge}+\Lcal_{ghost}$ reduces to
\begin{eqnarray}
\Lcal^{red}_{blk}[\Phi,X^3,A_5]\q\q\q\q\q\q\q\q\q\q\q\q\q\q\q\q\q\q\nn
=\tr \left\{ -\pl_M\Phi\pl^M\Phi+X^3X^3-\pl_MA_5\pl^MA_5
+2g(\pl_5\Phi\times A_5)\Phi-g^2(A_5\times\Phi)(A_5\times\Phi)\right\}\nn
+\Lcal^{red}_{ghost}[c,\cbar,A_5]
+\mbox{irrel. terms}
\com\nn
\Lcal^{red}_{ghost}[c,\cbar,A_5]=-2\tr \{
\pl_m\cbar\cdot\pl^m c+\pl_5\cbar\cdot (\pl^5 c+ig[A^5,c])
\}\com
\label{ep3}
\end{eqnarray}
where we have dropped terms of $X^1_\al X^1_\al, X^2_\al X^2_\al$
as 'irrelevant terms' because they decouple from other fields. 
As for the boundary part, we may impose the conditions:
\begin{eqnarray}
\mbox{a. Chiral matter : }
A^m=0\com\q \psi=0\com\q \la_L=0\ ;\nn
\mbox{b. vector-like matter : }
A^m=0\com\q \psi_S=\psi_R=0\com\q \la_L=0
\pr
\label{ep2}
\end{eqnarray}
$\Lcal_{bnd}$ reduces to
\begin{eqnarray}
\mbox{a. Chiral matter}
\nn
\Lcal^{red}_{bnd(a)}[\phi,\phi^\dag,\Dcal=X^3-\na_5\Phi]=
\nn
-\pl_m\phi^\dag\pl^m\phi
+g(X^3_\al-\pl_5\Phi_\al+gf^{\al\be\ga}A_{5\be}\Phi_\ga)
\phi^\dag_\bep (T^\al)_{\bep\gap}\phi_\gap+F^\dag F
\nn
+
\left\{
\frac{m_{\alp\bep}}{2}(\phi_\alp F_\bep+F_\alp\phi_\bep)
+\frac{\la_{\alp\bep\gap}}{3!}(\phi_\alp\phi_\bep F_\gap
+\phi_\alp F_\bep\phi_\gap+F_\alp\phi_\bep\phi_\gap)+\mbox{h.c.}
\right\}\pr
\nn
\mbox{b. Vector-like matter}
\nn
\Lcal^{red}_{bnd(b)}[\phi_S,\phi_S^\dag,\phi_R,\phi_R^\dag,
\Dcal=X^3-\na_5\Phi]=
-\pl_m\phi_S^\dag\pl^m\phi_S-\pl_m\phi_R^\dag\pl^m\phi_R
\nn
+g(X^3_\al-\na_5\Phi_\al)(T^\al)_{\bep\gap}
(\phi^\dag_{S\bep} \phi_{S\gap}-\phi^\dag_{R\bep} \phi_{R\gap})
+F^\dag_S F_S+F^\dag_R F_R+m(\phi_S F_R+\phi_R F_S+\mbox{h.c.})
\com
\label{ep4}
\end{eqnarray}
where 
$\alp, \bep$ are the suffixes of the fundamental representation.
In the same way, the boundary Lagrangians at $\xf=l$, 
$\Lcal'_{bnd(a)}$ and $\Lcal'_{bnd(b)}$, reduce to
$\Lcal^{'red}_{bnd(a),(b)}=(\phi\ra\phi', F\ra F' 
\mbox{in (\ref{ep4})} )$. 
Now we expand all scalar fields 
($\Phi, X^3, A_5; \phi, F,\phi', F'$)
, except ghost fields, 
into the {\it quantum fields} (which are denoted again by the same symbols) 
and the {\it background fields}
($\vp, \chi^3, a_5; \eta, f, \eta', f'$).
\begin{eqnarray}
\Phi\ra\vp+\Phi\com\q
X^3\ra \chi^3+X^3\com\q
A_5\ra a_5+A_5\com\nn
\left\{
\begin{array}{cc}
\phi\ra\eta+\phi, F\ra f+F \mbox{  and primed ones}
                               & \mbox{chiral matter} \\
\phi_S\ra\eta_S+\phi_S, \phi_R\ra\eta_R+\phi_R,
F_S\ra f_S+F_S, F_R\ra f_R+F_R & \mbox{vector-like}\\
\mbox{  and primed ones}       & \mbox{matter}
\end{array}
\right.
\label{ep5}
\end{eqnarray}
We treat the ghost fields as quantum ones. 

In Sec.2, 
for the purpose of obtaining the effective potential
we consider the case that the background fields are constant
(in order to pick up the non-derivative part of the effective action). 
In the present case of
the 5D space-time, we have the extra coordinate $\xf$.
Because 4D($x^m$-space) scalar property is independent
of the extra space, we take into account the $\xf$-dependency
of the vacuum configuration. The distribution along the extra 
coordinate is important to make the localization configuration.
We require that
{\it the background fields may be constant only in 4D world, not necessarily
in 5D world}. 
We may allow the background field to depend on the extra coordinate $x^5$. 
This gives us an interesting possibility to the extra space model
. (See also the beginning paragraph of Sec.6 where the necessity of the present
treatment is explained using an explicitly-$\xf$-dependent solution (\ref{det6}).)

When the background fields ($\vp, \chi^3, a_5; \eta, f, \eta', f'$) satisfy 
the field equations derived from 
$\Lcal^{red}_{blk}+\del(\xf)\Lcal^{red}_{bnd}+\del(\xf-l)\Lcal^{'red}_{bnd}$
, using (\ref{ep3}) and (\ref{ep4}),
the situation is called "on-shell". 
The equations (on-shell condition) are given as, 
\begin{eqnarray}
\del\Phi_\al\q;\nn
{\pl_5}^2\vp_\al+gf_{\be\ga\al}\pl_5\vp_\be a_{5\ga}
-gf_{\ab\ga}\pl_5(a_{5\be}\vp_\ga)
-g^2f_{\be\al\tau}f_{\ga\del\tau}a_{5\be}a_{5\ga}\vp_\del
\nn
+g\pl_5\del(\xf)\,\eta^\dag T^\al\eta+g\pl_5\del(\xf-l)\,\eta^{'\dag} T^\al\eta'
+g^2(\del(\xf)\eta^\dag T^\ga\eta+\del(\xf-l)\eta^{'\dag} T^\ga\eta')f^{\be\al\ga}a_{5\be}\nn
=-\pl_5Z_\al-g(Z\times a_5)_\al=0,
\nn\nn
\del A_{5\al}\q;\nn
{\pl_5}^2a_{5\al}+gf_{\be\al\ga}\pl_5\vp_\be\, \vp_\ga
-g^2f_{\ab\tau}f_{\ga\del\tau}\vp_\be a_{5\ga}\vp_\del
+g^2(\del(\xf)\eta^\dag T^\ga\eta+\del(\xf-l)\eta^{'\dag} T^\ga\eta')f^{\al\be\ga}\vp_\be\nn
={\pl_5}^2a_{5\al}-g(\vp\times Z)_\al=0\ ,
\nn\nn
\del X^3_\al\q;\nn
\chi^3_\al+g(\del(\xf)\eta^\dag T^\al\eta+\del(\xf-l)\eta^{'\dag} T^\al\eta')=0\ ,
\nn\nn
\del\phi^\dag_\alp\q(\del\phi^{'\dag}_\alp)\q ;\nn
d_\be(T^\be\eta)_\alp
+m_{\alp\bep}f^\dag_\bep+\half\la_{\alp\bep\gap}\eta^\dag_\bep f^\dag_\gap
=0\ ,\q (\eta\ra\eta', f\ra f' \ \mbox{ in the left eq.})
\nn\nn
\del F^\dag_\alp\q(\del F^{'\dag}_\alp)\q;\nn
f_\alp+m_{\alp\bep}\eta^\dag_\bep+\half\la_{\alp\bep\gap}
\eta^\dag_\bep\eta^\dag_\gap=0
\com\q (\eta\ra\eta', f\ra f' \ \mbox{ in the left eq.})\com
\label{ep5b}
\end{eqnarray}
where $d_\al=(\chi^3-\pl_5\vp+ga_5\times\vp)_\al$ 
is the background (4 dimensional) D-field and 
$Z_\al\equiv -g(\del(\xf)\eta^\dag T^\al\eta+\del(\xf-l)\eta^{'\dag} T^\al\eta')
-\pl_5\vp_\al+gf_{\ab\ga}a_{5\be}\vp_\ga$.
In deriving the above equations, 
we assume, based on the statement of the previous paragraph
on the background field,
$\vp=\vp(\xf), \chi^3=\chi^3(\xf), a_5=a_5(\xf), \eta=\mbox{const},
\eta'=\mbox{const}$. 
The total symmetricity of $m_{\alp\bep}$ and $\la_{\alp\bep\gap}$
with respect to the suffixes is also assumed. 
In the above derivation, we use the fact that 
total divergences vanish from the {\it periodicity condition}. 
As stated below (\ref{wz4b}), we do {\it not} assume the above
on-shell condition except when we state its use.
We regard the background fields as general external fields
or as off-shell fields.

The {\it quadratic} part w.r.t. the quantum fields 
($\Phi, X^3, A_5; \phi, F,\phi', F'$)
gives us
the 1-loop quantum effect. That part of $\Lcal^{red}_{blk}$
is given as
\begin{eqnarray}
\Lcal^2_{blk}[\Phi,A_5,X^3]
=\tr\,\{ -\pl_M\Phi\pl^M\Phi
+ X^3 X^3 -\pl_M A_{5}\pl^M A_{5}\}\nn
+2g\,\tr\left[ (\pl_5\vp\times A_5)\Phi+(\pl_5\Phi\times a_5)\Phi
+(\pl_5\Phi\times A_5)\vp \right]\nn
-2g^2\tr\left[ (a_5\times\vp)(A_5\times \Phi)\right]
-g^2\tr (a_5\times\Phi+A_5\times\vp)^2+
\Lcal^{2}_{ghost}[c,\cbar]\com\nn
\Lcal^{2}_{ghost}[c,\cbar]=-2\tr \{
\pl_m\cbar\cdot\pl^m c+\pl_5\cbar\cdot (\pl^5 c+ig[a^5,c])\}\pr
\label{ep6}
\end{eqnarray}
The quadratic part of $\Lcal^{red}_{bnd}$ is given by
\begin{eqnarray}
\Lcal^2_{bnd(a)}[\phi,\phi^\dag,F,F^\dag;\Phi,A_5,X^3]
=-\pl_m\phi^\dag\pl^m\phi
+g d_\al\phi^\dag T^\al\phi-ig^2[A_5,\Phi]_\al\eta^\dag T^\al\eta
\nn
+g(X^3_\al-\pl_5\Phi_\al-ig[a_5,\Phi]_\al-ig[A_5,\vp]_\al)
(\eta^\dag T^\al\phi+\phi^\dag T^\al\eta)
                   +F^\dag F
\nn
+\{m_{\alp\bep}\phi_\alp F_\bep
+\frac{\la_{\alp\bep\gap}}{2}(\phi_\alp\phi_\bep f_\gap
+2\phi_\alp\eta_\bep F_\gap)+\mbox{h.c.}\}
                   \com
\nn
d_\al\equiv (\chi^3-\pl_5\vp-ig[a_5,\vp])_\al\com
\nn
\Lcal^2_{bnd(b)}=-\pl_m\phi_S^\dag\pl^m\phi_S-\pl_m\phi_R^\dag\pl^m\phi_R
+g\{
d_\al(\phi^\dag_{S}T^\al\phi_{S}-\phi^\dag_{R}T^\al\phi_{R})   
-ig[A_5,\Phi]_\al (\eta^\dag_S T^\al\eta_S-\eta^\dag_R T^\al\eta_R)
\nn
+(X^3_\al-\pl_5\Phi_\al-ig[a_5,\Phi]-ig[A_5,\vp])(\eta^\dag_{S}T^\al\phi_{S}+\phi^\dag_{S} T^\al\eta_{S}
-\eta^\dag_{R}T^\al\phi_{R}-\phi^\dag_{R}T^\al\eta_{R})
                   \}
\nn
+F^\dag_S F_S+F^\dag_R F_R+m(\phi_SF_R+\phi_R F_S+\mbox{h.c.})
\com
\label{ep7}
\end{eqnarray}
where 
$\phi^\dag T^\ga\phi\equiv \phi^\dag_\alp(T^\ga)_{\alp\bep}\phi_\bep$. 
$\Lcal^{'2}_{bnd(a)}$ and $\Lcal^{'2}_{bnd(b)}$ are the same as
$\Lcal^{2}_{bnd(a)}$ and $\Lcal^{2}_{bnd(b)}$ except the replacement:
$\phi\ra\phi', F\ra F', \eta\ra\eta', f\ra f'$. 
Now we can integrate out the auxiliary field $X^3_\al$ in
$\Lcal^2_{blk}+\del(x^5)\Lcal^2_{bnd}+\del(x^5-l)\Lcal^{'2}_{bnd}$. 
Using a "squaring" equation
\footnote{
Note the relation:
$\del(\xf)\del(\xf-l)=0$.
}
:
\begin{eqnarray}
\half X^3_\al X^3_\al+g\{ 
\del(x^5)(\eta^\dag T^\al\phi+\phi^\dag T^\al\eta)
+\del(x^5-l)(\eta^{'\dag} T^\al\phi'+\phi^{'\dag} T^\al\eta')  \} X^3_\al
\nn
=\half \{
X^3_\al+g\del(x^5)(\eta^\dag T^\al \phi+\phi^\dag T^\al\eta)
+g\del(x^5-l)(\eta^{'\dag} T^\al \phi'+\phi^{'\dag} T^\al\eta')
\}^2
\nn
-\half g^2\del(x^5)\del(0)(\eta^\dag T^\al \phi+\phi^\dag T^\al\eta)^2
-\half g^2\del(x^5-l)\del(0)(\eta^{'\dag} T^\al \phi'+\phi^{'\dag} T^\al\eta')^2
\com\label{ep8}
\end{eqnarray}
we obtain
the final 1-loop Lagrangian, necessary for the present purpose, as
\begin{eqnarray}
S^2_a[\Phi,A_5;\phi,F]=\int d^5X                     \left[
\Lcal_{blk}^2|_{X^3=0}                               \right.
\nn                              
+\del(x^5)      \left\{ -\pl_m\phi^\dag \pl^m\phi
 +gd_\al (\phi^\dag T^\al\phi)
-g\pl_5\Phi_\al (\eta^\dag T^\al\phi+\phi^\dag T^\al\eta)
                 \right.
\nn
                  \left.
+\left(
m_{\alp\bep}\phi_\alp F_\bep
+\frac{\la_{\alp\bep\gap}}{2}(\phi_\alp\phi_\bep f_\gap
+2\phi_\alp\eta_\bep F_\gap)+\mbox{h.c.}
\right)
+F^\dag F
-\frac{g^2}{2}
\del(0)(\eta^\dag T^\al\phi+\phi^\dag T^\al\eta)^2
                  \right\}
\nn
                                                        \left.                
+\del(\xf-l)\{ \phi\ra\phi', \eta\ra\eta', F\ra F'  \}
                                                         \right]
\com
\nn
\Lcal_{blk}^2|_{X^3=0}
=\Lcal^2_{ghost}[c,\cbar]+
\tr\,\left\{ 
-\pl_M\Phi\pl^M\Phi-\pl_M A_{5}\pl^M A_{5}
+2g\,\left( (\pl_5\vp\times A_5)\Phi+(\pl_5\Phi\times a_5)\Phi
+(\pl_5\Phi\times A_5)\vp 
      \right)\right.
\nn
     \left.
-2g^2(a_5\times\vp)(A_5\times \Phi)
-g^2(a_5\times\Phi+A_5\times\vp)^2
     \right\}
\com\label{ep9}
\end{eqnarray}
for the chiral matter model.
\footnote{
Here we may omit the terms,$
-ig^2[A_5,\Phi]_\al\eta^\dag T^\al\eta,\ 
-ig^2([a_5,\Phi]_\al+[A_5,\vp]_\al)(\eta^\dag T^\al\phi+\phi^\dag T^\al\eta)$
in $\Lcal^2_{bnd(a)}$ of (\ref{ep7}). 
From the Z$_2$-odd property of $a_5, A_5, \vp$ and $\Phi$, 
the above terms, with the term 
$\del(x^5)$ or $\del(x^5-l)$ multiplied, have no contribution to
the 1-loop effect.
The same thing is used in (\ref{ep10}). 
}
Similarly we obtain,
for the vector-like matter, as
\begin{eqnarray}
S^2_b[\Phi,A_5;\phi_S, \phi_R]=\int d^5X \left[
\Lcal_{blk}^2|_{X^3=0}+
\del(\xf)   \left\{ -(\pl_m\phi_S^\dag \pl^m\phi_S  
+\pl_m\phi_R^\dag \pl^m\phi_R) 
            \right.
                                          \right.
\nn
+gd_\al (\phi_S^\dag T^\al\phi_S-\phi_R^\dag T^\al\phi_R)
-g\pl_5\Phi_\al
(\eta_S^\dag T^\al\phi_S+\phi_S^\dag T^\al\eta_S
-\eta_R^\dag T^\al\phi_R-\phi_R^\dag T^\al\eta_R)
\nn
+F^\dag_SF_S+F^\dag_RF_R+m(\phi_SF_R+\phi_RF_S+\mbox{h.c.})
\nn
                                           \left.
           \left.
-\frac{g^2}{2}\del(0)(  
\eta_S^\dag T^\al\phi_S+\phi_S^\dag T^\al\eta_S
-\eta_R^\dag T^\al\phi_R-\phi_R^\dag T^\al\eta_R
                        )^2
           \right\}
+\del(\xf-l)\{\phi\ra\phi', \eta\ra\eta', F\ra F' \}                                            \right]                
\pr\label{ep10}
\end{eqnarray}
For simplicity we consider the case of no superpotential:\ 
$m_{\alp\bep}=\la_{\alp\bep\gap}=0$, hereafter.
\footnote{
We examine the case with the superpotential in ref.\cite{IMplb593}.
}
\section{Bulk and Boundary Quantum Effects}
\begin{figure}
\centerline{ \psfig{figure=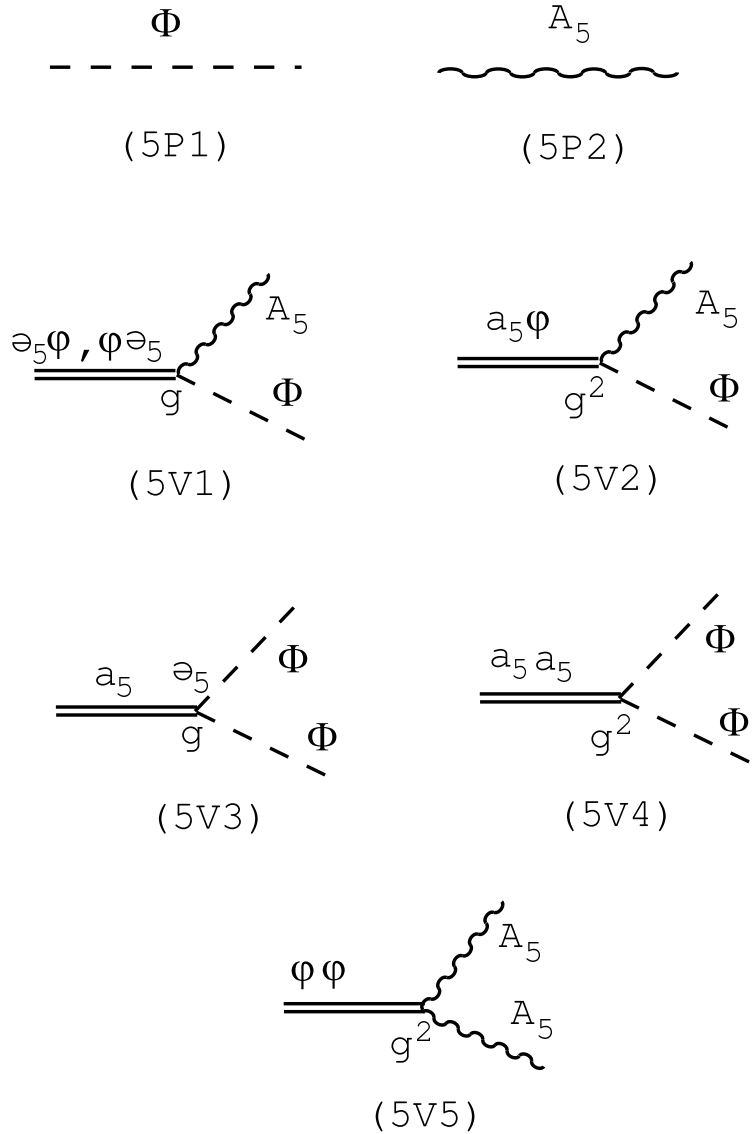,height=15cm,angle=0}}
\caption{Feynman graphs for the bulk fields. (5P1),...(5V5) of (\ref{qe1}).
}
\label{fig:Feynman5}
\end{figure}

\begin{figure}
\centerline{ \psfig{figure=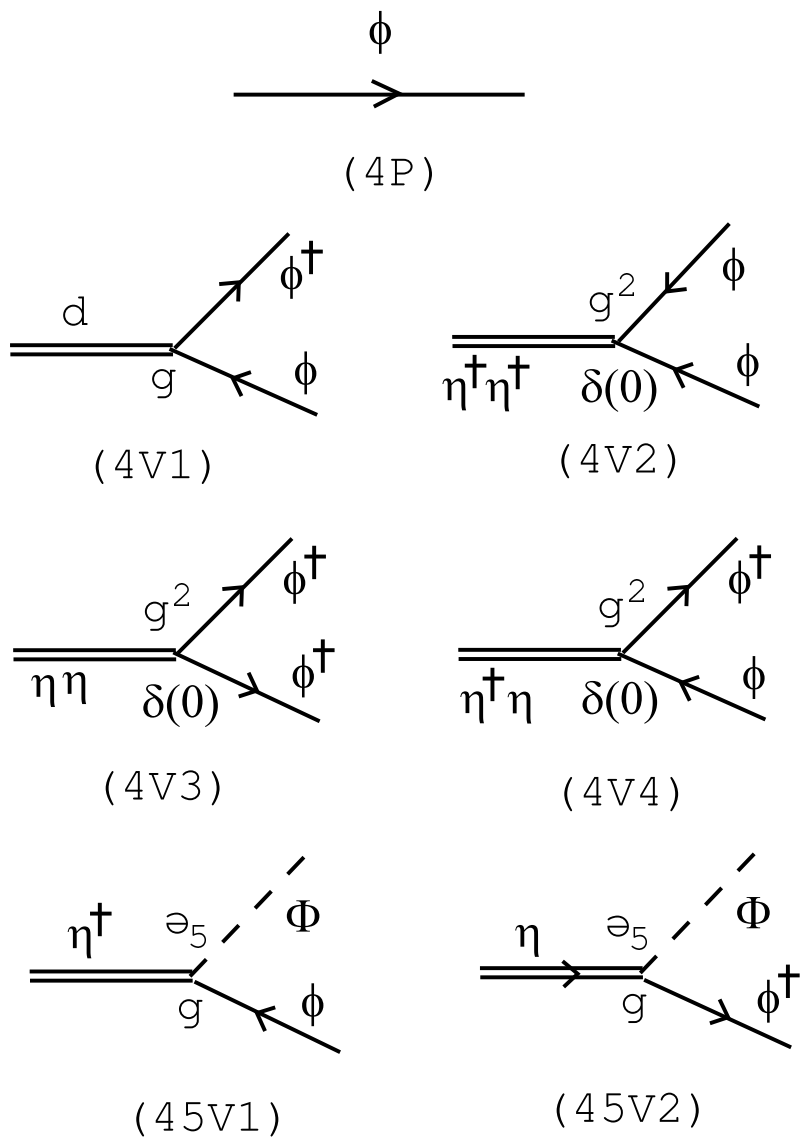,height=15cm,angle=0}}
\caption{ Feynman graphs for the boundary at $\xf=0$ and 
bulk-boundary-mixed fields. (4P),...(45V2) of (\ref{qe2}).
For all vertex graphs, the overall factor,
$\del(\xf)$, is omitted. See (\ref{qe2})
for detail.
}
\label{fig:Feynman4}
\end{figure}

Before the {\it full} 1-loop calculation of the next section,
it is useful to look at some important diagrams
appearing in the perturbation w.r.t. the coupling $g$. We can 
express propagators and vertices
 as in Fig.\ref{fig:Feynman5} (for the bulk part) 
 and Fig.\ref{fig:Feynman4} (for the boundary and mixed parts).
All double lines express the background fields.
The corresponding terms in the Lagragian,
from which the Feynman rules can be easily read,
are given by, for the bulk part(Fig.\ref{fig:Feynman5}),
\begin{eqnarray}
\mbox{(5P1)}\ :\ 
-\half\pl_M\Phi_\al\pl^M\Phi_\al\com\q
\mbox{(5P2)}\ :\ 
-\half\pl_M A_{5\al}\pl^M A_{5\al}\com\nn
\mbox{(5V1)}\ :\ 
gf_{\ab\ga}\{  \pl_5\vp_\al\cdot A_{5\be}\Phi_\ga 
+\pl_5\Phi_\al\cdot A_{5\be}\vp_\ga  \} \com\nn
\mbox{(5V2)}\ :\ 
-g^2f_{\ab\tau}f_{\ga\del\tau}a_{5\al}\vp_\be A_{5\ga}\Phi_\del
-g^2 f_{\ab\tau}f_{\ga\del\tau}a_{5\al}\Phi_\be A_{5\ga}\vp_\del \com\nn
\mbox{(5V3)}\ :\ 
gf_{\ab\ga}\pl_5\Phi_\al a_{5\be}\Phi_\ga\ ,\ 
\mbox{(5V4)}\ :\ 
-\frac{g^2}{2}f_{\ab\tau}f_{\ga\del\tau}a_{5\al}\Phi_\be a_{5\ga}\Phi_\del \com\nn
\mbox{(5V5)}
-\frac{g^2}{2}f_{\ab\tau}f_{\ga\del\tau}A_{5\al}\vp_\be A_{5\ga}\vp_\del
\pr
\label{qe1}
\end{eqnarray}
They can be read from terms in $\Lcal^2_{blk}|_{X^3=0}$ 
of (\ref{ep9}).
Those for the boundary at $\xf=0$
and mixed parts (Fig.\ref{fig:Feynman4}) are given by "$\del$-function parts"
of (\ref{ep9}).
\begin{eqnarray}
\mbox{(4P)}\ :\ 
-\del(\xf)\pl_m\phi^\dag\pl^m\phi\com
\nn
\mbox{(4V1)}\ :\ 
\del(x^5)gd_\al\phi^\dag T^\al\phi\com\q
\mbox{(4V2)}\ :\ 
-\half g^2\del(x^5)\del(0)(\eta^\dag T^\al \phi)^2\com
\nn
\mbox{(4V3)}\ :\ 
-\half g^2\del(x^5)\del(0)(\phi^\dag T^\al\eta)^2\com\q
\nn
\mbox{(4V4)}\ :\ 
-g^2\del(x^5)\del(0)(\eta^\dag T^\al \phi)(\phi^\dag T^\al\eta)\com
\nn
\mbox{(45V1)}\ :\ 
-\del(x^5)g\pl_5\Phi_\al \eta^\dag T^\al\phi\com\q
\mbox{(45V2)}\ :\ 
-\del(x^5)g\pl_5\Phi_\al \phi^\dag T^\al\eta
\pr
\label{qe2}
\end{eqnarray}
Those for the boundary at $\xf=l$ and mixed parts are the same as above
except the replacement:\ $\del(\xf)\ra\del(\xf-l), \phi\ra\phi', \eta\ra\eta'$.

(i) Boundary (4D) Quantum Effect

\begin{figure}
\centerline{ \psfig{figure=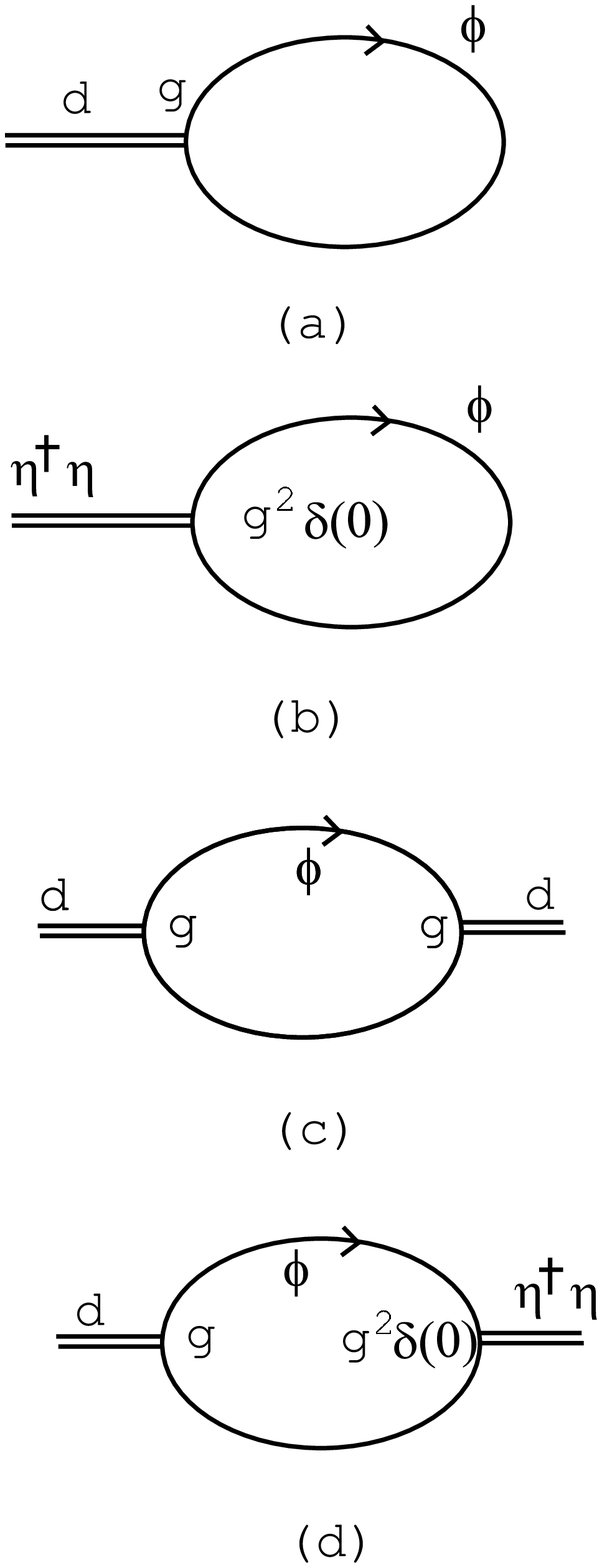,height=15cm,angle=0}}
\caption{ Divergent Feynman graphs for the boundary part
up to the order of $g^3$.
}
\label{fig:BndGraph}
\end{figure}

\begin{figure}
\centerline{ \psfig{figure=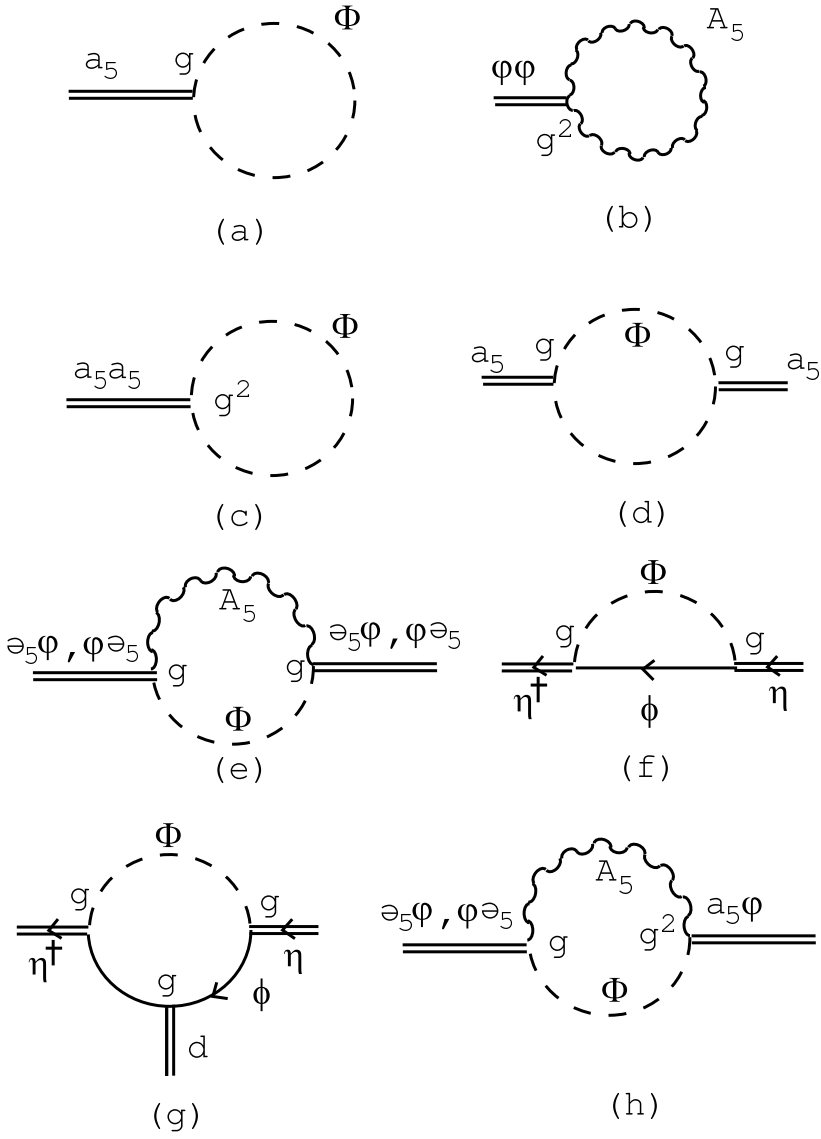,height=15cm,angle=0}}
\caption{ Divergent Feynman graphs for the bulk part
up to the order of $g^3$.
}
\label{fig:BlkGraph}
\end{figure}

All divergent diagrams (for the chiral matter model) 
up to the order of $g^3$ are listed up in Fig.\ref{fig:BndGraph}.
All are 1-loop diagrams within the $\xf=0$ brane. 
\nl 
The diagram (a) is interesting because
its presence says Fayet-Iliopoulos D-term 
appears in the boundary due to 
the {\it radiative correction}. 
It is {\it quadratically} divergent. 
The term is proportional to $\Tr T^\al$, hence
it exists {\it only when the gauge group $G$ involves $U(1)$}. 
If the appearance really happens
it could give a dynamical SUSY breaking (see a textbook\cite{WB92}).
(Note that the tadpole diagram of massless field in 4D
vanishes in the {\it dimensional regularization}
\cite{tH73}
\footnote{
On the other hand, the dimensional regularization
is generally considered non-appropriate for the
SUSY theories because the totally anti-symmetric
tensor $\ep_{\mn\ls}$ is essentially involved
with SUSY symmetry\cite{JJ97}.
This looks to obscure the presence of the valid calculation
of the tadpole diagram.
}
.
Hence the presence of the FI D-term is rather subtle.
)
This D-term does {\it not} appear in the vector-like 
matter case, $\Lcal^{(b)}_{bnd}$,
because the $\phi_S$ and $\phi_R$ contributions
cancel each other. 
(The situation is the same as the super QED.)\nl 
The diagram (b) was considered in ref.\cite{MP97}.
It contributes, with (f) of Fig.\ref{fig:BlkGraph} explained later, 
to the self-energy of the scalar matter.
This diagram (b) is independent of $d$, hence
 does {\it not} contribute to the effective potential 
under the SUSY boundary condition. \nl
The diagram (c) gives the renormalization of
D-field. The tree part is in the bulk as $\tr (X^3X^3-\pl_5\Phi\pl^5\Phi)$.
(The corresponding part appears in Super QED. See $d^2$-term
of eq.(\ref{qed14}).)\nl
The diagram (d) gives, with (g) of Fig.\ref{fig:BlkGraph},  the renormalization of the gauge
coupling $g$, and contributes to the $\be$-function
$\be(g)$. 
(This part is very contrasting with the corresponding part
of Super QED ($da\abar$-term). We will discuss it
in the final part of this section as (g)/Fig.\ref{fig:BlkGraph}+(d)/Fig.\ref{fig:BndGraph} part 
)

The contribution to the effective potential
of each diagram (of Fig.\ref{fig:BndGraph}) is given by
\begin{eqnarray}
\mbox{(a) and (b)}\ :\ 
i\{gd_\al (T^\al)_{\bep\bep}-g^2\del(0)
(\eta^\dag T^\al)_\bep (T^\al\eta)_\bep\}
\int\frac{d^4k}{(2\pi)^4}\frac{1}{k^2}\com\nn
\mbox{(c)}\ :\ 
\frac{i^2}{2!}g^2\cdot\half d_\al d_\al
\int\frac{d^4k}{(2\pi)^4}\frac{1}{(k^2)^2}\com\nn
\mbox{(d)}\ :\ 
\frac{i^2}{2!}\cdot 2\cdot (-g^3\del(0))
d_\al (\eta^\dag T^\be T^\al T^\be\eta)
\int\frac{d^4k}{(2\pi)^4}\frac{1}{(k^2)^2}
\pr
\label{qe3}
\end{eqnarray}

(ii) Bulk (5D) Quantum Effect

Among bulk quantum propagations, we do not, at present, consider
those which propagate between one brane and the other brane.
Those diagrams play an important role as the gauge mediation
mechanism in ref.\cite{MP97}. The present interest, however, 
is not the SUSY breaking mechanism.  
This simplification is admitted because we can control
the ignored contribution by adjusting the length $l$
between the two branes.

All divergent diagrams are listed up in Fig.\ref{fig:BlkGraph}
up to the order of $g^3$. Only the diagram (g)
contributes, others do not in the SUSY boundary
condition. 
\footnote{
$\Dcal=X^3-\na_5\Phi$ is defined in the bulk. It plays
the role of D-field on the boundary theory $\Lcal_{bnd}$.
Its background field $d=\chi^3-\na_5(a)\vp$ can be taken
independt of $-\na_5(a)\vp$. The purely bulk diagrams
(e) and (h) (of Fig.\ref{fig:BlkGraph}), which contains $\pl_5\vp$, 
are treated as d-independent ones. They do not
contribute the effective potential in the SUSY
boundary condition.  
}
The diagram (f) was analysed in Ref.\cite{MP97}
for the calculation of the matter-field self energy.
The diagram
(g) gives the bulk contribution to the $\be$-function
of the coupling $g$. 

The contribution to the effective potential
of diagrams (f) and (g) are given by
\begin{eqnarray}
\mbox{(f)}\ :\ 
g^2(\eta^\dag T^\al T^\al\eta)
\frac{i^2}{2!}\cdot 2\int_{k5}\frac{(k^5)^2}{k^2+(k^5)^2}
\frac{1}{k^2}\com\nn
\mbox{(g)}\ :\ 
g^3(\eta^\dag T^\al T^\ga T^\al\eta)d_\ga
(-1)\int_{k5}\frac{(k^5)^2}{-k^2-(k^5)^2}
\frac{1}{(k^2)^2}\com
\label{qe4}
\end{eqnarray}
where 
\begin{eqnarray}
\int_{k5}\equiv \int\frac{d^4k}{(2\pi)^4} \frac{1}{2l}
\sum_{k^5\in {\frac{\pi}{l}} {\bf Z} }
\pr
\label{qe5}
\end{eqnarray}
The $k^5$-summation comes from the KK-expansion
for the bulk field $\Phi$, which will be explained in (\ref{det3}). 
The result (f) of (\ref{qe4}) is consistent with
the corresponding term in (24) and (25) of ref.\cite{MP97}.

Using a formula
\begin{eqnarray}
\sum_{n\in \bfZ}\frac{1}{x^2+n^2}=\frac{\pi}{x}
\coth (\pi x)
\com
\label{qe6}
\end{eqnarray}
the $k^5$ summation part in (f) and (g) can be
evaluated as
($k^0=ik^4,k^2=(k^1)^2+(k^2)^2+(k^3)^2+(k^4)^2\equiv \kbar^2$)
\begin{eqnarray}
\sum_{k^5\in {\frac{\pi}{l}} {\bf Z} }
\frac{(k^5)^2}{-k^2-(k^5)^2}
=-\sum_{k^5\in {\frac{\pi}{l}} {\bf Z} }
\frac{(k^5)^2}{\kbar^2+(k^5)^2}\nn
=-(\sum_{n\in\bfZ}1)+\kbar^2(\frac{l}{\pi})^2
\sum_{n\in\bfZ}\frac{1}{\kbar^2(\frac{l}{\pi})^2+n^2}\nn
=-2l\del(0)+l\sqrt{\kbar^2}\coth (l\sqrt{\kbar^2})
\com
\label{qe7}
\end{eqnarray}
where the {\it Wick-rotation} of $k^0$-axis is done and
the relation 
\begin{eqnarray}
\sum_{n\in\bfZ}1=2l\del(0)
\com
\label{qe7b}
\end{eqnarray}
is used. 
The above result tells us the contribution from the
{\it singular} parts in the boundary, 
that is, $\del(0)$ parts in (b) and (d)
of eq.(\ref{qe3})( see also Fig.\ref{fig:BndGraph}), {\it cancel} those in 
the bulk, that is, (f) and (g) of (\ref{qe4})( see also Fig.\ref{fig:BlkGraph}).
This phenomenon was pointed out for the self-energy
diagrams in \cite{MP97}. 
In App.B, it is shown that the cancellation phenomenon
more generally 
(at the full order of the coupling within 1-loop)
occurs in the effective potential. 
\footnote{
In ref.\cite{GW0104},
$\del(\xf)$ is called "classical singularity" and the treatment of
its divergence $\del(0)$ is discussed using a "generalized" renormalization
group.
}
The final results are obtained as
\begin{eqnarray}
\mbox{(f)/Fig.\ref{fig:BlkGraph}+(b)/Fig.\ref{fig:BndGraph}}\ :\ 
-g^2(\eta^\dag T^\al T^\al\eta)
\int\frac{id^4\kbar}{(2\pi)^4}\half\sqrt{\kbar^2}
\coth(l\sqrt{\kbar^2})\,\frac{1}{-\kbar^2}\com\nn
\mbox{(g)/Fig.\ref{fig:BlkGraph}+(d)/Fig.\ref{fig:BndGraph}}\ :\ 
-g^3(\eta^\dag T^\al T^\ga T^\al\eta)d_\ga
\int\frac{id^4\kbar}{(2\pi)^4}\half\sqrt{\kbar^2}
\coth(l\sqrt{\kbar^2})\,\frac{1}{(-\kbar^2)^2}
\pr
\label{qe8}
\end{eqnarray}
We note the above results give correct 4D expressions
in the limit of $l\sqrt{\kbar^2}\ll 1$:\ 
$l\sqrt{\kbar^2}\coth(l\sqrt{\kbar^2})\ra 1$. 
(See the $da\abar$-part of Super QED, (\ref{qed14}).)
Hence the present bulk-boundary system can
be regarded as some
"deformation" of the corresponding 4D theory.
\footnote{
The similar propagator form appears also
in 5D bulk (AdS$_5$) approach at some limit\cite{RS0108b}. 
}

If we write the main "deformation" factor as follows
\begin{eqnarray}
\coth (l\sqrt{\kbar^2})=1-2\sum_{n=0}^{\infty}
\e^{-2nl\sqrt{\kbar^2}}
\com
\label{qe9}
\end{eqnarray}
the {\it role of the extra space} becomes clear.
Because the spectrum above 
($"E_n"=n\sqrt{\kbar^2}$)
shows that of the harmonic
oscillator in the temperature $T=1/(2l)$, it can be
translated as 
"the whole (4D Euclidean) system is exposed to the heat-bath
 and in the equilibrium
state with temperature $T=1/(2l)$". 
\footnote{
The same thing is commented in ref.\cite{RS0108b}
from the AdS$_5$ approach.
}
The size of the extra space
gives the {\it inverse temperature}.
The $l\sqrt{\kbar^2}\ll 1$ limit in the previous paragraph
corresponds to the {\it high temperature limit}.

Here the role of the {\it singular} term becomes clear. It is
a "{\it counterterm}" to cancel the divergences
coming from the {\it KK-mode summation}. 
The (f)+(b) part is independent
of $d$, hence it does {\it not} contribute to the SUSY effective potential.
\footnote{
This is consistent with $\Ncal=1$ SUSY non-renormalization theorem.
}
(We expect (f)+(b) part is cancelled by the vector- and spinor-loop 
contribution. ) This fact implies the above "smoothing" phenomenon
takes place {\it independently of the SUSY requirement}.
We should note that, after the above cancellation,
divergences, due to 4D-momentum integral, still remain. They correspond to the
ordinary divergences due to the (SUSY) local interaction.

Let us obtain renormalization group quantities from the previous
result, $\eta^\dag T^\ga \eta \, d_\ga$-term. 
The "deformed" propagator still
makes the ultraviolet
behaviour {\it worse} than the usual 4D propagator.
\begin{eqnarray}
l\times
\int\frac{d^4\kbar}{(2\pi)^4}\sqrt{\kbar^2}
\coth(l\sqrt{\kbar^2})\frac{1}{(\kbar^2)^2}
=\frac{1}{16\pi^2}\{ l\La-\ln 2-\ln\sinh l\ep\}
\com
\label{qe10}
\end{eqnarray}
where $\La$ and $\ep$ is the ultra-violet and infra-red cut-offs
respectively, $\ep\leq |\kbar|\leq \La$. 
(We do not care about the infra-red divergence
because it can be cured by taking the massive matter multiplet.)
It is {\it linearly} divergent. This result is reasonable from
the power counting. 
Now we consider the scaling behaviour of the gauge coupling
in the renormalization procedure. 
This problem is generally hard because the 5D
gauge theory is regarded (perturbatively) {\it nonrenormalizable}.
In the present case, however, we expect this model reduces
to a 4D SUSY gauge theory in the limit $l\ra 0$. 
%
%
We precisely define the limit as
\begin{eqnarray}
\frac{g^2}{l}\equiv \al\q (\mbox{fixed})\ll 1\com\nn
l\ra 0\com\q g\ra 0\com\q g^2\ll l
\com
\label{qe11}
\end{eqnarray}
where the {\it dimensionless} coupling $\al$ is introduced
instead of $g^2$. 
From the relation $g^2\La=\al\cdot l\La$, 
we are naturally led to introduce {\it another cut-off} $\Latil$ 
instead of $\La$ as
\begin{eqnarray}
l\La\equiv \ln \Latil
\pr
\label{qe12}
\end{eqnarray}
This relation connects two transformations, {\it scaling} and 
{\it translation}:
$\Latil\ra\Latil\e^\nu$(scaling) versus $\La\ra \La+\nu/l$(translation).
Then the renormalization group $\be$-function of the dimensionless coupling
$\al$ is obtained as
\begin{eqnarray}
g_b=g+\Del g=g(1+\frac{1}{8\times 16\pi^2}\al\ln \Latil)\com\nn
\al_b=\frac{1}{l}(g+\Del g)^2=\al (1+\frac{1}{4\times 16\pi^2}\al\ln\Latil+O(\al^2))\com\nn
0=\frac{d}{d(\ln\Latil)}\ln\al_b=(1+O(\al))\frac{d}{d(\ln\Latil)}\ln\al
+\frac{1}{4\times 16\pi^2}\al+O(\al^2)\com\nn
\be_\al=\frac{d}{d(\ln\Latil)}\ln\al=-\frac{1}{4\times 16\pi^2}\al
\com
\label{qe13}
\end{eqnarray}
where $g_b$ and $\al_b$ are bare quantities and G=SU(2) is taken
($T^\al T^\ga T^\al=-T^\ga/4$, see eq.(\ref{B3})).
The above result coincides with that of 
the ordinary 4D chiral-gauge SUSY theory (See textbooks\cite{West90,Freund86}).
We confirm here that the correct 4D renormalization works although
5D quantum loop expression (\ref{qe10}) is linearly divergent. 

The previous paragraph confirms that
the renormalization procedure works well at the 4D limit.
In this paragraph, 
we argue a "renormalization" procedure for the general
case of $l$ (not limited to the $l\ra 0$ case) and propose
a practical approach to define a finite quantity
from a divergent one ( such as (\ref{qe8}), (\ref{qe10}) ) coming from
the 5D quantum effect. 
In the present standpoint the extra axis is regarded as 
a {\it regularization} axis. 
\footnote{
The standpoint is the same as that appeared in the 
domain wall fermion of the lattice field theory.\cite{Kap92,Jan92,Sha93}
}
We have already pointed out that 
the {\it macro} parameter $1/l$ plays a role of 
the temperature $T=1/(2l)$ to smooth the UV behaviour. 
We recall a historically-famous fact in the beginning
of the quantum mechanics. In the Planck
distribution of the energy spectrum in a cavity, 
the light behaves like a wave in the
high temperature region compared with its own energy
(Rayleigh-Jeans's region, $k_B T\geq h\n$, where $k_B$ and $h$ are the 
Boltzman constant and the Planck constant respectively)
 whereas it behaves like a particle in the
low temperature region (Wien's region, $k_B T\leq h\n$). 
In the present case, we are treating 5D quantum dynamics which
can be regarded as a system composed of 4D Euclidean system of "light"
 (oscillator)
with energy $\sqrt{\kbar^2}$. They are thermally distributed in
the heat-bath with the temperature $T=1/(2l)$. 
Now we take a natural restriction on the present treatment of the 
5D quantum field theory:\ 
We cannot treat it in the Wien's region because, in this region,  
"the light" behaves like a particle 
and some new mass (energy) unit (probably the Planck mass)
should exist in the theory. At present, however, 
we do not have such mass unit and it implies 
the present {\it field theory} treatment breaks down
in the Wien's region. 
(In other words, "quantization" in the "phase" space of
$\La$ and $l$ is lacking. )
We must switch to an unknown treatment in order to obtain
a meaningful quantity from the divergent ones (\ref{qe8}), (\ref{qe10}).
Let us propose a condition on the 4D momentum cut-off $\La$. 
We should choose $\La$ in such a way that {\it the structure of the extra space} 
(the circle in the present case) {\it can not be recognized in the 4D world}.
\begin{eqnarray}
\La\ \lsim \ T=\frac{1}{2l}
\pr
\label{qe14}
\end{eqnarray}
If we adopt this idea
\footnote{
A comparative treatment was proposed by Randall and Schwarz
\cite{RS0108a,RS0108b}.
They examined the UV-divergence problem in 5D Yang-Mills theory
on the AdS$_5$ geometry. In the analysis, 4D-momentum/extra-space-coordinate
propagator is taken. They take such a regularization that the 4D-momentum
cut-off depends on the extra coordinate. They claim the linear divergence
reduces to log-divergence.
}
, the integral (\ref{qe10}) becomes well-defined( $l\La\lsim 1$). 
Here we propose a sort of "renormalization" for the present 5D quantum
system. Note here that the {\it UV cut-off} $\La$, of the {\it 4D momentum} $\kbar$, 
is essentially given by the inverse
of the {\it IR cut-off} parameter $l$ of the {\it extra space}. 
In this way, we have the following
relation
\begin{eqnarray}
|\kbar|\leq \La\lsim \frac{1}{2l}=T
\pr
\label{qe15}
\end{eqnarray}
This is a sort of "mass hierarchy" relation which appears in extra-dimensional
models.
The relation (\ref{qe15}) reminds us of the similar one that appears in the regularization
of fermion determinant (the domain wall fermion or the overlap formalism) 
in the lattice field theory. 
(See (32) of ref.\cite{SIpr00}, eq.(29) of ref.\cite{SInpb00}, and 
eq.(26) of ref.\cite{SIptp02}.)
The chiral symmetry in the fermion system corresponds
to $Z^2$-symmetry in the present case.


In this section, we have confirmed that the renormalization works well
as far as the 4D world is concerned. Aside from the (5D) renormalization
problem, we next examine the vacuum structure.

\section{Vacuum in the Brane World and Mass Matrix}
\begin{figure}
\centerline{ \psfig{figure=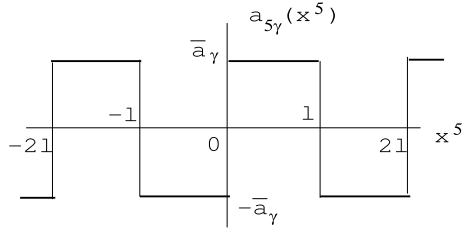,height=3cm,angle=0}}
\caption{ Behaviour of  the background field $a_{5\ga}(x^5)$
(the extra component of the bulk vector).
}
\label{fig:SignFunc}
\end{figure}

\begin{figure}
\centerline{ \psfig{figure=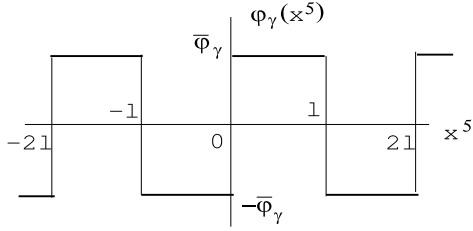,height=3cm,angle=0}}
\caption{ Behaviour of the background field
 $\vp_\ga(x^5)$(the bulk scalar).
}
\label{fig:SignFun2}
\end{figure}

First we examine the vacuum
in the present 5D approach. The relevant scalars
are $a_5$ and $\vp$, that is, the background fields
of $A_5$ (the extra component of the bulk vector) and 
$\Phi$ (the bulk scalar) respectively. 
They should be, in principle, given by solving
the (renormalized) equation of motion.
They describe the vacuum. 
We usually take the following procedure
in order to obtain a vacuum.\nl
\nl
[{\it Ordinary} procedure for the vacuum search\cite{SI84NP}]\nl
1) First we obtain the effective potential
assuming the {\it scalar} property of the vacuum
(as described in (\ref{ep1},\ref{ep2}))
and the {\it constancy} of the scalar vacuum
expectation values. \nl
2) Take the minimum of the effective
potential.\nl
\nl
At the present case, however, we should
take into account the {\it $x^5$-dependency}
and the {\it Z$_2$-property} of the vacuum expectation value.
We take the following forms of $a_5(\xf)$ and $\vp(\xf)$,
which describe the {\it localized} (around $x^5=0,l$) configuration and
a natural generalization
\footnote{
The condition of {\it constant} is generalized to
{\it piece-wise constant}. This is required from
the necessity of a non-trivial vacuum and 
the consistency with the $Z_2$ odd property. 
We stress here the present generalization, that is, 
the allowance of $\xf$-dependence on the vacuum scalars
($a_5$ and $\vp$), makes it possible to naturally introduce
the piece-wise constant ($\ep(\xf)$) in the theory. 
It is consistent with SUSY because the configuration (\ref{det6})
is obtained as a solution of the present SUSY theory. See App.C.
This situation should be compared with that appeared in the work
by Bergshoeff, Kallosh and Van Proeyen\cite{BKP0007}. 
They {\it replace} some constants (masses, couplings, $\cdots$) 
with supersymmetric singlet fields which behave as piece-wise
constants. They have to 
newly add (D-1)-form field in order to keep SUSY.
}
 of the ordinary treatment stated above.
\begin{eqnarray}
a_{5\ga}(\xf)=\abar_\ga\,\ep (\xf)\com\q \vp_\ga(\xf)=\vpbar_\ga\ep (\xf)
\com\nn
\ep(\xf)=\left\{
\begin{array}{cc}
+1 & \mbox{for }2nl<\xf<(1+2n)l \\
0  & \mbox{for } \xf=nl\\
-1 & \mbox{for }(2n-1)l<\xf<2nl \end{array}
\right.\q n\in {\bf Z}\com
\label{det6}
\end{eqnarray}
where $\ep(\xf)$ is the {\it periodic sign function}
with the periodicity $2l$. $\abar_\ga$ and $\vpbar_\ga$ are 
positive constants. See Fig.\ref{fig:SignFunc} and Fig.\ref{fig:SignFun2}.
It is shown, 
in App.C, that the above forms of $a_5(\xf)$ and 
$\vp(\xf)$ {\it satisfy the field equation} of the present model. 
The periodic sign function 
can be regarded as the {\it thin-wall limit} of a {\it kink} solution
and shows the {\it localization} of 
the bulk scalar and the extra component of the bulk vector. 
This generalization is also natural from the viewpoint
that the present theory starts with the {\it singular}
interaction ($\delta$-function term of (\ref{mp1}).).
We may use the {\it piecewise-continuous} 
or {\it piecewise-smooth} functions
as the theoretical materials, 
which is required from Z$_2$-property\cite{SI0008}.

We now begin to prepare for the full ( with respect to the coupling order) 
calculation of the 1-loop effective potential. 
The "1-loop" action, (\ref{ep9}),  can be expressed as
\begin{eqnarray}
S^2_a=S^{free}_a+S^{ghost}+\int d^5X
\nn
\times\half \left(\begin{array}{cccc}
\phi^\dag_{\al'} & \phi_{\al'} & \Phi_\al & A_{5\al} 
              \end{array}\right)
       {\left(\begin{array}{cc}
\left(\begin{array}{cc}
M_{\phi^\dag\phi} & M_{\phi^\dag\phi^\dag}  \\
M_{\phi\phi} & M_{\phi\phi^\dag}  \\
\end{array}
\right)_{\alp\bep}         &
\left(\begin{array}{cc}
M_{\phi^\dag\Phi} & 0 \\
 M_{\phi\Phi} & 0 
\end{array}
\right)_{\alp\be}          \\
\left(\begin{array}{cc}
M_{\Phi\phi} & M_{\Phi\phi^\dag} \\
0 & 0  
\end{array}
\right)_{\al\bep}         &
\left(\begin{array}{cc}
M_{\Phi\Phi}  & M_{\Phi A_5} \\
M_{A_5\Phi} & M_{A_5 A_5}
\end{array}
\right)_{\al\be}
             \end{array}\right)}
%
\left(\begin{array}{c}
\phi_{\be'} \\ \phi^\dag_{\be'} \\ \Phi_\be \\ A_{5\be}
              \end{array}\right) \nn
+(\phi'\mbox{ and }\phi^{'\dag} \mbox{involving terms}) , \nn
S^{free}_a=\intdX
\left[ \tr\,\{ -\pl_M\Phi\pl^M\Phi
-\pl_M A_{5}\pl^M A_{5}\}
-\del(\xf)\pl_m\phi^\dag\pl^m\phi\right],\nn
S^{ghost}=-\int d^5X\left[
\pl_M\cbar_\al\cdot\pl^Mc_\al
+igf_{\ab\ga}\pl_5\cbar_\al\cdot a_{5\be}c_\ga
\right]\com
\label{det1}
\end{eqnarray}
where each component is read from (\ref{ep9}) as
\begin{eqnarray}
(M_{\phi^\dag\phi})_{\alp\bep}=
g\del(\xf)d_\ga (T^\ga)_{\alp\bep}
-g^2\del(0)\del(\xf)(T^\ga\eta)_\alp (\eta^\dag T^\ga)_\bep\com
\nn
(M_{\phi\phi^\dag})_{\alp\bep}=
g\del(\xf)d_\ga (T^\ga)_{\bep\alp}
-g^2\del(0)\del(\xf)(\eta^\dag T^\ga)_\alp(T^\ga\eta)_\bep \com
\nn
(M_{\phi\phi})_{\alp\bep}=
-g^2\del(0)\del(\xf)(\eta^\dag T^\ga)_\alp (\eta^\dag T^\ga)_\bep 
\nn
(M_{\phi^\dag\phi^\dag})_{\alp\bep}=
-g^2\del(0)\del(\xf)(T^\ga\eta)_\alp (T^\ga\eta)_\bep \ ,
\nn
(M_{\Phi\phi})_{\al\bep}=(M_{\phi\Phi})_{\bep\al}=
g\pl_5\del(\xf)\cdot (\eta^\dag T^\al)_\bep 
\ ,
\nn
(M_{\Phi\phi^\dag})_{\al\bep}=(M_{\phi^\dag\Phi})_{\bep\al}=
g\pl_5\del(\xf)\cdot (T^\al\eta)_\bep
\ ,
\nn
%
%
(M_{\Phi\Phi})_{\al\be}=
-g\lpl_5f_{\ab\ga}a_{5\ga}+gf_{\ab\ga}a_{5\ga}\rpl_5
-g^2f_{\al\del\tau}f_{\be\ga\tau}a_{5\del}a_{5\ga}\ ,
\nn
(M_{A_5 A_5})_{\al\be}=
-g^2f_{\al\ga\tau}f_{\be\del\tau}\vp_\ga\vp_\del\ ,
\nn
(M_{A_5\Phi})_{\al\be}=
gf_{\ab\ga}\pl_5\vp_\ga-gf_{\ab\ga}\vp_{\ga}\rpl_5
-g^2f_{\ab\tau}f_{\ga\del\tau}a_{5\ga}\vp_{\del}
-g^2f_{\ga\be\tau}f_{\al\del\tau}a_{5\ga}\vp_{\del}\ ,
\nn
(M_{\Phi A_5})_{\al\be}=
-gf_{\ab\ga}\pl_5\vp_\ga+gf_{\ab\ga}\lpl_5\vp_{\ga}
+g^2f_{\ab\tau}f_{\ga\del\tau}a_{5\ga}\vp_{\del}
-g^2f_{\ga\al\tau}f_{\be\del\tau}a_{5\ga}\vp_{\del}
\pr\label{det2}
\end{eqnarray}
\vs 1

In the present analysis, as mentioned in Sect.5, 
we ignore the quantum propagation between the two branes.
We consider only the case that the quantum-loops propagate
between the $\xf=0$ brane and the bulk or purely
within the $\xf=0$ brane. 
Hence we may ignore the $\del(\xf-l)$-terms in the above expression.

From the periodicity ($\xf\ra\xf+2l$) and the Z$_2$-odd
property, the bulk fields $\Phi(X), A_5(X)$
can be KK-expanded as
\begin{eqnarray}
\Phi(x,\xf)=
\frac{1}{\sql}\sum_{n=1}^{\infty}\Phi_n(x)\sin(\npl\xf)\com\nn
A_5(x,\xf)=
\frac{1}{\sql}\sum_{n=1}^{\infty}A_n(x)\sin(\npl\xf)\com
\label{det3}
\end{eqnarray}
where the normalization is taken in the way:\ 
$
\int_{-l}^{l}\Phi^2d\xf=\sum_{n=1}^{\infty}{\Phi_n(x)}^2,\ 
\int_{-l}^{l}{A_5}^2d\xf=\sum_{n=1}^{\infty}{A_n(x)}^2.
$

We evaluate the action term by term.

\vs {.3}

(i) Free Part of the bulk and boundary system\nl

The free part $S^{free}_a$ can be obtained as
\begin{eqnarray}
S^{free}_a=\int d^4x\nn
\left[
\sum_{k=1}^\infty\tr
  \left\{
-\pl_m\Phi_k\pl^m\Phi_k-(\frac{k\pi}{l})^2\Phi_k^2
-\pl_mA_k\pl^mA_k-(\frac{k\pi}{l})^2A_k^2
  \right\}
  -\pl_m\phi^\dag\pl^m\phi
\right]\ .
\label{det4}
\end{eqnarray}
From the Z$_2$-odd property, zero KK-mode does not appear.
All quantum modes are massive with the order of $l^{-1}$. 

\vspace{10mm}

(ii) $M_{\phi^\dag\phi}, M_{\phi^\dag\phi^\dag},
M_{\phi\phi},  M_{\phi\phi^\dag}$ 
(boundary part)\nl

The boundary part can be read from (\ref{det2}).

\vspace{10mm}

(iii)$M_{\phi^\dag\Phi}, M_{\phi\Phi} ,M_{\Phi\phi}, M_{\Phi\phi^\dag}$
(bulk-boundary mixed part)
\begin{eqnarray}
\half\intdX\phi_\alp (M_{\phi\Phi})_{\alp\be}\Phi_\be=
\half\intdX\Phi_\al (M_{\Phi\phi})_{\al\bep}\phi_\bep=\nn
-\frac{g}{2\sql}
\intfx (\eta^\dag T^\be\phi)\sum_{n=1}^\infty\npl\Phi_{n\be}(x)\com\nn
\half\intdX\phi^\dag_\alp (M_{\phi^\dag\Phi})_{\alp\be}\Phi_\be=
\half\intdX\Phi_\al (M_{\Phi\phi^\dag})_{\al\bep}\phi^\dag_\bep=\nn
-\frac{g}{2\sql}
\intfx (\phi^\dag T^\al\eta)\sum_{n=1}^\infty\npl\Phi_{n\al}(x)\pr
\label{det5}
\end{eqnarray}

\vspace{3mm}

The remaining ones are bulk-bulk contribution.

\vs 1

(iv)$M_{\Phi\Phi}$\nl
The one part of 
$\half\intdX\,\Phi_\al (M_{\Phi\Phi})_\ab\Phi_\be\equiv S^{int}_{iv}$
is evaluated as
\begin{eqnarray}
S^{int}_{iv1}\equiv
\frac{-g^2}{2}\intdX f_{\al\del\tau}f_{\be\ga\tau}a_{5\del}a_{5\ga}\Phi_\al\Phi_\be\nn
=-\frac{g^2}{2}f_{\al\del\tau}f_{\be\ga\tau}\abar_{\del}\abar_{\ga}
\intfx\sum_{n=1}^\infty\Phi_{n\al}(x)\Phi_{n\be}(x)
\com
\label{det7}
\end{eqnarray}
where we use $a_{5\del}a_{5\ga}=\abar_{5\del}\abar_{5\ga}\ep(\xf)^2=
\abar_{5\del}\abar_{5\ga}$. The other part can be expressed as
\begin{eqnarray}
S^{int}_{iv2}\equiv
\frac{g}{2}\intdX f_{\ab\ga}a_{5\ga}(
-\pl_5\Phi_\al\,\Phi_\be+\Phi_\al\pl_5\Phi_\be)\nn
=-gf_{\ab\ga}\abar_{\ga}\intfx\int_{-l}^{l}d\xf \ep(\xf)\pl_5
\Phi_\al\cdot\Phi_\be
\pr
\label{det8}
\end{eqnarray}
Using the Fourier expansion of the periodic sign function,
\begin{eqnarray}
\ep(x)=\frac{4}{\pi}\sum_{n=0}^\infty\frac{1}{2n+1}
\sin\{\frac{(2n+1)\pi}{l}x\}
\com
\label{det9}
\end{eqnarray}
we obtain
\begin{eqnarray}
S^{int}_{iv2}=\frac{2g}{l}f_{\ab\ga}\abar_\ga
\intfx\sum_{m=1}^\infty\sum_{n=1}^\infty
\Phi_{m\al}m Q_{mn}\Phi_{n\be}\com\nn
\int_{-l}^{l}d\xf\ep(\xf)\cos(\mpl\xf)\sin(\npl\xf)
=-\frac{2l}{\pi}Q_{mn}\com\nn
Q_{mn}=\left\{
\begin{array}{cc}
\frac{1}{m-n}\q & m-n=\mbox{odd} \\
0    &  m-n=\mbox{even}
\end{array}
\right.
=\left\{
\begin{array}{cc}
\half\{1-(-1)^{m-n}\}\frac{1}{m-n}\q & m\neq n \\
0    &  m=n
\end{array}
\right.\pr
\label{det10}
\end{eqnarray}
We note the anti-symmetricity:\ $Q_{mn}=-Q_{nm}$. 

\vs 1

(v)$M_{A_5 A_5}$\nl

\begin{eqnarray}
S^{int}_{v}\equiv
\half\intdX A_{5\al}(X)(M_{A_5A_5})_\ab A_{5\be}(X)\nn
=-\frac{g^2}{2l}f_{\al\ga\tau}f_{\be\del\tau}
\sum_{n=1}^\infty\sum_{k=1}^\infty\intfx A_{n\al}(x)A_{k\be}(x)\times\nn
\int_{-l}^{l}d\xf \,\vp_\ga(\xf)\vp_\del(\xf)\sin(\npl\xf) 
\sin(\kpl\xf)
\pr
\label{det11}
\end{eqnarray}

Using the localized form of $\vp(\xf)$ given in (\ref{det6}), 
we obtain
\begin{eqnarray}
S^{int}_{v}=
-\frac{g^2}{2}f_{\al\ga\tau}f_{\be\del\tau}\,
\sum_{n=1}^\infty\vpbar_\ga\vpbar_\del
\intfx A_{n\al}(x)A_{n\be}(x)
\pr
\label{det13}
\end{eqnarray}

\vs 1

(vi)$M_{\Phi A_5}, M_{A_5\Phi}$\nl

\begin{figure}
\centerline{ \psfig{figure=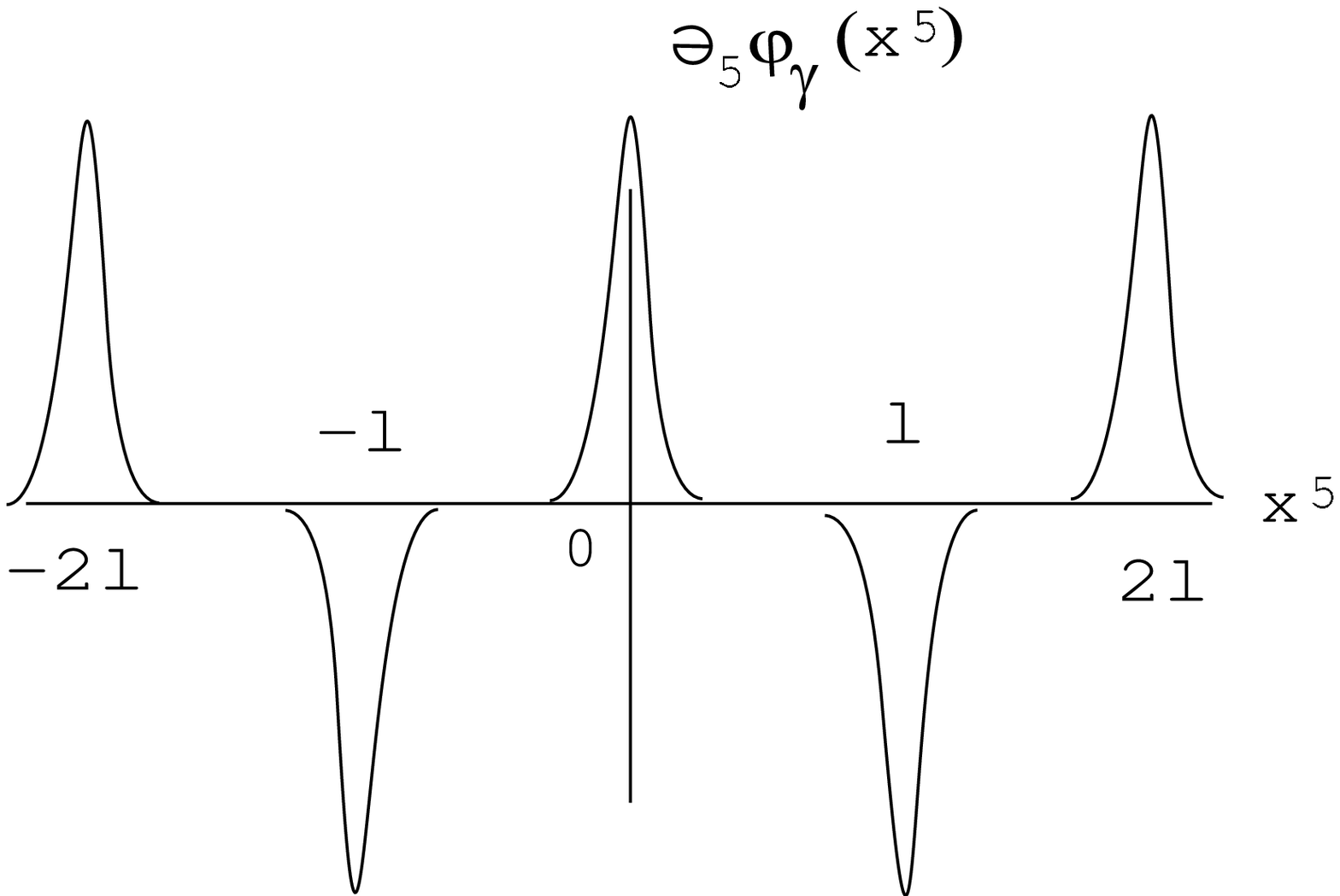,height=3cm,angle=0}}
\caption{ Behaviour of 
 $\pl_5\vp_\ga(x^5)$.
}
\label{fig:DelFunc}
\end{figure}

This group consists of four terms. 
\begin{eqnarray}
S^{int}_{vi}\equiv\half\intdX\Phi_\al (M_{\Phi A_5})_\ab A_{5\be}
=\half\intdX A_{5\al}(M_{A_5\Phi})_\ab \Phi_\be       \nn
=S^{int}_{vi1}+S^{int}_{vi2}+S^{int}_{vi3}+S^{int}_{vi4}
\com\nn
S^{int}_{vi1}=-\half\intdX gf_{\ab\ga}\Phi_\al\pl_5\vp_\ga\, A_{5\be}
\ ,\ 
S^{int}_{vi2}=\half\intdX gf_{\ab\ga}\pl_5\Phi_\al\,\vp_\ga A_{5\be}
\ ,\nn
S^{int}_{vi3}=\half\intdX g^2f_{\ab\tau}f_{\ga\del\tau}\Phi_\al
a_{5\ga}\vp_\del A_{5\be}
\ ,\ 
S^{int}_{vi4}=-\half\intdX g^2f_{\ga\al\tau}f_{\be\del\tau}\Phi_\al
a_{5\ga}\vp_\del A_{5\be}
\ .
\label{det14}
\end{eqnarray}
Here we note the relation
\begin{eqnarray}
\pl_5\vp_\ga=2\vpbar_\ga\{\del(\xf)-\del(\xf-l)\}
\com
\label{det15}
\end{eqnarray}
which expresses the {\it localization} of the bulk scalar. 
$\del(x)$ is the periodic (periodicity $2l$) delta function. 
See Fig.\ref{fig:DelFunc}.
Using the above equation, we can evaluate the first term as follows.
\begin{eqnarray}
S^{int}_{vi1}=-gf_{\ab\ga}\vpbar_\ga\intfx
\left[\left.\Phi_\al A_{5\be}\right|_{\xf=0}-
\left.\Phi_\al A_{5\be}\right|_{\xf=l}\right]=0
\pr
\label{det16}
\end{eqnarray}
The third and fourth terms are evaluated as
\begin{eqnarray}
S^{int}_{vi3}=\frac{g^2}{2}f_{\ab\tau}f_{\ga\del\tau}
\abar_{\ga}\vpbar_\del
\sum_{n=1}^\infty\intfx\,\Phi_{n\al}(x)A_{n\be}(x)\com
\nn
S^{int}_{vi4}=-\frac{g^2}{2}f_{\ga\al\tau}f_{\be\del\tau}
\abar_{\ga}\vpbar_\del
\sum_{n=1}^\infty\intfx\,\Phi_{n\al}(x)A_{n\be}(x)\pr
\label{det17}
\end{eqnarray}
Using the relation (\ref{det10}), we obtain
\begin{eqnarray}
S^{int}_{vi2}=-\frac{g}{l}f_{\ab\ga}\vpbar_\ga
\sum_{m=1}^\infty\sum_{n=1}^\infty\Phi_{m\al}m Q_{mn}A_{n\be}
\pr
\label{det18}
\end{eqnarray}

The background fields we take, (\ref{det6}), 
satisfy the required boundary condition. They also satisfy
the on-shell condition (\ref{ep5b}) for an appropriate
choice of $\abar, \vpbar, \eta, \eta^\dag$ and $\chi^3$.
Explanation is given in App.C.

\vs 1

We summarize the results of (i)-(vi) as follows.
\begin{eqnarray}
S^2_a=S^{ghost}+\int d^4x\times\nn
\half \left(\begin{array}{cccc}
\phi^\dag_{\al'} & \phi_{\al'} & \Phi_{m\al} & A_{m\al} 
              \end{array}\right)
       {\left(\begin{array}{cc}
\left(\begin{array}{cc}
\Mcal_{\phi^\dag\phi} & \Mcal_{\phi^\dag\phi^\dag}  \\
\Mcal_{\phi\phi} & \Mcal_{\phi\phi^\dag}  \\
\end{array}
\right)_{\alp\,\bep}         &
\left(\begin{array}{cc}
\Mcal_{\phi^\dag\Phi} & 0 \\
\Mcal_{\phi\Phi} & 0 
\end{array}
\right)_{\alp \, n\be}          \\
\left(\begin{array}{cc}
\Mcal_{\Phi\phi} & \Mcal_{\Phi\phi^\dag} \\
0 & 0  
\end{array}
\right)_{m\al \, \bep}         &
\left(\begin{array}{cc}
\Mcal_{\Phi\Phi}  & \Mcal_{\Phi A} \\
\Mcal_{A\Phi} & \Mcal_{A A}
\end{array}
\right)_{m\al \, n\be}
             \end{array}\right)}
%
\left(\begin{array}{c}
\phi_{\be'} \\ \phi^\dag_{\be'} \\ \Phi_{n\be} \\ A_{n\be}
              \end{array}\right)\nn
+(\phi'\mbox{ and }\phi^{'\dag} \mbox{involving terms})
\com\label{det19}
\end{eqnarray}
where the integer suffixes $m$ and $n$ runs from 1 to $\infty$, and 
each component is described as
\begin{eqnarray}
\Mcal_{\phi^\dag_\alp\phi_\bep}=
\pl^2\del_{\alp\bep}+
gd_\ga (T^\ga)_{\alp\bep}-g^2\del(0)(T^\ga\eta)_\alp
(\eta^\dag T^\ga)_\bep\com
\nn
\Mcal_{\phi^\dag_\alp\phi^\dag_\bep}=
-g^2\del(0)(T^\ga\eta)_\alp (T^\ga\eta)_\bep\com\q
\Mcal_{\phi_\alp\phi_\bep}= 
-g^2\del(0)(\eta^\dag T^\ga)_\alp (\eta^\dag T^\ga)_\bep\com
\nn
\Mcal_{\phi_\alp\phi^\dag_\bep}= 
\pl^2\del_{\alp\bep}+
gd_\ga (T^\ga)_{\bep\alp}-g^2\del(0)(\eta^\dag T^\ga)_\alp (T^\ga\eta)_\bep
\nn
\Mcal_{\phi^\dag_\alp\Phi_{n\be}}=
-\frac{g}{\sql}(T^\be\eta)_\alp\npl=\Mcal_{\Phi_{n\be}\phi^\dag_\alp}\com\q
\Mcal_{\phi_\alp\Phi_{n\be}}=
-\frac{g}{\sql}(\eta^\dag T^\be)_\alp\npl=\Mcal_{\Phi_{n\be}\phi_\alp}\com
\nn
\Mcal_{\Phi_{m\al}\Phi_{n\be}}=
-\{-\pl^2+(\npl)^2\}\del_{mn}\del_\ab
-g^2f_{\al\del\tau}f_{\be\ga\tau}\abar_\del\abar_\ga\del_{mn}
+\frac{4g}{l}f_{\ab\ga}\abar_\ga mQ_{mn}\com
\nn
\Mcal_{\Phi_{m\al} A_{n\be}}=
g^2f_{\ab\tau}f_{\ga\del\tau}\abar_\ga\vpbar_\del\del_{mn}
-g^2f_{\ga\al\tau}f_{\be\del\tau}\abar_\ga\vpbar_\del\del_{mn}
-\frac{2g}{l}f_{\ab\ga}\vpbar_\ga mQ_{mn}
=\Mcal_{A_{n\be}\Phi_{m\al}} ,
\nn
\Mcal_{A_{m\al} A_{n\be}}=
-\{-\pl^2+(\npl)^2\}\del_{mn}\del_\ab
-g^2 f_{\al\ga\tau}f_{\be\del\tau}\vpbar_\ga\vpbar_\del\del_{mn}\pr
\label{det20}
\end{eqnarray}
where the kinetic (free) part, $S^{free}_a$, is included 
($\pl^2\equiv \pl_m\pl^m$ is the 4D Laplacian)
in the ``Mass'' matrix. The repeated indices imply the Einstein's 
summation convention. 
\footnote{
For the convenience, we list the physical dimensions of various
quantities. 
$
[\eta]=[\pl_m]=M, [\abar]=[\vpbar]=M^{3/2}, [d]=M^{5/2}, 
[g]=M^{-1/2}, [l]=M^{-1}, [\del(0)]=M, 
[\phi]=[\phi^\dag]=[\Phi_n]=[A_n]=M.
$
}

\vs 1
\section{Effective Potential of Bulk-Boundary System}
As shown in Sec.2 and App.A for simple models, the effective potential
is obtained from the eigenvalues of the relevant mass-matrix
obtained by the background expansion. 
Let us obtain the effective potential from the mass matrix
$\Mcal$ of (\ref{det19}) and (\ref{det20}).
It is composed of three field values $\eta,\eta^\dag,d_\al$,
two wall "heights", $\abar_\al,\vpbar_\al$, the gauge coupling, $g$
and the boundary parameter $l$. 
The full explicit calculation, even at 1-loop level,
is technically hard. We obtain some interesting "sections" of the full result:
Case (A), $\eta=0,\eta^\dag=0$;\ Case (B), $\abar=0, \vpbar=0$.
Detailed explanation is given in App.B.
\nl\nl
Case (A): $\eta=0,\eta^\dag=0$\nl
We look the potential with the suppression of the scalar matter
dependence. 
(Or we may say we look the potential from the $\eta=\eta^\dag=0$
point. )
In this case the mass matrix $\Mcal$ has the following
properties:
\ (1) In $\Mcal$, the boundary part and the bulk one {\it decouple} each other;
\ (2) All $\del(0)$-terms disappear.  The boundary-loop quantum effect
gives rise to the following potential before the renormalization procedure:
\begin{eqnarray}
V^{eff}_{1-loop}
=\int\frac{d^4k}{(2\pi)^4}
\ln \{1-\frac{g^2}{4}\frac{d^2}{(k^2)^2}\}
\com
\label{EP1}
\end{eqnarray}
where we take G=SU(2) as the internal gauge group. The behaviour is
similar to the super QED explained in App.A. 
The above expression, when treated perturbatively, is
{\it logarithmically} divergent. 
Noting the relation:\ $d_\al=(\chi^3-\pl_5\vp+g a_5\times\vp)_\al,$ 
we realize the renormalization
procedure connects the boundary and the bulk phenomena through 
the field renormalization of $X^3$ and $\Phi$ although we do not touch on
the renormalization of the bulk fields. 
\footnote{
The importance of the "communication" between the bulk
and boundary renormalizations was stressed by
Goldberger and Wise\cite{GW0104}.
}

The bulk-loop quantum effect does not give the $d_\al$-dependence
to the vacuum energy. Hence it does {\it not} 
contribute to the effective potential
after the use of the SUSY boundary condition. It gives, however, an important
result:\  the scalar-loop contribution to the vacuum energy
which depends on the "wall heights" $\abar$ and $\vpbar$. 
\footnote{
No Casimir energy in the SUSY invariant theory is reasonable
from the general result about 
the vanishing energy of the SUSY vacuum.
}
(We expect some part of the contribution appear when the boson-fermion
balance breaks down due to some SUSY breaking mechanism.
) 
We can regard it as a {\it new type Casimir energy}, because $\abar$ and $\vpbar$
can be regarded as 
different-type boundary parameters from $l$.\cite{IMres03} 
For the large circle limit $l\ra\infty$, 
the final result of the
new Casimir energy, per one KK-mode, is
\begin{eqnarray}
\frac{1}{l}V^{eff}_{Casimir}
=\frac{g^2}{l^3}(\al_1\vpbar^2+\al_2\abar^2
+\al_3\abar\cdot\vpbar)+O(g^4)
\com
\label{EP2}
\end{eqnarray}
where $\al_1, \al_2$ and $\al_3$ are some constants.
The new points, compared with the ordinary Casimir energy\cite{AC83},
are 1) the potential depends on the circle radius as $l^{-3}$;\ 
2) the potential depends on the gauge coupling $g$;\ 
3) the potential depends on the "wall heights", $\vpbar$ and $\abar$.
We expect the above quantity (\ref{EP2}) does not depend on 
the gauge we have chosen
\cite{SI85PL}. 
This contribution from the scalar loop, however, 
is expected to be cancelled by those from the fermion
and vector loops in the present SUSY-invariant setting.\nl

Case (B): $\abar=0, \vpbar=0$\nl
 In this case the brane structure disappears. The situation
is similar to the case of Appelquist and Chodos(AC). From the bulk
modes of $\Phi$ and $A_5$, we have AC-type eigenvalues.
\begin{eqnarray}
\la_n=-k^2-(\npl)^2\com\q n=1,2,3,\cdots
\com
\label{EP3}
\end{eqnarray}
(See (\ref{B10}) and (\ref{B18})).
This gives the famous form of the Casimir energy.
\begin{eqnarray}
\frac{1}{l}V^{eff}_{Casimir}
=\frac{\mbox{const}}{l^5}
\pr
\label{EP4}
\end{eqnarray}
This is the scalar-loop contribution and is expected to be cancelled
by other non-scalar fields effect.

The eigenvalues for the boundary part is obtained as 
a complicated expression involving the following terms:
\begin{eqnarray}
S\equiv\eta^\dag\eta\ ,\ d^2=d_\al d_\al\ ,\ 
d\cdot V\equiv d_\al \, \eta^\dag T^\al\eta\ ,\ 
V^2\equiv (\eta^\dag T^\al\eta)^2 
\pr
\label{B20a}
\end{eqnarray}
We have the full expression in the computer file.
In the manipulation of eigenvalues search
(determinant calculation), 
we face
the following combination of terms.
\begin{eqnarray}
\del(0)+\frac{1}{l}\sum_{m=1}^{\infty}
\frac{(\pi m/l)^2}{-\la-k^2-(\pi m/l)^2}
\pr
\label{B19x}
\end{eqnarray}
The first term comes from the singular terms
in $\Mcal$, the second from the KK-mode sum.
Using the relation$\sum_{m\in\bfZ}1=2l\del(0)$, 
the above sum leads to a regular quantity.
\begin{eqnarray}
{\del(0)}|_{sm}=
\frac{1}{2l}\sum_{m\in\bfZ}
\frac{\la+k^2}{\la+k^2+(\pi m/l)^2}
=\left\{
\begin{array}{cc}
\half\sqrt{\la+k^2}\coth\{l\sqrt{\la+k^2}\} & \la>-k^2 \\
\half\sqrt{-\la-k^2}\cot\{l\sqrt{-\la-k^2}\} & -k^2>\la 
\end{array}
\right.
\pr
\label{B20x}
\end{eqnarray}
We have confirmed this "smoothing" phenomenon
occurs at the 1-loop {\it full} level.

The effective potential induced on the boundary comes from
the eigenvalues depending on the field $d_\al$. When we look at
$d^2$-part, the following ones are obtained as the dominant part.
\begin{eqnarray}
\la_\pm=-k^2\pm \frac{g}{2}\sqrt{d^2}
\pr
\label{EP5}
\end{eqnarray}
This is the same as Case A. 

\begin{figure}
\centerline{ \psfig{figure=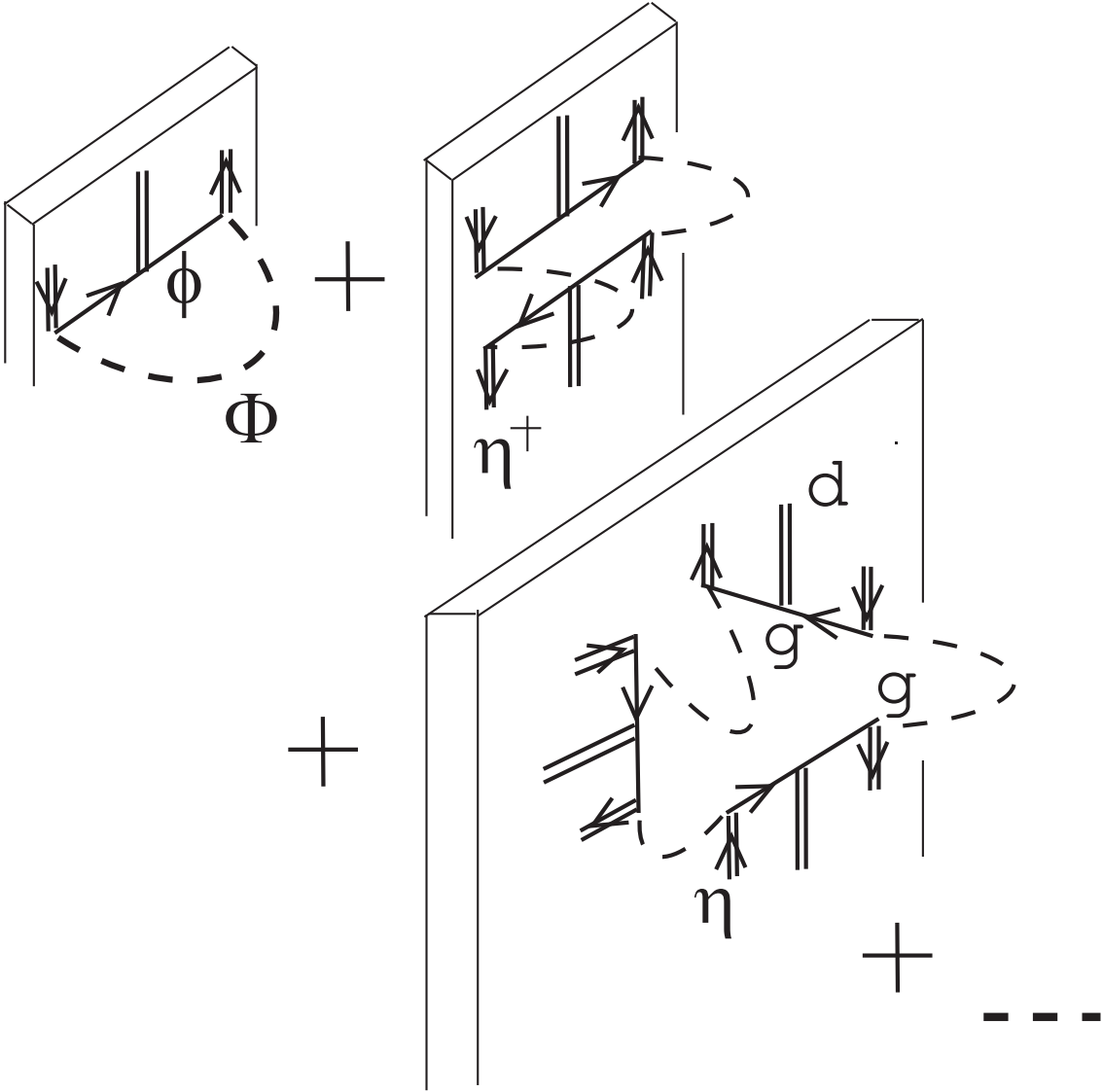,height=15cm,angle=0}}
\caption{ 
The diagrams contributing to the effective potential
at the full (w.r.t. $g$) 1-loop level.
The dotted lines represent quantum propagation of the
bulk scalar $\Phi$, the directed single lines represent
that of the boundary (4D) scalars $\phi, \phi^\dag$.
The double lines represent boundary background fields
$\eta, \eta^\dag$ and $d$. The "wall" represents the $\xf=0$
brane. 
The corresponding eigenvalues are given by (\ref{EP6}).
This figure is the bulk-boundary generalization of FIG.2
of ref.\cite{CW73}.
}
\label{fig:dV1loop}
\end{figure}

When we look at the $d\cdot V \equiv d_\al\eta^\dag T^\al\eta$ part,
the eigenvalues are dominated by the solutions of
the following equation.
\begin{eqnarray}
 (\la+k^2)^2-\frac{g^3}{2}d\cdot V\frac{\sqrt{\la+k^2}}{2}
\coth l\sqrt{\la+k^2}=0
\pr
\label{EP6}
\end{eqnarray}
In the perturbative approach, this equation gives, 
in the $O(g^3)$ order, two eigenvalues $\la_1, \la_2$
which satisfy
\begin{eqnarray}
\la_1\la_2=
(k^2)^2\left( 1-\frac{g^3}{4}d\cdot V
\frac{\sqrt{k^2}\coth l\sqrt{k^2}}{(k^2)^2}\right)
\com
\label{EP7}
\end{eqnarray}
(see (\ref{B26})). This is the same as (\ref{qe8}). 
The eigenvalues obtained as the full
solutions of (\ref{EP6}) gives 
the effective potential at the 1-loop {\it full} level.
The corresponding diagrams are shown in Fig.\ref{fig:dV1loop}.
The figure is a bulk-boundary generalization of the
Coleman-Weinberg's case\cite{CW73}.
\footnote{
If we take the 4D-limit, $l\sqrt{k^2}\ll 1$, in (\ref{EP7}),
we see the result essentially reduces to the ordinary type
appearing in 4D theory (such as a term, (\ref{qed14}), in 4D Super QED).
}
\nl

We succeed in obtaining the full 1-loop eigenvalues
induced by the bulk-boundary quantum effect.

\vs 1
\section{Conclusion}

We have analyzed the quantum structure of a bulk-boundary
system by taking the example of the Mirabelli-Peskin model.
The analysis is newly formulated by the background
field method. Feynman rules for the perturbative calculation
are presented. 
We focus on the (1-loop) effective potential and the
vacuum energy. 
It is confirmed that the singular terms
well behave with 
 the Kaluza-Klein modes summation. 
The whole effect can be regarded as
some {\it deformation} of the 4D quantization.  
Its 4D reduction by $l\ra 0$ is confirmed
in the renormalization group calculation. 
The characteristic relation among the 4D-momentum $\kbar$,
UV-cutoff $\La$, and the IR-cutoff($S^1$ radius) $l$ appears. 
It comes from the requirement to escape from the linear divergence.
The relation is the same one as in the lattice domain wall fermion.
In addition to
the bulk scalar $\Phi$, the extra component of the bulk
vector $A^5$ plays an important role in determining
the vacuum.
Especially their localized configurations are exploited. 
In the treatment, the vacuum is generalized in the sense that
scalars may depend on the extra coordinate $\xf$. 
In the intermediate stage, we have obtained a new type Casimir energy
in addition to the ordinary type by Appelquist and Chodos. 
The obtained result of the effective potential includes
the bulk-boundary generalization of the Coleman-Weinberg's case.

We hope the present analysis advances further development
of the brane world physics.

\vs 1
\begin{flushleft}
{\bf Acknowledgment}
\end{flushleft}
The authors thank N. Sakai for valuable comments 
when this work, still at the primitive stage, 
was presented at the Chubu Summer School 2002 (Tsumagoi, Gunma, Japan,
2002.8.30-9.2). 
Parts of the present results were presented
at the seminar of DAMTP, Univ. of Cambridge (2003.1.24). 
The encouraging comment by G.W. Gibbons is much appreciated. 
Some results were also presented
at the annual meetings of the Physical Society of Japan
(Rikkyo Univ.,Tokyo, Japan, 2002.9.13-16;
Thohoku Gakuin Univ.,Sendai,Japan, 2003.3.28-31;
Miyazaki World Convention Center Summit,Miyazaki,Japan,2003.9.9-12)
 and SUSY04(Jun.17-23, Tsukuba, Japan). 
The authors thank the audience for stimulated comments. 

\vs 1
\section{Appendix A:\ Effective Potential of Super QED}
Now we consider the super QED. The action is most concisely described by one vector

superfield $V$ and two chiral ones $S$ (charge $g$) and $R$ (charge $-g$):\ 
\begin{eqnarray}
S=\int d^4xd^2\theta d^2\thbar ({\bar S}\e^{gV}S+{\bar R}\e^{-gV}R)+
\left\{ \int d^4xd^2\theta(\fourth WW+mSR)+\mbox{h.c.}\right\}\com\nn
W^\al=-\fourth {\bar D}^2D^\al V\com\q
V=-\theta\si^m\thbar~v_m+i\theta\theta\thbar~\labar-i\thbar\thbar\theta~\la
+\half\theta\theta\thbar\thbar~D\com\nn
S=A_S(y)+\sqrt{2}\theta~\psi_S(y)+\theta\theta~F_S(y)\ ,\ 
R=A_R(y)+\sqrt{2}\theta~\psi_R(y)+\theta\theta~F_R(y)\ ,\nn
y^m=x^m+i\theta\si^m\thbar
\pr
\label{qed1}
\end{eqnarray}
We focus on the scalar sector
of the effective potential, based on the following points:\ 
1) Lorentz invariance of the vacuum, 
2) the 1-loop contribution from the non-scalar fields 
(spinors, vectors) can be recovered
by taking "{\it supersymmetric boundary condition}". 
We put the condition.
\begin{eqnarray}
v_m=0\com\q \la=\labar=0\com\q \psi_S=\psi_R=0
\pr
\label{qed2}
\end{eqnarray}
Then the Lagragian of the super QED reduces to the simple form.
\begin{eqnarray}
\Lcal[A_S,F_S;A_R,F_R;D:\ m,g]=\nn
\{\Abar_S\pl_m\pl^m A_S+\Fbar_S F_S
+m(A_RF_S+\Abar_R\Fbar_S)+R\change S\}\nn
+\{\half D^2+\half g D(\Abar_SA_S-\Abar_RA_R)\}
\pr
\label{qed3}
\end{eqnarray}
Now we expand all scalar fields around the background {\it constants}
($a_S,f_S,\cdots$). 
\begin{eqnarray}
A_S\ra a_S+A_S\ ,\ F_S\ra f_S+F_S\ ,\ D\ra d+D\ ,\ 
\Abar_S\ra {\bar a}_S+\Abar_S\ ,\ \Fbar\ra \fbar_S+\Fbar_S
\com
\label{qed4}
\end{eqnarray}
and similarly for $A_R,\Abar_R,F_R,\Fbar_R$. 

The effective potential $V^{eff}[a,\abar,f,\fbar,d]$ is defined as
\begin{eqnarray}
\exp\{ -i\intfx V^{eff}[a,\abar,f,\fbar,d]\}= 
\int(\Dcal A\Dcal F\Dcal \Abar\Dcal \Fbar)_{S,R}\Dcal D\nn
\times \exp i\intfx\left\{ \Lcal[a+A,\abar+\Abar,f+F,\fbar+\Fbar,d+D]
-\left.\frac{\del\Lcal}{\del\Phi^I} \right|_b\Phi^I
             \right\}     \nn
=\exp i\{ -\intfx V^{eff}_0\}
\times
\int(\Dcal A\Dcal F\Dcal \Abar\Dcal \Fbar)_{S,R}\Dcal D \nn
\times\exp i\intfx\left\{ \Lcal_2+(\mbox{quantum field})^3\ 
\mbox{and higher-order}\ 
             \right\}\com
\label{qed4b}
\end{eqnarray}
where $(\Phi^I)\equiv (A_{S,R}\,,F_{S,R}\,,\Abar_{S,R}\,,\Fbar_{S,R}\,,D)$ 
are treated as the quantum fields 
and 
$(\Phi^I)|_b\equiv (a_{S,R}\,,f_{S,R}\,,\abar_{S,R}\,,\fbar_{S,R}\,,d)$
are as the background fields. 
$-V^{eff}_0$ is the tree (zero-th order) part 
and is given below. $\Lcal_2$ 
is the quadratic part and will be given in (\ref{qed9}). 
The zero-th order is
\begin{eqnarray}
\Lcal_0=-V^{eff}_0=\{\fbar_Sf_S+m(a_Rf_S+\abar_R\fbar_S)+R\change S\}
+\half d^2+\frac{g}{2}d(\abar_S a_S-\abar_R a_R)
\pr\label{qed5}
\end{eqnarray}
The first order is given by
\begin{eqnarray}
\Lcal_1
=\left.\frac{\del\Lcal}{\del\Phi^I} \right|_b\Phi^I
=\{ \Fbar_S(f_S+m\abar_R)+(\fbar_S+ma_R)F_S+R\change S\}\nn
+\{A_R(mf_S-\frac{g}{2}d\abar_R)+A_S(mf_R+\frac{g}{2}d\abar_S)+\mbox{h.c.}\}
+D\{d+\frac{g}{2}(\abar_S a_S-\abar_R a_R)\}
\ .\label{qed6}
\end{eqnarray}
There is a special choice of the background constants, 
the {\it on-shell condition}:\ 
\begin{eqnarray}
f_S+m\abar_R=0\com\q f_R+m\abar_S=0\com\q
mf_S-\frac{g}{2}d\abar_R=0\com\q mf_R+\frac{g}{2}d\abar_S=0\com\nn
d+\frac{g}{2}(\abar_S a_S-\abar_R a_R)=0
\com\label{qed7}
\end{eqnarray}
and their complex conjugate. 
This is the solution of 
the field equation: $\Lcal_1=0$. 
When the background constants satisfy the above equations, 
the tree effective potential $V^{eff}_0$ takes
\begin{eqnarray}
V^{eff}_0|_{\mbox{on-shell}}=\fbar_S f_S+\fbar_R f_R+\half d^2\geq 0
\pr\label{qed8}
\end{eqnarray}
From the results (\ref{qed7}) and (\ref{qed8}), 
the (classical) vacuum is given by the solution:\ 
$f_S=f_R=0,\ d=0,\ a_R=a_S=0$\ 
 where $m\neq 0$ is assumed. 
In the following analysis, however, 
we consider the general case of the background constants.
(We do {\it not} require the on-shell condition: $\Lcal_1=0$.)

The second order part $\Lcal_2$ is given by taking
the quadratic terms with respect to the quantum fields
($A_{S,R}\,,F_{S,R}\,,\Abar_{S,R}\,,\Fbar_{S,R}\,,D$). 
$\Lcal_2$ can be expressed in the following form,
where $F,\Fbar$-involved terms are separated.
\footnote{
$f$ and $\fbar$ disappear at this stage because the
$F$ and $\Fbar$-auxiliary fields appear in (\ref{qed3})
only as quadratic terms.
}
\begin{eqnarray}
\Lcal_2=\half \left(\begin{array}{ccccc}
\Abar_S & A_S & \Abar_R & A_R & D
              \end{array}\right)
{\bf A}
\left(\begin{array}{c}
A_S \\ \Abar_S \\ A_R \\ \Abar_R \\ D
              \end{array}\right)  
+\{\Fbar_SF_S+m(A_RF_S+\Abar_R\Fbar_S)+R\change S\}\ ,\nn
{\bf A}=
 \left(\begin{array}{ccccc}
\Box+dg/2 & 0 & 0 & 0 & ga_S/2\\
0 & \Box+dg/2  & 0 & 0 & g\abar_S/2\\
0 & 0 & \Box-dg/2  & 0 & -ga_R/2\\
0 & 0 & 0 & \Box-dg/2  & -g\abar_R/2\\
g\abar_S/2 & ga_S/2 & -g\abar_R/2 & -ga_R/2 & 1
       \end{array}\right)
\pr\label{qed9}
\end{eqnarray}
The above matrix ${\bf A}$ is the same as that in Ref.\cite{Mil83NP}.
\footnote{
In the paper, however, the contribution from $F$ and $\Fbar$ is not
taken into account.}
Integrating out all auxiliary fields
$D,F_S,F_R,\Fbar_S,\Fbar_R$ using "squaring equations":
\begin{eqnarray}
\left\{ \half g (a_S\Abar_S-a_R\Abar_R)D+\mbox{c.c.}\right\}
+\half D^2\nn
=\half \{D+\frac{g}{2}(a_S\Abar_S-a_R\Abar_R+\mbox{c.c.})\}^2
-\frac{g^2}{8}(a_S\Abar_S-a_R\Abar_R+\mbox{c.c.})^2\com\nn
\Fbar_SF_S+m(A_RF_S+\Abar_R\Fbar_S)+(S\change R)\nn
=(\Fbar_S+mA_R)(F_S+m\Abar_R)-m^2A_R\Abar_R+(S\change R)
\com\label{qed10}
\end{eqnarray}
$\Lcal_2$ reduces to
\begin{eqnarray}
\Lcal_2'=
\Abar_S\Box A_S+\Abar_R\Box A_R+
\half \left(\begin{array}{cccc}
\Abar_S & A_S & \Abar_R & A_R 
              \end{array}\right)
{\bf M}
\left(\begin{array}{c}
A_S \\ \Abar_S \\ A_R \\ \Abar_R 
              \end{array}\right)  \ ,\ \Box=\pl_m\pl^m \ ,\nn
{\bf M}=
 \left(\begin{array}{cccc}
-Ga_S\abar_S+\dtil-m^2 & -G\abar_S\abar_S & Ga_R\abar_S   & G\abar_R\abar_S \\
-Ga_S a_S   & -G\abar_Sa_S+\dtil-m^2  & Ga_R a_S   & G\abar_Ra_S \\
Ga_S\abar_R   & G\abar_S\abar_R  & -Ga_R\abar_R-\dtil-m^2 & -G\abar_R\abar_R \\
Ga_S a_R    & G\abar_Sa_R     & -Ga_R a_R        & -G\abar_Ra_R-\dtil-m^2  
 \end{array}\right)
.\nn
\label{qed11}
\end{eqnarray}
where $G\equiv g^2/4,\ \dtil\equiv dg/2$. The four eigenvalues 
of ${\bf M}$ are obtained as
\begin{eqnarray}
\la_1=\dtil-m^2=\frac{g}{2}d-m^2\com\q
\la_2=\dtil-m^2-2Ga_S\abar_S=\frac{g}{2}d-m^2-\frac{g^2}{2}a_S\abar_S\com\nn
\la_3=-\dtil-m^2=-\frac{g}{2}d-m^2\com\q
\la_4=-\dtil-m^2-2Ga_R\abar_R=-\frac{g}{2}d-m^2-\frac{g^2}{2}a_R\abar_R
\pr\label{qed12}
\end{eqnarray}

\begin{figure}
\centerline{
\psfig{figure=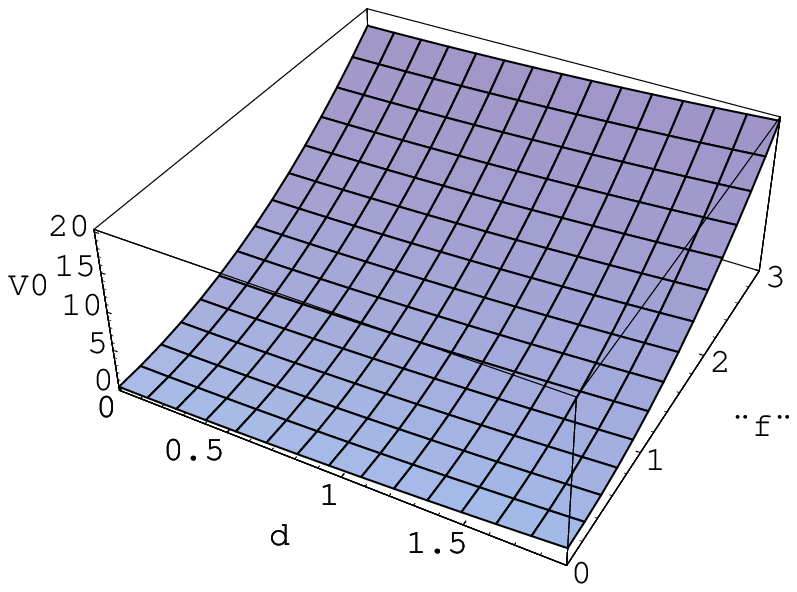,height=6cm,angle=0}
}
\centerline{
\psfig{figure=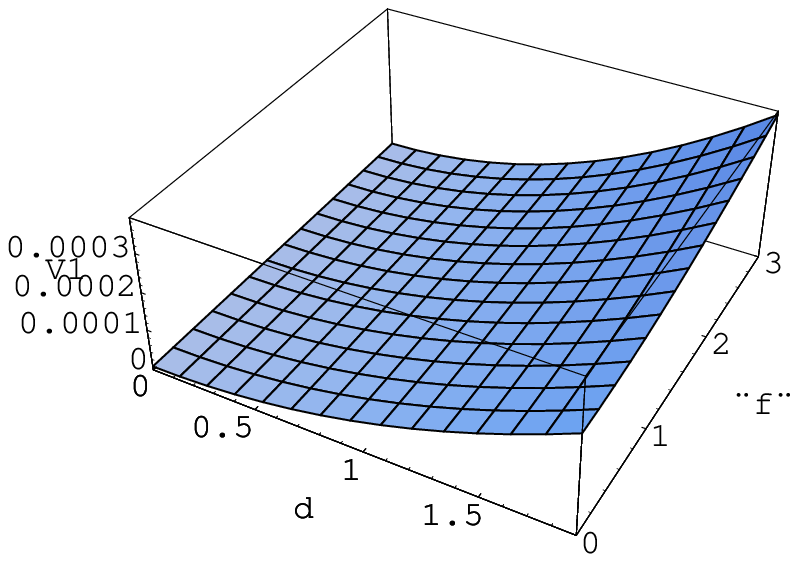,height=6cm,angle=0}
\psfig{figure=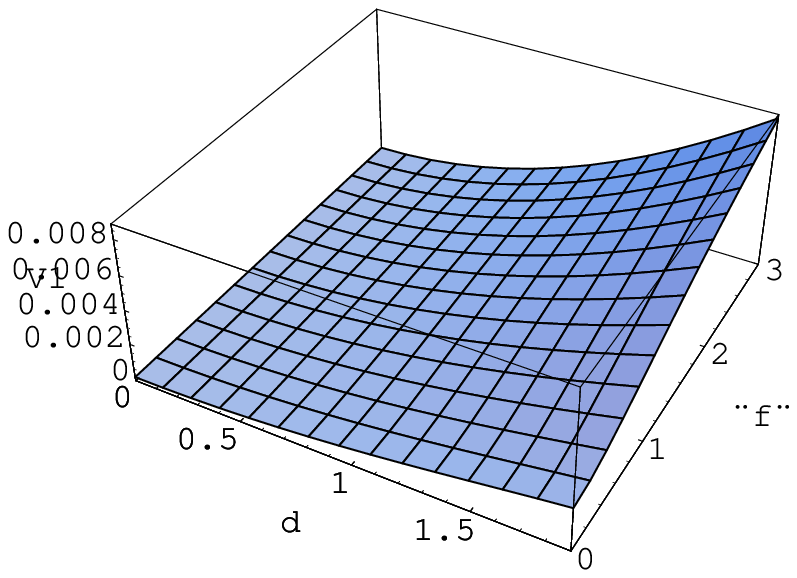,height=6cm,angle=0}
}
\caption{The effective potential of SQED(\ref{qed16}) for the case:\ 
$m=1, \fbar_S f_S=\fbar_R f_R\equiv \fbar f$. The tree part 
($V^{eff}_0|_{\mbox{on-shell}}\equiv V_0$, above) and 
the 1-loop correction part 
($V^{1-loop}_R|_{\mbox{on-shell}}\equiv V_1$, below) are depicted for
$g=0.3$(left), $g=1$(right). The axis-ranges are  
$0\leq d\leq 1.9, 0\leq |f|\equiv \sqrt{\fbar f}\leq 3$.
}
\label{fig:SQED1}
\end{figure}

Then the 1-loop contribution is given as
\begin{eqnarray}
\int (\Dcal A\Dcal\Abar)_{S,R}\exp i\intfx \Lcal_2'\nn
=\left[\det (\Box+\la_1)(\Box+\la_2)(\Box+\la_3)(\Box+\la_4)\right]^{-\half}
=\exp -i\intfx V^{eff}_{1-loop}\ ,\ \nn
V^{eff}_{1-loop}=\half \Tr \sum_{i=1}^{4}\ln (\Box+\la_i)
\pr\label{qed13}
\end{eqnarray}
Normalizing $V^{eff}_{1-loop}$ at $d=0$, 
from the requirement of the {\it supersymmetric boundary condition} (Sect.2), 
we finally obtain
\begin{eqnarray}
V^{eff}_{1-loop}-V^{eff}_{1-loop}|_{d=0}=
\half \int \frac{d^4k}{(2\pi)^4}\left\{
\ln \left(
1-(\frac{g}{2})^2\frac{d^2}{(k^2+m^2)^2}
    \right)    \right.                              \nn
+\ln \left(
1- (\frac{g}{2})^2\frac{d^2}{(k^2+m^2+g^2a_S\abar_S/2)(k^2+m^2+g^2a_R\abar_R/2)}
     \right.\nn
   \left. \left.
+\frac{g}{2}\frac{g^2}{2}\frac{d(a_S\abar_S-a_R\abar_R)}
{(k^2+m^2+g^2a_S\abar_S/2)(k^2+m^2+g^2a_R\abar_R/2)}
          \right) \right\}      \nn
\approx \left(
-\frac{g^4}{4}d^2+\frac{g^3}{8}d(a_S\abar_S-a_R\abar_R)
\right)\int \frac{d^4k}{(2\pi)^4}\frac{1}{(k^2+m^2)^2}
+O(g^4)
\pr\label{qed14}
\end{eqnarray}
The last approximate form is {\it logarithmically} divergent.
We introduce a counterterm $\Del\Lcal$ as in the
following form.
\begin{eqnarray}
V_R=V^{eff}_0+V^{eff}_{1-loop}-V^{eff}_{1-loop}|_{d=0}-\Del\Lcal\com\nn
V^{eff}_0=-\half d^2-\half g d(\abar_Sa_S-\abar_Ra_R)+\cdots\com\nn
\Del\Lcal=\half\Del Z\, d^2+\half\Del g\, d(\abar_Sa_S-\abar_Ra_R)\com
\label{qed14b}
\end{eqnarray}
where $V^{eff}_0$ is the tree part of the potential (\ref{qed5}). 
$Z=1+\Del Z$ and $g_b=g+\Del g$ are 
 the {\it wave-function renormalization} constant of $D$ and 
the {\it bare coupling constant}, respectively. 
We fix $\Del Z$ and $\Del g$ by imposing the following
{\it renormalization condition}.
\begin{eqnarray}
\left.\frac{dV_R}{d(d^2)}\right|_{d=0,a=\abar=0}=-\half\com\nn
\left.\frac{dV_R}{d[d(\abar_Sa_S-\abar_Ra_R)]}\right|_{d=0,a=\abar=0}=-\half g
\pr\label{qed14c}
\end{eqnarray}
Hence $\Del Z$ and $\Del g$ are fixed as
\begin{eqnarray}
\Del Z=-\frac{g^2}{2}\int_{|k|\leq \La} \frac{d^4k}{(2\pi)^4}
\frac{1}{(k^2+m^2)^2}=(-\frac{g^2}{2})\frac{-1}{16\pi^2}
\ln\frac{\La^2}{m^2}\com\nn
\Del g=\frac{g^3}{4}\int_{|k|\leq \La} \frac{d^4k}{(2\pi)^4}
\frac{1}{(k^2+m^2)^2}=(\frac{g^3}{4})\frac{-1}{16\pi^2}
\ln\frac{\La^2}{m^2}\com
\label{qed14d}
\end{eqnarray}
where $\La$ is the momentum cutoff, $|k^2|\leq \La^2$. 
Then the $\be$-function of the coupling and the anomalous
dimension $\ga$ of the $D$ field are given as
\begin{eqnarray}
g_b=g+\Del g=g(1-\frac{g^2}{4}\frac{1}{16\pi^2}
\ln\frac{\La^2}{m^2})\com\nn
0\equiv\frac{d}{d(\ln\La)}\ln g_b=\frac{d\ln g}{d(\ln\La)}(1+O(g^2))
-\frac{g^2}{2}\frac{1}{16\pi^2}+O(g^2)\com\nn
\be(g)\equiv\frac{1}{g}\frac{dg}{d\ln\La}=\frac{g^2}{2}\frac{1}{16\pi^2}\com\nn
Z=1+\Del Z\com\q
\ga=\frac{\pl}{\pl\ln\La}\ln Z=\frac{g^2}{16\pi^2}+O(g^4)
\pr
\label{qed15}
\end{eqnarray}
Finally the on-shell potential is obtained as
\begin{eqnarray}
V^{1-loop}_R\equiv V^{eff}_{1-loop}-V^{eff}_{1-loop}|_{d=0}-\Del\Lcal\com
\nn
V_R|_{on-shell}=(V^{eff}_0+V^{1-loop}_R)|_{on-shell}\com
\nn
V^{eff}_0|_{\mbox{on-shell}}=\fbar_S f_S+\fbar_R f_R+\half d^2\q (\mbox{eq.(\ref{qed8})})\com
\nn
V^{1-loop}_R|_{on-shell}=\frac{1}{64\pi^2}
\left[
-g^2 d^2+(m^4+\frac{g^2}{4}d^2)\ln (1+\frac{g}{2m^2}d)(1-\frac{g}{2m^2}d)
\right.
\nn
-g m^2 d\,\ln\frac{1-\frac{g}{2m^2}d}{1+\frac{g}{2m^2}d}
-m^4(1+\frac{g^2}{2m^4}\fbar_Sf_S)^2
\ln\frac{1+\frac{g^2}{2m^4}\fbar_Sf_S}
{1+\frac{g^2}{2m^4}\fbar_Sf_S+\frac{g}{2m^2}d}
\nn
-m^4(1+\frac{g^2}{2m^4}\fbar_Rf_R)^2
\ln\frac{1+\frac{g^2}{2m^4}\fbar_R f_R}
{1+\frac{g^2}{2m^4}\fbar_R f_R-\frac{g}{2m^2}d}
\nn
+\frac{g^2}{4}d^2\,\ln \left\{(1+\frac{g^2}{2m^4}\fbar_Sf_S+\frac{g}{2m^2}d)
(1+\frac{g^2}{2m^4}\fbar_Rf_R-\frac{g}{2m^2}d)
                      \right\}
\nn
+gm^2 d\,(1+\frac{g^2}{2m^4}\fbar_Sf_S) 
\ln(1+\frac{g^2}{2m^4}\fbar_Sf_S+\frac{g}{2m^2}d)
\nn
\left.
-gm^2 d\,(1+\frac{g^2}{2m^4}\fbar_Rf_R) 
\ln(1+\frac{g^2}{2m^4}\fbar_Rf_R-\frac{g}{2m^2}d)
\right]
\pr\label{qed16}
\end{eqnarray}
For the case: $m=1, \fbar_S f_S=\fbar_R f_R\equiv \fbar f$; 
the above potentials, 
$V^{eff}_0|_{\mbox{on-shell}}$ and 
$V^{1-loop}_R|_{\mbox{on-shell}}$, are depicted in Fig.\ref{fig:SQED1}.
The shape of the 1-loop correction is {\it not}
the Coleman-Weinberg type.
Positive definiteness is preserved after the renormalization.
The potential minimum does not change. 
The minimum ($f_S=f_R=0, d=0, a_S=a_R=0$) corresponds to
the SUSY invariant vacuum. 
This shows a characteristic feature of the SUSY theory, that is,
it is stable against the quantum effect.
The form of the potential $V_R$ does not essentially change 
from the tree one (\ref{qed8}). 
The stableness of the vacuum 
 was already pointed out, in the counter-term calculation, 
by Barbieri et al\cite{BFMPS82}. We confirm it by the
explicit form of the renormalized potential.

\vs 1
\section{Appendix B:\ Eigenvalues of Mass Matrix $\Mcal$ and
Effective Potential of the Mirabelli-Peskin Model}
The effective potential of the present bulk-boundary model 
can be obtained from the eigenvalues of the mass matrix
$\Mcal$ of (\ref{det19}) and (\ref{det20}).
It is made of three field values $\eta,\eta^\dag,d_\al$,
two wall "heights", $\abar_\al,\vpbar_\al$, the gauge coupling, $g$
and the boundary parameter $l$. 
The general case is hard to analyze explicitly. 
Here we consider two interesting "sections":\ 
A)\ $\eta=0, \eta^\dag=0$ (bulk-boundary decoupled case);\ 
B)\ $\vpbar=0, \abar=0$ (bulk-boundary coupled case).

\vs{0.2}
\subsection{Effective Potential From Matter Field Vanishing Point
\q ---\q Case A)\ $\eta=0, \eta^\dag=0$\q ---}
\vs{0.2}
In this configuration, the interaction term
$g\Dcal_\al\phi^\dag T^\al\phi$ of (\ref{mp7}) 
does not contribute to the bulk-boundary loop. 
The bulk and boundary are {\it decoupled} in the quantum fluctuation. 
It turns out, however, that the {\it renormalization} procedure
to deal with the divergences due to the boundary-loop
makes connection between the bulk and boundary. 
(See the explanation below (\ref{B7}).)
$\Mcal$ has the following form:
\begin{eqnarray}
       {\left(\begin{array}{cc}
\left(\begin{array}{cc}
\Mcal_{\phi^\dag\phi} & 0  \\
0 & \Mcal_{\phi\phi^\dag}  \\
\end{array}
\right)_{\alp\,\bep}         &
\left(\begin{array}{cc}
0 & 0 \\
0 & 0 
\end{array}
\right)_{\alp \, n\be}          \\
\left(\begin{array}{cc}
0 & 0 \\
0 & 0  
\end{array}
\right)_{m\al \, \bep}         &
\left(\begin{array}{cc}
\Mcal_{\Phi\Phi}  & \Mcal_{\Phi A} \\
\Mcal_{A\Phi} & \Mcal_{A A}
\end{array}
\right)_{m\al \, n\be}
             \end{array}\right)}
\pr
\label{B1}
\end{eqnarray}
The components are given, from (\ref{det20}),  as
\begin{eqnarray}
\Mcal_{\phi^\dag_\alp\phi_\bep}=
\pl^2\del_{\alp\bep}+gd_\ga (T^\ga)_{\alp\bep}\com\ 
\nn
\Mcal_{\phi_\alp\phi^\dag_\bep}= 
\pl^2\del_{\alp\bep}+gd_\ga (T^\ga)_{\bep\alp}\com\ 
\nn
%
%
\Mcal_{\Phi_{m\al}\Phi_{n\be}}=
-\{-\pl^2+(\npl)^2\}\del_{mn}\del_\ab
-g^2f_{\al\del\tau}f_{\be\ga\tau}\abar_\del\abar_\ga\del_{mn}
+\frac{4g}{l}f_{\ab\ga}\abar_\ga mQ_{mn}\com
\nn
\Mcal_{\Phi_{m\al} A_{n\be}}=
g^2f_{\ab\tau}f_{\ga\del\tau}\abar_\ga\vpbar_\del\del_{mn}
-g^2f_{\ga\al\tau}f_{\be\del\tau}\abar_\ga\vpbar_\del\del_{mn}
-\frac{2g}{l}f_{\ab\ga}\vpbar_\ga mQ_{mn}
=M_{A_{n\be}\Phi_{m\al}} ,
\nn
\Mcal_{A_{m\al} A_{n\be}}=
-\{-\pl^2+(\npl)^2\}\del_{mn}\del_\ab
-g^2 f_{\al\ga\tau}f_{\be\del\tau}\vpbar_\ga\vpbar_\del\del_{mn}\com
\label{B2}
\end{eqnarray}
where the integer indices $m,n$ run from 1 to $\infty$. 
$Q_{mn}$ is defined in (\ref{det10}). 
The singular terms, $\del(0)$-terms, disappear. 
The bulk and the boundary are decoupled, hence the eigen
values can be obtained separately.

For simplicity we take $SU(2)$ as the gauge group G 
($f_{\ab\ga}=\ep^{\ab\ga}$) 
and
the doublet representation for the matter fields\ 
$\phi_\alp\ (\alp=1,2)$.
\begin{eqnarray}
T^\al=\half \si^\al\com\q
[T^\al,T^\be]=i\ep^{\ab\ga}T^\ga\com\q
\Tr (T^\al T^\be)=\half\del^\ab
\com\label{B3}
\end{eqnarray}
where $\si^\al (\al=1,2,3)$ is the Pauli sigma matrices, and
$\ep^{123}=1$.

\vs{0.2}
(Ai)\ boundary part \nl
\vs{0.2}
The eigenvalues of
\begin{eqnarray}
\left(\Mcal_{\phi^\dag_\alp\phi_\bep}\right)= 
       \left(
\begin{array}{cc}
-k^2+\frac{g}{2}d_3 & \frac{g}{2}(d_1-id_2) \\
\frac{g}{2}(d_1+id_2) & -k^2-\frac{g}{2}d_3 
\end{array}
        \right)
\com
\label{B4}
\end{eqnarray}
are
\begin{eqnarray}
\la_\pm=-k^2\pm\frac{g}{2}\sqrt{d^2}\com\q
d^2\equiv {d_1}^2+{d_2}^2+{d_3}^2
\com
\label{B5}
\end{eqnarray}
The same ones are obtained for $(\Mcal_{\phi_\alp\phi^\dag_\bep})$.
The effective potential can be obtained as
\begin{eqnarray}
{V^{eff}_{1-loop}}'=\half\Tr\ln (\la_+)^2(\la_-)^2
=\Tr\ln \{(k^2)^2-\frac{g^2}{4}d^2\}\nn
=\int\frac{d^4k}{(2\pi)^4}
[\ln (k^2)^2+\ln \{1-\frac{g^2}{4}\frac{d^2}{(k^2)^2}\}]
\pr
\label{B6}
\end{eqnarray}
Compare this result with the super QED case (the first line
of (\ref{qed14})). Taking the SUSY condition (see the explanation
given above (\ref{wz8})), we reach the final answer.
\begin{eqnarray}
V^{eff}_{1-loop}
=\int\frac{d^4k}{(2\pi)^4}
\ln \{1-\frac{g^2}{4}\frac{d^2}{(k^2)^2}\}\nn
=-\frac{g^2}{4}\int\frac{d^4k}{(2\pi)^4}\frac{d^2}{(k^2)^2}
+O(g^4)
\pr
\label{B7}
\end{eqnarray}
The last approximated form corresponds to (c) of (\ref{qe3}),
and is {\it logarithmically} divergent. 
Because $d_\al$ is given by $d_\al=(\chi^3-\pl_5\vp+g a_5\times\vp)_\al$, 
the UV divergence of (\ref{B7}) is renormalized
by the {\it bulk} wave function of $X^3$ and $\Phi$.  Here the 4D world's 
connection to the bulk world appears. The quantum fluctuation
within the boundary influence the bulk world through the renormalization.
The boundary dynamics does {\it not} close within the brane.
We do not touch on the renormalization of the bulk fields. 
After an appropriate renormalization, we expect the effective potential
(\ref{B7}) leads to a similar potential to that given in App.A
for the case $m=0, \fbar f=0$.
(Note that the boundary theory treated in this section is chiral,
whereas the Super QED treated in App.A is vector-like. )
We may conclude that, in the vacuum specified by $\eta=0, \eta^\dag=0$,
the renormalization works well as far as the boundary world is 
concerned. The 4D theory is well-defined.

\vs{0.2}
(Aii) Bulk Part\nl
Let us evaluate the eigenvalues from the bulk part.
Because it does not depend on $d_\al$, this part
does {\it not} contribute to the effective potential 
in the SUSY boundary condition.
However it is important to see what terms 
are quantumly induced by the scalar fields. (Those terms
are expected to be cancelled by
the fermion and vector fields contribution.)
The result depends on the "heights" of the 4D scalars,
$\abar$ and $\vpbar$, in addition
to the periodicity $2l$. 
Generally
that part of the vacuum energy  which depends on 
the boundary parameters is called "Casimir energy".
We regard $\abar$ and $\vpbar$, besides $l$,  as those parameters.
They correspond to the brane tension and the brane thickness
in the brane world. 
One of most important points of the brane model is
how to treat the KK-modes. 
We can see such a point in this calculation.

The eigenvalue equation can be written as 
\begin{eqnarray}
\left(\begin{array}{cc}
\Mcal_{\Phi_{m\al}\Phi_{n\be}}  & \Mcal_{\Phi_{m\al} A_{n\be}} \\
\Mcal_{A_{m\al}\Phi_{n\be}} & \Mcal_{A_{m\al} A_{n\be}}
\end{array}
\right)
\left(\begin{array}{c}
\Phihat_{n\be}\\
\Ahat_{n\be}
\end{array}\right)
=\la
\left(\begin{array}{c}
\Phihat_{m\al}\\
\Ahat_{m\al}
\end{array}\right)
\pr
\label{B7.2}
\end{eqnarray}
From the symmetry, we can take the following general form
as an eigen vector.
\begin{eqnarray}
\Phihat_{n\be}=f_1(n)\abar_\be+f_2(n)\vpbar_\be
+f_3(n)f_{\be\ga\del}\abar_\ga\vpbar_\del\com\nn
\Ahat_{n\be}=g_1(n)\abar_\be+g_2(n)\vpbar_\be
+g_3(n)f_{\be\ga\del}\abar_\ga\vpbar_\del
\com
\label{B7.3}
\end{eqnarray}
where $f_i(n)$ and $g_i(n)$ are scalar quantities 
(with respect to the internal group transformation)
which may depend on $\abar$ and $\vpbar$. 
\footnote{
The physical dimensions of the "coefficients" functions
are as follows; \ 
$[f_1]=[f_2]=M^{-1/2}, [f_3]=M^{-2}, 
[g_1]=[g_2]=M^{-1/2}, [g_3]=M^{-2}.
$
}
\footnote{
The change of the vector space
of the eigen functions makes the number of eigenvalues change. 
We can, however, choose proper values from the consistency
with the perturbative results of Sec.5.
}
Through the above relation, the eigenvalue equation
for $\Phihat$ and $\Ahat$ is replaced by that
for $f_i(n)$ and $g_i(n)$. 
The eigenvalues are obtained from the zeros of the determinant
of the following matrix.
\begin{eqnarray}
       \begin{array}{c|c|c}
&
\begin{array}{ccc}
f_1(n) & f_2(n) & f_3(n) 
\end{array}         
                             &
\begin{array}{ccc}
g_1(n) & g_2(n) & g_3(n) 
\end{array}         
                             \\
\hline
\begin{array}{c}
\abar,m  \\
\vpbar,m\\
\abar\times\vpbar,m
\end{array}
         & \mbox{{\Huge a}} & \mbox{{\Huge c}} \\
\hline
\begin{array}{c}
\abar,m  \\
\vpbar,m\\
\abar\times\vpbar,m
\end{array}
         & \mbox{{\Huge d}} & \mbox{{\Huge b}}  
                 \end{array}
\pr
\label{B7.4}
\end{eqnarray}
The above 4 matrices are given as follows.

The first row equation of (\ref{B7.2}), 
$\Mcal_{\Phi_{m\al}\Phi_{n\be}}\Phihat_{n\be}+
\Mcal_{\Phi_{m\al} A_{n\be}}\Ahat_{n\be}=\la \Phihat_{m\al}$, 
gives two matrices $a$ and $c$ as
\begin{eqnarray}
\mbox{{\Huge a}}=       
{\left(\begin{array}{ccc}
(-\la-k^2-(m\pi/l)^2)\del_{mn} & g^2\abar\cdot\vpbar\del_{mn} &
                              -4gmQ_{mn}\abar\cdot\vpbar/l\\
0 & (-\la-k^2-(m\pi/l)^2-g^2\abar^2)\del_{mn} &
                              4gmQ_{mn}\abar^2/l\\
0 & -4gmQ_{mn}/l^2 &  (-\la-k^2-(m\pi/l)^2-g^2\abar^2)\del_{mn} 
\end{array}\right)}
\com\nn
\mbox{{\Huge c}}=       
{\left(\begin{array}{ccc}
2g^2\abar\cdot\vpbar\del_{mn} & g^2\vpbar^2\del_{mn} &
                             2gmQ_{mn}\vpbar^2/l \\
-2g^2\abar^2\del_{mn} & -g^2\abar\cdot\vpbar\del_{mn} &
                             -2gmQ_{mn}\abar\cdot\vpbar/l \\
-2gmQ_{mn}/l^4  &  0  & g^2\abar\cdot\vpbar\del_{mn}
\end{array}\right)}
\pr
\label{B7.5}
\end{eqnarray}
The second row equation of (\ref{B7.2}), 
$\Mcal_{A_{m\al}\Phi_{n\be}}\Phihat_{n\be}+
\Mcal_{A_{m\al} A_{n\be}}\Ahat_{n\be}=\la\Ahat_{m\al}$, 
gives two matrices $d$ and $b$ as
\begin{eqnarray}
\mbox{{\Huge d}}=       
{\left(\begin{array}{ccc}
-g^2\abar\cdot\vpbar\del_{mn} & -2g^2\vpbar^2\del_{mn} &
                             -2gnQ_{nm}\vpbar^2/l \\
g^2\abar^2\del_{mn} & 2g^2\abar\cdot\vpbar\del_{mn} &
                             2gnQ_{nm}\abar\cdot\vpbar/l \\
2gnQ_{nm}/l^4  &  0  & g^2\abar\cdot\vpbar\del_{mn}
\end{array}\right)}
\com\nn
\mbox{{\Huge b}}=       
{\left(\begin{array}{ccc}
(-\la-k^2-(m\pi/l)^2-g^2\vpbar^2)\del_{mn} & 0 & 0    \\
g^2\abar\cdot\vpbar\del_{mn} & (-\la-k^2-(m\pi/l)^2)\del_{mn} & 0 \\
0 & 0 &  (-\la-k^2-(m\pi/l)^2-g^2\vpbar^2)\del_{mn} 
\end{array}\right)}
\pr
\label{B7.6}
\end{eqnarray}
For convenience, let us introduce two quantities $x_m, y_m$;\ 
\begin{eqnarray}
x_m\equiv -\la-k^2-(m\pi/l)^2-g^2\vpbar^2\com\q
y_m\equiv -\la-k^2-(m\pi/l)^2\com\nn
x_m-y_m=-g^2\vpbar^2
\pr
\label{B7.7}
\end{eqnarray}
Then the following relations are obtained.
\begin{eqnarray}
\mbox{{\Huge b}}=       
\left(
\begin{array}{ccc}
x_m\del_{mn} & 0 & 0 \\
g^2\abar\cdot\vpbar\del_{mn} & y_m\del_{mn} & 0 \\
0  &  0  & x_m\del_{mn}
\end{array}
\right)
\com\nn
\mbox{{\Huge b}}^{-1}=       
{\left(\begin{array}{ccc}
\frac{1}{x_m}\del_{mn} & 0 & 0    \\
-\frac{g^2\abar\cdot\vpbar}{x_my_m}\del_{mn} & \frac{1}{y_m}\del_{mn} & 0 \\
0 & 0 &  \frac{1}{x_m}\del_{mn} 
\end{array}\right)}
\com\nn
\mbox{{\Huge det\, b}}=\prod_{m=1}^{\infty}({x_m}^2y_m)
\pr
\label{B7.8}
\end{eqnarray}
Using a useful formula about general matrices $\atil, \btil, \ctil$, 
and $\dtil$
($\det \btil\neq 0, \det \atil\neq 0$); 
\begin{eqnarray}
\det\, 
\left( \begin{array}{cc}
     \atil & \ctil \\
     \dtil & \btil
       \end{array}\right)
=\det (\atil-\ctil{\btil}^{-1}\dtil)\det \btil=\det \atil
\det (\btil-\dtil{\atil}^{-1}\ctil)
\com
\label{B7.9}
\end{eqnarray}
we can decompose the determinant of the matrix (\ref{B7.4}). 
The components of the matrix $\Acal\equiv a-c{b}^{-1}d$ are explicitly
written as
\begin{eqnarray}
 \Acal\q=\q
       \begin{array}{c|ccc}
  & f_1(n) & f_2(n) & f_3(n)    \\      
\hline
\abar,m & \Acal_{11} & \Acal_{12} & \Acal_{13}  \\
\vpbar,m &  \Acal_{21} & \Acal_{22} & \Acal_{23}\\
\abar\times\vpbar,m  &  \Acal_{31} & \Acal_{32} & \Acal_{33}
       \end{array}
\com
\nn
\nn
(\Acal_{11})_{mn}=y_m\del_{mn}
-g^4\left\{  
(\abar\cdot\vpbar)^2\frac{1}{x_m}(-2+\frac{g^2\vpbar^2}{y_m})
+\frac{\vpbar^2\abar^2}{y_m}
\right\}\del_{mn}
-4\frac{g^2\vpbar^2}{l^2}mn\sum_{j}\frac{Q_{mj}Q_{nj}}{x_j}\com
\nn
(\Acal_{12})_{mn}=g^2\abar\cdot\vpbar\left\{1+2g^2\frac{\vpbar^2}{x_m}\right\}
\del_{mn}\com\q
(\Acal_{13})_{mn}=-\frac{\abar\cdot\vpbar}{l} Q_{mn}\left\{
4gm+2g^3\vpbar^2(\frac{n}{x_m}+\frac{m}{x_n})\ \right\}\com
\nn
(\Acal_{21})_{mn}=-g^4\frac{\abar\cdot\vpbar}{x_m}\left\{
\abar^2+\frac{g^2}{y_m}(\abar^2\vpbar^2-(\abar\cdot\vpbar)^2)
                                                  \right\}\del_{mn}
+4\frac{g^2\abar\cdot\vpbar}{l^2} mn
\sum_{j}\frac{Q_{mj}Q_{nj}}{x_j}\ ,
\nn
(\Acal_{22})_{mn}=\left\{
y_m-g^2\abar^2-\frac{2g^4}{x_m}(2\abar^2\vpbar^2-(\abar\cdot\vpbar)^2)
                                                  \right\}\del_{mn}\com
\nn
(\Acal_{23})_{mn}=\frac{Q_{mn}}{l}\left\{
4g\abar^2 m+2g^3(\frac{n}{x_m}-\frac{m}{x_n})(\abar^2\vpbar^2-(\abar\cdot\vpbar)^2)
                                                  \right\}\com
\nn
(\Acal_{31})_{mn}=2\frac{g^3\abar\cdot\vpbar}{l^4} 
Q_{mn}(\frac{n}{x_m}-\frac{m}{x_n})\com
\q
(\Acal_{32})_{mn}=-\frac{Q_{mn}}{l^4}\left(
4gm+4g^3\vpbar^2\frac{m}{x_n}\right)\com
\nn
(\Acal_{33})_{mn}=
\left\{
y_m-g^2\abar^2-g^4\frac{(\abar\cdot\vpbar)^2}{x_m}
\right\}\del_{mn}
-4\frac{g^2\vpbar^2}{l^2}mn\sum_{j}\frac{Q_{mj}Q_{nj}}{x_j}
\com
\label{B7.10}
\end{eqnarray}
where the repeated integer suffixes do {\it not} mean the summation.
\footnote{
Because of this, the product $Q_{mn}(\frac{n}{x_m}+\frac{m}{x_n})$ 
appearing in $(\Acal_{13})_{mn}$ of (\ref{B7.10}), 
does not vanish in spite of the antisymmetricity of $Q_{mn}$. 
}
The summation should be taken only where the symbol $\sum_j$ appears.

Let us evaluate the eigenvalues $\la$ from the zeros of $\det \Acal$.
General case is technically difficult. We consider the following special cases.


We consider the following limit:
\begin{eqnarray}
\gh^2\equiv \frac{g^2}{l}=\mbox{fixed}\ll 1\com\q
\ahat=\sqrt{l}\abar=\mbox{fixed}\com\q
\vphat=\sqrt{l}\vpbar=\mbox{fixed}\com\nn
l\ra\infty\pr\qqq\qqq\qqq
\label{B7.11}
\end{eqnarray}
This is the situation where the circle is large
compared with the inverse of the domain wall height.
($\ahat$ and $\vphat$ have the dimension of $M$. )
Then the elements of $\Acal$ reduce to
\begin{eqnarray}
(\Acal_{11})_{mn}=y\del_{mn}
-\gh^4\left\{  
(\ahat\cdot\vphat)^2\frac{1}{x}(-2+\frac{\gh^2\vphat^2}{y})
+\frac{\vphat^2\ahat^2}{y}
\right\}\del_{mn}\com
\nn
(\Acal_{12})_{mn}=\gh^2\ahat\cdot\vphat\left\{1+2\gh^2\frac{\vphat^2}{x}\right\}\del_{mn}\com\q
(\Acal_{13})_{mn}=0\com
\nn
(\Acal_{21})_{mn}=-\gh^4\frac{\ahat\cdot\vphat}{x}\left\{
\ahat^2+\frac{\gh^2}{y}(\ahat^2\vphat^2-(\ahat\cdot\vphat)^2)
                                                  \right\}\del_{mn}\ ,
\nn
(\Acal_{22})_{mn}=\left\{
y-\gh^2\ahat^2-\frac{2\gh^4}{x}(2\ahat^2\vphat^2-(\ahat\cdot\vphat)^2)
                                                  \right\}\del_{mn}\com
\nn
(\Acal_{23})_{mn}=0\com
\nn
(\Acal_{31})_{mn}=0
\com
\q
(\Acal_{32})_{mn}=0
\com
\nn
(\Acal_{33})_{mn}=
\left\{
y-\gh^2\ahat^2-\gh^4\frac{(\ahat\cdot\vphat)^2}{x}
\right\}\del_{mn}
\com
\label{B7.12}
\end{eqnarray}
where $x\equiv -\la-k^2-\gh^2\vphat^2,\ y\equiv -\la-k^2$. 
We notice, in this limit, $Q_{mn}$-terms disappear. In the
"propagator" terms $x_m, y_m$, KK-mass terms $m^2\pi^2/l^2$
disappear. All KK-modes equally contribute to the vacuum energy.
The condition
$0=\det\Acal=\det\Acal_{33}\det\Acal_{22}
\det (\Acal_{11}-\Acal_{12}{\Acal_{22}}^{-1}\Acal_{21})
$
gives us the following eigenvalues.\nn
(i) $\det \Acal_{33}=0$ gives two eigenvalues $\la_1, \la_2$
whose product is given by
\begin{eqnarray}
\la_1\la_2=(k^2+\gh^2\vphat^2)(k^2+\gh^2\ahat^2)
-\gh^4(\ahat\cdot\vphat)^2
\pr\label{B7.13}
\end{eqnarray}
(ii) $\det\Acal_{22}=0$ gives $\la_3, \la_4$ whose product is
\begin{eqnarray}
\la_3\la_4=(k^2)^2+
k^2\gh^2(\vphat^2+\ahat^2)
+\gh^4(-3\vphat^2\ahat^2+(\ahat\cdot\vphat)^2)
\pr\label{B7.14}
\end{eqnarray}
(iii) $\det(\Acal_{11}-\Acal_{12}{\Acal_{22}}^{-1}\Acal_{21})=0$
gives $\la_5, \la_6, \la_7, \la_8, \la_9$ whose product
is given by
\begin{eqnarray}
\la_5\la_6\la_7\la_8\la_9
=-(k^2+\gh^2\vphat^2)\{(k^2)^2+\gh^4
((\ahat\cdot\vphat)^2-\ahat^2\vphat^2)\}\nn
\{(k^2)^2+\gh^2k^2(\ahat^2+\vphat^2)+3\gh^4
((\ahat\cdot\vphat)^2-\ahat^2\vphat^2)  \}
\pr\label{B7.15}
\end{eqnarray}
In particular, for the special case $\ahat=0$, the nontrivial
factor is only $k^2+\gh^2\vphat^2$. Hence each KK-mode
equally contribute to the vacuum energy as
\begin{eqnarray}
V^{eff}_{1KK-mode}
\propto\int\frac{d^4k}{(2\pi)^4}
\ln \{1+\gh^2\frac{\vphat^2}{k^2}\}
\pr
\label{B7.16}
\end{eqnarray}
This quantity is {\it quadratically} divergent. 
After an appropriate normalization,
which we do not know precisely, the final form should become, 
based on the dimensional analysis, the following.
\begin{eqnarray}
\frac{1}{l}\,V^{eff}_{1-loop}
=\gh^2(c_1\frac{\vphat^2}{l^3}+c_2\frac{\ahat^2}{l^3}
+c_3\frac{\ahat\cdot\vphat}{l^3})+O(\gh^4)
\com
\label{B7.17}
\end{eqnarray}
where $c_1, c_2$ and $c_3$ are some finite constants.
This is a {\it new} type Casimir energy.
Comparing the ordinary one (\ref{B10b}) explained soon,
it is new in the following points:\ 
1) it depends on the brane parameters $\vphat$ and $\ahat$
besides the extra-space size $l$;\ 
2) it depends on the gauge coupling;\ 
3) it is proportional to $1/l^3$.


\vs {0.5}

\subsection{Effective Potential With No Brane Structure\q
---\q Case (B)\ $\vpbar=0,\abar=0$\q ---}
Let us evaluate the case B), $\vpbar=0,\abar=0$. 
In this case 5D vacuum does not have the brane structure. 
The situation is similar to the case of Appelquist and Chodos's work. 
The matrix $\Mcal$ has the form:
\begin{eqnarray}
       {\left(\begin{array}{cc}
\left(\begin{array}{cc}
\Mcal_{\phi^\dag\phi} & \Mcal_{\phi^\dag\phi^\dag}  \\
\Mcal_{\phi\phi} & \Mcal_{\phi\phi^\dag}  \\
\end{array}
\right)_{\alp\,\bep}         &
\left(\begin{array}{cc}
\Mcal_{\phi^\dag\Phi} & 0 \\
\Mcal_{\phi\Phi} & 0 
\end{array}
\right)_{\alp \, n\be}          \\
\left(\begin{array}{cc}
\Mcal_{\Phi\phi} & \Mcal_{\Phi\phi^\dag} \\
0 & 0  
\end{array}
\right)_{m\al \, \bep}         &
\left(\begin{array}{cc}
\Mcal_{\Phi\Phi}  & 0 \\
0 & \Mcal_{A A}
\end{array}
\right)_{m\al \, n\be}
             \end{array}\right)}
,
\label{B8}
\end{eqnarray}
where each component is described as
\begin{eqnarray}
\Mcal_{\phi^\dag_\alp\phi_\bep}=
\pl^2\del_{\alp\bep}+
gd_\ga (T^\ga)_{\alp\bep}-g^2\del(0)(T^\ga\eta)_\alp
(\eta^\dag T^\ga)_\bep\com
\nn
\Mcal_{\phi^\dag_\alp\phi^\dag_\bep}=
-g^2\del(0)(T^\ga\eta)_\alp (T^\ga\eta)_\bep\com\q
\Mcal_{\phi_\alp\phi_\bep}= 
-g^2\del(0)(\eta^\dag T^\ga)_\alp (\eta^\dag T^\ga)_\bep\com
\nn
\Mcal_{\phi_\alp\phi^\dag_\bep}= 
\pl^2\del_{\alp\bep}+
gd_\ga (T^\ga)_{\bep\alp}-g^2\del(0)(\eta^\dag T^\ga)_\alp (T^\ga\eta)_\bep
\nn
\Mcal_{\phi^\dag_\alp\Phi_{n\be}}=
-\frac{g}{\sql}(T^\be\eta)_\alp\npl=\Mcal_{\Phi_{n\be}\phi^\dag_\alp}\com\q
\Mcal_{\phi_\alp\Phi_{n\be}}=
-\frac{g}{\sql}(\eta^\dag T^\be)_\alp\npl=\Mcal_{\Phi_{n\be}\phi_\alp}\com
\nn
\Mcal_{\Phi_{m\al}\Phi_{n\be}}=
-\{-\pl^2+(\npl)^2\}\del_{mn}\del_\ab\com
\nn
\Mcal_{A_{m\al} A_{n\be}}=
-\{-\pl^2+(\npl)^2\}\del_{mn}\del_\ab\com
\label{B9}
\end{eqnarray}
where the integer indices $m$ and $n$ run from 1 to $\infty$.
$Q_{mn}$-terms disappear. 
Let us find the eigenvalues of the above matrix.
\nl
\nl
 $\Mcal_{AA}$ part is decoupled with others, hence the eigenvalues
of the part is obtained as 
\begin{eqnarray}
\la_n=-k^2-(\npl)^2\com\q n=1,2,3,\cdots
\pr
\label{B10}
\end{eqnarray}
These correspond to the massive KK-modes of the fifth component
of the bulk vector. The eigenvalues (\ref{B10}) and 
another ones (\ref{B18}) explained soon, have the same form
as that appearing in the work by Appelquist and Chodos\cite{AC83}.
They contribute to the Casimir energy.
\begin{eqnarray}
\frac{1}{l}\,V^{eff}_{Casimir}
=\frac{\mbox{const}}{l^5}
\pr
\label{B10b}
\end{eqnarray}
\nl
\nl
The eigenvalue equation for the other parts can be written as
\begin{eqnarray}
\Mcal_1
\left(\begin{array}{c}
\phihat \\ \phihat^\dag \\ \Phihat 
              \end{array}\right) 
\equiv
       {\left(\begin{array}{cc}
\left(\begin{array}{cc}
\Mcal_{\phi^\dag\phi} & \Mcal_{\phi^\dag\phi^\dag}  \\
\Mcal_{\phi\phi} & \Mcal_{\phi\phi^\dag}  \\
\end{array}
\right)_{\alp\,\bep}         &
\left(\begin{array}{c}
\Mcal_{\phi^\dag\Phi}  \\
\Mcal_{\phi\Phi}       
\end{array}
\right)_{\alp \, n\be}          \\
\left(\begin{array}{cc}
\Mcal_{\Phi\phi} & \Mcal_{\Phi\phi^\dag} 
\end{array}
\right)_{m\al \, \bep}         &
\left(\begin{array}{c}
\Mcal_{\Phi\Phi}   
\end{array}
\right)_{m\al \, n\be}
             \end{array}\right)}
\left(\begin{array}{c}
\phihat_{\be'} \\ \phihat^\dag_{\be'} \\ \Phihat_{n\be} 
              \end{array}\right) 
=\ \la\ 
\left(\begin{array}{c}
\phihat_{\al'} \\ \phihat^\dag_{\al'} \\ \Phihat_{m\al}
              \end{array}\right) 
\ .\label{B10.2}
\end{eqnarray}
We can take
the following form as the eigen vector, from the transformation
property.
\begin{eqnarray}
\phihat_\bep=h_1\eta_\bep+
(h_2d_\ga+h_3V^\ga+h_4\ep^{\ab\ga}d_\al V^\be)
(T^\ga \eta)_\bep\com
\nn
\phihat^\dag_\bep=k_1\eta^\dag_\bep+
(k_2 d_\ga+k_3 V^\ga+k_4\ep^{\ab\ga}d_\al V^\be)
(\eta^\dag T^\ga)_\bep\com
\nn
\Phihat_{n\be}=f_1(n)d_\be+f_2(n)V^\be
+f_3(n)\ep^{\be\ga\del}d_\ga V^\del\com
\label{B11}
\end{eqnarray}
where $V^\al\equiv \eta^\dag T^\al\eta$ 
and $h_i, k_i$ and $f_i(n)$ are functions
which are made of $\eta_\alp, \eta^\dag_\alp$ and $d_\al$.
\footnote{
Their physical dimensions are as follows:\ 
$
([h_1],[h_2],[h_3],[h_4])=([k_1],[k_2],[k_3],[k_4])=
(M^0, M^{-5/2}, M^{-2}, M^{-9/2}),\ 
([f_1],[f_2],[f_3])=(M^{-3/2}, M^{-1}, M^{-7/2}).
$
}
The eigenvalue equation for $(\phihat, \phihat^\dag, \Phihat)$,
$\det (\Mcal_1-\la I)=0$, can be rewritten by that for
$(h_i, k_i, f_j(n))$. 
The eigenvalues are obtained from the zeros of the determinant
of the following matrix $\Mcal_2$.
\begin{eqnarray}
       \begin{array}{c|c|c|c}
&
\begin{array}{cccc}
h_1 & h_2 & h_3 & h_4
\end{array}         
                             &
\begin{array}{cccc}
k_1 & k_2 & k_3 & k_4
\end{array}         
                             &
\begin{array}{ccc}
f_1(n) & f_2(n) & f_3(n) 
\end{array}         
                             \\
\hline
\begin{array}{c}
\eta_\alp  \\
d_\al (T^\al\eta)_\alp\\
V_\al (T^\al\eta)_\alp\\
\ep_{\ab\ga}d_\al V^\be (T^\ga\eta)_\alp  
\end{array}
         & \mbox{{\Huge a}} & \mbox{{\Huge c}} & \mbox{{\Huge c}}1 \\
\hline
\begin{array}{c}
\eta^\dag_\alp  \\
d_\al (\eta^\dag T^\al)_\alp\\
V_\al (\eta^\dag T^\al)_\alp\\
\ep_{\ab\ga}d_\al V^\be (\eta^\dag T^\ga)_\alp  
\end{array}
         & \mbox{{\Huge d}} & \mbox{{\Huge b}} & \mbox{{\Huge c}}2 \\
\hline
\begin{array}{c}
d_\al, m  \\
V_\al, m\\
\ep_{\ab\ga}d_\be V^\ga, m  
\end{array}
         & \mbox{{\Huge d}}1 & \mbox{{\Huge d}}2 & \mbox{{\Huge b}}1 \\
                 \end{array}
.
\label{B12}
\end{eqnarray}
The components in each "box" are displayed in the following.
For the purpose, 
we introduce here the following quantities which turn out to
constitute the final result of the effective potential.
\begin{eqnarray}
\mbox{4-dim scalar mass term:  }
S=\eta^\dag\eta\com\q
\mbox{D mass term:  }
d^2=d_\al d_\al\com\nn
\mbox{3-body term:  }d\cdot V=d_\al V^\al\com\q
\mbox{4-dim 4-body term:  }V^2=V^\al V^\al\com\nn
\mbox{where  }V^\al=\eta^\dag T^\al\eta\pr
\label{B12.2}
\end{eqnarray}
The 9 matrices in (\ref{B12})  are given by as follows.
\\\\\\\\\\\\
The first row equation of (\ref{B10.2}),
$
\Mcal_{\phi^\dag_\alp\phi_\bep}\phihat_\bep+
\Mcal_{\phi^\dag_\alp\phi^\dag_\bep}\phihat^\dag_\bep+
\Mcal_{\phi^\dag_\alp\Phi_{n\be}}\Phihat_{n\be}
=\la \phihat_\alp$\ , 
gives three matrices $a,c,c_1$ as
\begin{eqnarray}
\mbox{{\Huge a}}=       
{\left(\begin{array}{cccc}
-\la-k^2 & \frac{g}{4}d^2 & \frac{g}{4}d\cdot V& 0 \\
g & -\la-k^2-\frac{g^2}{4}\del(0)S&  0  & 
                       \frac{i}{2}(gd\cdot V+g^2\del(0)V^2) \\
-g^2\del(0) & 0 & -\la-k^2-\frac{g^2}{4}\del(0)S &
                      -\frac{i}{2}(gd^2+g^2\del(0)d\cdot V) \\
0 & -\frac{i}{2}g^2\del(0) & \frac{ig}{2} & -\la-k^2-\frac{g^2}{4}\del(0)S
\end{array}\right)}
\com\nn
\mbox{{\Huge c}}=       
{\left(\begin{array}{cccc}
0 & 0 & 0 & 0 \\
0 & -\frac{g^2}{4}\del(0)S &  0  & -\frac{i}{2}g^2\del(0)V^2 \\
-g^2\del(0) & 0 &-\frac{g^2}{4}\del(0)S &
                      \frac{i}{2}g^2\del(0)d\cdot V \\
0 & \frac{i}{2}g^2\del(0) & 0 & -\frac{g^2}{4}\del(0)S
\end{array}\right)}
\com\nn
\mbox{{\Huge c}}1=\frac{1}{l\sqrt{l}}       
{\left(\begin{array}{ccc}
0 & 0 & 0  \\
-g\pi n & 0 &  0   \\
 0 & -g\pi n  & 0   \\
0 &  0 & -g\pi n 
\end{array}\right)}
\pr
\label{B13}
\end{eqnarray}

The second row equation of (\ref{B10.2}),
$
\Mcal_{\phi_\alp\phi_\bep}\phihat_\bep+
\Mcal_{\phi_\alp\phi^\dag_\bep}\phihat^\dag_\bep+
\Mcal_{\phi_\alp\Phi_{n\be}}\Phihat_{n\be}=
\la\phihat^\dag_\alp
$,\ 
gives three matrices $d,b,c_2$ as  

\begin{eqnarray}
\mbox{{\Huge d}}=       
{\left(\begin{array}{cccc}
0 & 0 & 0 & 0 \\
0 & -\frac{g^2}{4}\del(0)S &  0  & \frac{i}{2}g^2\del(0)V^2 \\
-g^2\del(0) & 0 &-\frac{g^2}{4}\del(0)S &
                      -\frac{i}{2}g^2\del(0)d\cdot V \\
0 & -\frac{i}{2}g^2\del(0) & 0 & -\frac{g^2}{4}\del(0)S
\end{array}\right)}
\com\nn
\mbox{{\Huge b}}=       
{\left(\begin{array}{cccc}
-\la-k^2 & \frac{g}{4}d^2 & \frac{g}{4}d\cdot V& 0 \\
g & -\la-k^2-\frac{g^2}{4}\del(0)S&  0  & 
                       -\frac{i}{2}(gd\cdot V+g^2\del(0)V^2) \\
-g^2\del(0) & 0 & -\la-k^2-\frac{g^2}{4}\del(0)S &
                      \frac{i}{2}(gd^2+g^2\del(0)d\cdot V) \\
0 & \frac{i}{2}g^2\del(0) & -\frac{ig}{2} & -\la-k^2-\frac{g^2}{4}\del(0)S
\end{array}\right)}
\com\nn
\mbox{{\Huge c}}2= \frac{1}{l\sqrt{l}}
{\left(\begin{array}{ccc}
0 & 0 & 0  \\
-g\pi n & 0 &  0   \\
 0 & -g\pi n  & 0   \\
0 &  0 & -g\pi n 
\end{array}\right)}
\pr
\label{B14}
\end{eqnarray}
We note the relations $a=b^*, c=d^*, c_1=c_2$. 

The third row equation of (\ref{B10.2}),
$
\Mcal_{\Phi_{m\al}\phi_\bep}\phihat_\bep+
\Mcal_{\Phi_{m\al}\phi^\dag_\bep}\phihat^\dag_\bep+
M_{\Phi_{m\al}\Phi_{n\be}}\Phihat_{n\be}=
\la \Phihat_{m\al}
$,\ 
gives three matrices $d_1,d_2,b_1$ as
\begin{eqnarray}
\mbox{{\Huge d}}1= \frac{1}{l\sqrt{l}}
{\left(\begin{array}{cccc}
0 & -\frac{g}{4}\pi mS & 0 & i\frac{g}{2}\pi mV^2 \\
-g\pi m & 0 & -\frac{g}{4}\pi mS & -i\frac{g}{2}\pi md\cdot V \\
0 & -i\frac{g}{2}\pi m & 0 & -\frac{g}{4}\pi m S 
\end{array}\right)}
\com\nn
\mbox{{\Huge d}}2= \frac{1}{l\sqrt{l}}
{\left(\begin{array}{cccc}
0 & -\frac{g}{4}\pi mS & 0 & -i\frac{g}{2}\pi mV^2 \\
-g\pi m & 0 & -\frac{g}{4}\pi mS & i\frac{g}{2}\pi md\cdot V \\
0 & i\frac{g}{2}\pi m & 0 & -\frac{g}{4}\pi m S 
\end{array}\right)}
\com\nn
\mbox{{\Huge b}}1=       
{\left(\begin{array}{ccc}
\left(-\la-k^2-(\frac{\pi m}{l})^2)^2\right)\del_{mn} & 0 & 0 \\
0 & \left(-\la-k^2-(\frac{\pi m}{l})^2)^2\right)\del_{mn} & 0  \\
0 & 0 & \left(-\la-k^2-(\frac{\pi m}{l})^2)^2\right)\del_{mn} 
\end{array}\right)}
\pr
\label{B15}
\end{eqnarray}
We note $d_1={d_2}^*$.

Using the formula (\ref{B7.9}),
the determinant of $\Mcal_2$ (\ref{B12}) decomposes as follows.
\begin{eqnarray}
A\equiv\left(
     \begin{array}{cc}
     a & c \\
     d & b
     \end{array}\right)
     \com\q
\Mcal_2=\left(
                        \begin{array}{cc}
     A & \begin{array}{c}
         c_1 \\
         c_2
         \end{array}
                                 \\
     \begin{array}{cc}
     d_1 & d_2 
     \end{array}
                   & b_1
                          \end{array}\right)
     \com\nn
\det (\Mcal_1-\la I)\sim \det\Mcal_2=\det \left(A-
\left(\begin{array}{c} c_1 \\ c_2\end{array}\right)
{b_1}^{-1}
\left(\begin{array}{cc} d_1 & d_2 \end{array}\right)
\right)
\times \det b_1
\pr
\label{B17}
\end{eqnarray}
The last expression is a product of two determinants. 
The eigenvalues from the right determinant $\det b_1=0$ gives
\begin{eqnarray}
\la_m=-k^2-(\frac{m\pi}{l})^2\com\q m=1,2,3,\cdots
\com
\label{B18}
\end{eqnarray}
which correspond to the massive KK-modes of the
bulk scalar $\Phi$. 
As for the left determinant, the matrix in the inside
can be evaluated 
using the explicit expressions
of (\ref{B13}),(\ref{B14}) and (\ref{B15}). Here we find
a {\it smoothing} procedure of the singular term takes
place as follows. We write the matrix $A$ in (\ref{B17}) as $A(\del(0))$
to show the $\del(0)$ dependence explicitly.
Then we find the following renormalization-like relation
with respect to the singular quantity $\del(0)$. 
\begin{eqnarray}
A(\del(0))-
\left(\begin{array}{c} c_1 \\ c_2\end{array}\right)
{b_1}^{-1}
\left(\begin{array}{cc} d_1 & d_2 \end{array}\right)
=A( {\del(0)}|_{sm})\com
\nn
{\del(0)}|_{sm}=\del(0)+\frac{1}{l}\sum_{m=1}^{\infty}
\frac{(\pi m/l)^2}{-\la-k^2-(\pi m/l)^2}
\nn
=\del(0)+\frac{1}{2l}\left(
-\sum_{m\in \bfZ }1+
\sum_{m\in \bfZ}
\frac{\la+k^2}{\la+k^2+(\pi m/l)^2}\right)
\pr
\label{B19}
\end{eqnarray}
Using the relation
$\sum_{m\in\bfZ}1=2l\del(0)$, ${\del(0)}|_{sm}$
becomes a finite (regular) quantity.
\begin{eqnarray}
{\del(0)}|_{sm}=
\frac{1}{2l}\sum_{m\in\bfZ}
\frac{\la+k^2}{\la+k^2+(\pi m/l)^2}
=\left\{
\begin{array}{cc}
\half\sqrt{\la+k^2}\coth\{l\sqrt{\la+k^2}\} & \la>-k^2 \\
\half\sqrt{-\la-k^2}\cot\{l\sqrt{-\la-k^2}\} & -k^2>\la 
\end{array}
\right.
\pr
\label{B20}
\end{eqnarray}
The above relation manifestly shows that 
{\it the tower of the massive KK-modes smoothes the $\del(0)$
singularity} appearing in the boundary part of the mass matrix 
$\Mcal$,(\ref{B8}).
(In the perturbative analysis of Sec.5, the present
smoothing phenomenon corresponds to the
cancellation of singularity appearing in the equations
(\ref{qe4}-\ref{qe8}).) 
In the limit $|\la+k^2|\ra \infty$, $\del(0)|_{sm}$ 
reduces to $\del(0)$.

Next we evaluate $\det A'$ ($A'\equiv A(\del(0)|_{sm})$) in order to find
remaining 8 eigenvalues. We repeat the formula:\ 
$\det A'=\det (a'-c'(b')^{-1}d')\det b'$, 
where the primed quantities are defined by
those ones which are obtained by replacing 
$\del(0)$, in matrices $A=(a,c\ /\ d,b)$, 
by $\del(0)|_{sm}$.
Now we may deal with 4$\times$4 matrices
$a'-c'(b')^{-1}d'$ and $b'$. We can explicitly
calculate $\det A'$ (using an algebraic soft)
and indeed obtain the expression. 
$\det A'$ is the function
composed of the background quantities:\ $S, d^2, d\cdot V$ and
$V^2$ defined in (\ref{B12.2}).
It is better to see some "sections" rather than the full result
in order to see the structure of the effective potential.

\vs{0.2}

(i)\ $d_\al=0\ (d^2=0, d\cdot V=0)$\nl
This case gives 
the normalization value of the effective potential in the SUSY 
boundary condition.
\begin{eqnarray}
\det A'=(\la+k^2)^5\{ \la+k^2+\frac{g^2}{2}\del(0)|_{sm}S\}^3 \com\nn
\la_{1-5}=-k^2\ (\mbox{5-fold}),\nn
\la_6,\ \la_7,\ \la_8\ (3-fold) :\q 
\la+k^2+\frac{g^2}{4}S\sqrt{\la+k^2}
\coth l\sqrt{\la+k^2}=0
\pr
\label{B20b}
\end{eqnarray}
Let us look at the above full result from the perturbative approach
and relate it to the result of Sec.5. First we do the propagator
($1/k^2$) expansion because the perturbative approach is based on
the expansion around the free theory:\ $g=0$. 
\begin{eqnarray}
\la+k^2+\frac{g^2}{4}S\sqrt{k^2}\left\{
\coth l\sqrt{k^2}+\frac{\la}{k^2}  (\half\coth l\sqrt{k^2}
-\frac{l\sqrt{k^2}}{2(\coth l\sqrt{k^2})^2})
                     +O(\frac{1}{(k^2)^2})
                                \right\}=0
\ .
\label{B20c}
\end{eqnarray}
Secondly we restrict the coupling and the considered configuration as follows.
\begin{eqnarray}
\frac{g^2}{l}=\mbox{fixed}\ll 1\com\q l\sqrt{k^2}\leq 1
\pr
\label{B20d}
\end{eqnarray}
The second equation 
is required for the validity of $1/k^2$ expansion and it 
says the 4d momentum integral should have
the UV cutoff $1/l$. Taking into account the perturbative order
up to the 1st order w.r.t. $g^2/l$ and the 0-th order w.r.t.
$1/k^2$, we obtain 
\begin{eqnarray}
\la=-k^2\left(1+\frac{g^2}{4}S
\frac{\sqrt{k^2}\coth l\sqrt{k^2}}{k^2}\right)
\pr
\label{B20e}
\end{eqnarray}
This eigenvalue is consistent with
the first part of (\ref{qe8}). 
We must pick up one eigen value from $\la_1-\la_5$,
and three ones $\la_6, \la_7$ and $\la_8$(3-fold)
in order to be consistent with the perturbative result.

\vs{0.2}

(ii) $V^2\neq 0$, Others=0 ($d^2=0,d\cdot V=0, S=0$)\nl
We examine the part that is composed of purely
the 4-body interaction term operator $V^2$. 
\begin{eqnarray}
\det A'=(\la+k^2)^8
\pr
\label{B21}
\end{eqnarray}
The term $V^2$ does not appear. This is desirable from
the renormalization point of view.
The absence of the 4-body interaction term
in the SUSY normalization part implies the renormalization
of this term works well without SUSY.

\vs{0.2}

(iii) $d^2\neq 0,\ \mbox{Others}=0\ (d\cdot V=0, V^2=0, S=0)$\ \ 
[equivalently $\eta=\eta^\dag=0$]\nl
This is a special case of (A), the decoupled case.
\begin{eqnarray}
\det A'=((\la+k^2)^2-\frac{g^2}{4}d^2)^4\com\nn
\la_\pm=-k^2\pm \frac{g}{2}\sqrt{d^2}
\pr
\label{B22}
\end{eqnarray}
Both $\la_+$ and $\la_-$ are 4-fold eigenvalue.
We pick up two eigen values of $\la_+$(2-fold)
and another two ones $\la_-$(2-fold). 
This result is consistent with Case (A).

\vs{0.2}

(iv) $d\cdot V\neq 0$, Others=$0$ ($S=0, d^2=0, V^2=0$)\nl
We examine the part that is composed of purely
the 3-body interaction operator $d\cdot V$.
\begin{eqnarray}
\det A'=(\la+k^2)^6\{ (\la+k^2)^2-\frac{g^3}{2}\del(0)|_{sm}d\cdot V\} \com\nn
\la_{1-6}=-k^2\ (\mbox{6-fold}),\nn
\la_7,\ \la_8\ :
\q (\la+k^2)^2-\frac{g^3}{2}d\cdot V\frac{\sqrt{\la+k^2}}{2}
\coth l\sqrt{\la+k^2}=0
\pr
\label{B23}
\end{eqnarray}
The perturbative values are obtained as in (i). 
$1/k^2$-expansion gives,

\begin{eqnarray}
(\la+k^2)^2-\frac{g^3}{4}d\cdot V\sqrt{k^2}\left\{
\coth l\sqrt{k^2}+\frac{\la}{k^2}  (\half\coth l\sqrt{k^2}
-\frac{l\sqrt{k^2}}{2(\coth l\sqrt{k^2})^2})
                     +O(\frac{1}{(k^2)^2})
                                \right\}=0
\ .
\label{B24}
\end{eqnarray}
Taking the terms up to the 0-th order w.r.t. $1/k^2$ and 
up to the 1-st order w.r.t. $g^2/l$, we obtain
\begin{eqnarray}
(\la+k^2)^2-\frac{g^3}{4}d\cdot V\sqrt{k^2}
\coth l\sqrt{k^2}=0
\pr
\label{B25}
\end{eqnarray}
This is a quadratic equation w.r.t. $\la$. The two roots $\la_7, \la_8$
satisfy

\begin{eqnarray}
\la_7\la_8=
(k^2)^2\left( 1-\frac{g^3}{4}d\cdot V
\frac{\sqrt{k^2}\coth l\sqrt{k^2}}{(k^2)^2}\right)
\pr
\label{B26}
\end{eqnarray}
This is consistent with (\ref{qe8}).
As for the four eigenvalues, 
we pick up $\la_1, \la_2$(2-fold) and $\la_7, \la_8$.

\vs{0.2}

(v) $S\neq 0, \mbox{Others}=0$($d^2=0, d\cdot V=0, V^2=0$)\nl
We examine the part that is composed of purely
the mass term operator $S=\eta^\dag\eta$.
The form of $\det A'$ is the same as the case (i).
Hence the effective potential is the same as (i).
The 4D scalar mass term appears in the intermediate 
procedure, but it disappears in the SUSY boundary condition.
This shows the renormalization about the scalar mass term 
works  with the help of SUSY.

\vs 1
\section{Appendix C:\ Background Fields and On-Shell Condition}
We show the background fields taken in Sec.6 satisfy the
field equation of (\ref{ep5b}), the on-shell condition, 
for a special case given below. The assumed forms are
\begin{eqnarray}
\vp_\al(\xf)=\vpbar_\al\ep(\xf)\com\q a_{5\al}(\xf)=\abar_\al\ep(\xf)\com\nn
\eta_\alp=\mbox{const}\com\q \eta^\dag_\alp=\mbox{const}\com\q
d_\al=\chi^3_\al-\pl_5\vp_\al+g(a_5\times\vp)_\al=\mbox{const}
\com
\label{C1}
\end{eqnarray}
where $\ep(x)$ is the periodic sign function defined by (\ref{det6}).
First we stress that the total derivative terms, appearing in the
derivation of the field equation (\ref{ep5b}), can be safely
put to $0$ because of the periodicity property. Using the relation
(\ref{det15}) and the condition $m_{\alp\bep}=\la_{\alp\bep\gap}=0$, 
the equations in (\ref{ep5b}) can be expressed as

\begin{eqnarray}
2\vpbar_\al\pl_5(\del(\xf)-\del(\xf-l))
-g^2{\ep(\xf)}((\abar\times\vpbar)\times\abar)_\al
+g\pl_5\del(\xf)\cdot\eta^\dag T^\al\eta
+g\pl_5\del(\xf-l)\cdot{\eta'}^\dag T^\al\eta' \nn
+g^2[(\del(\xf)\eta^\dag T\eta
+\del(\xf-l){\eta'}^\dag T\eta')\times \abar]_\al\ep(\xf)
=0\com
\nn
2\abar\pl_5(\del(\xf)-\del(\xf-l))
-g^2{\ep(\xf)}(\vpbar\times(\abar\times\vpbar))_\al  \nn
-g^2[(\del(\xf)\eta^\dag T\eta
+\del(\xf-l){\eta'}^\dag T\eta')\times \vpbar]_\al\ep(\xf)
=0\com
\nn
\chi^3_\al+g(
\del(\xf)\eta^\dag T^\al\eta+\del(\xf-l){\eta'}^\dag T^\al\eta'
             )=0\com
\nn
g\{\chi^3_\be-2\vpbar_\be (\del(\xf)-\del(\xf-l))
+g(\abar\times\vpbar)_\be\}(T^\be\eta)_\alp=0\com
\nn
g\{\chi^3_\be-2\vpbar_\be (\del(\xf)-\del(\xf-l))
+g(\abar\times\vpbar)_\be\}(T^\be\eta')_\alp=0
\ .
\label{C2}
\end{eqnarray}
We note the following things.
\begin{enumerate}
\item
When $\abar_\al\propto\vpbar_\al$, the following relations hold:\ 
$(\abar\times\vpbar)_\al=f_{\ab\ga}\abar_\be\vpbar_\ga=0$.
\item
$\pl_5(\del(\xf)-\del(\xf-l))\times\mbox{const}=0$ with the Neumann boundary
condition:\ $\pl_5(\del A^5_\al)|_{\xf=0}=\pl_5(\del A^5_\al)|_{\xf=l}=0$.
\item
$\ep(\xf)^2=1, \ep(\xf)^3=\ep(\xf), 
\pl_5(\ep(\xf))=2(\del(\xf)-\del(\xf-l))$, 
$\half\pl_5\{\ep(\xf)^2\}=(\del(\xf)-\del(\xf-l))\ep(\xf)=0$
.
\end{enumerate}
Then we can conclude that (\ref{C1}) is a solution of the field
equation (\ref{ep5b}) for the following choice.
\begin{eqnarray}
\mbox{const}\times \abar_\al=\vpbar_\al
=-\frac{g}{2}\eta^\dag T^\al\eta
=\frac{g}{2}{\eta'}^\dag T^\al\eta'
\com\nn
\chi^3_\al=-g(\del(\xf)-\del(\xf-l))\eta^\dag T^\al\eta
\pr
\label{C3}
\end{eqnarray}
In this choice $d_\al=0$ is concluded. 
The more general solution is given in \cite{IMplb596}. 


\end{document}